\documentclass{svmult}
\usepackage{graphicx}     
\begin{document}

\title*{Quantum and Arithmetical Chaos}

\author{Eugene Bogomolny}

\institute{ Laboratoire de Physique Th\'eorique et Mod\`eles Statistiques\\
Universit\'e de Paris XI, B\^at. 100,
91405 Orsay Cedex, France\\
\texttt{bogomol@lptms.u-psud.fr}}

\maketitle

\begin{abstract}
The lectures are centered around three selected topics of quantum chaos:
 the Selberg trace formula, the two-point spectral correlation functions 
 of Riemann zeta function zeros, and of the Laplace--Beltrami operator for 
 the modular group. 
The lectures cover a wide range of quantum chaos applications and can 
serve as a non-formal introduction to mathematical methods of quantum
chaos.
\end{abstract}


\section*{Introduction}\label{Introduction}

Quantum chaos is a nickname for the investigation of quantum systems which
do not permit exact solutions. The absence of explicit formulas means that
underlying problems are so complicated that they cannot be expressed in
terms of known ($\simeq$ simple) functions.
The class of non-soluble systems is very large and practically any model
(except a small set of completely integrable systems) belongs to it. An
extreme case of quantum non-soluble problems appears naturally when one
considers the quantization of classically chaotic systems which explains the
word `chaos' in the title.

As, by definition, for complex systems exact solutions are not possible,
new analytical approaches were developed within the quantum chaos.
First, one may find relations between different non-integrable models,
hoping that for certain questions a problem will be more tractable than 
another. Second, one considers, instead of exact quantities, the 
calculation of their
smoothed values. In many cases such coarse graining appears naturally in
experimental settings and, usually, it is more easy to treat.
Third, one tries to understand statistical properties of quantum
quantities by organizing them in suitable ensembles. An advantage of such
approach is that many different models may statistically be
indistinguishable which leads to the notion of statistical universality. 

The ideas and methods of quantum chaos are not restricted  only to quantum 
models. They
can equally well be applied to any problem whose analytical solution either is
not possible or is very complicated. One of the most spectacular examples of
such interrelations is  the application of quantum chaos to number theory, 
in particular, to the zeros of the Riemann zeta function. Though a
hypothetical quantum-like system whose eigenvalues  coincide with the
imaginary part of  Riemann zeta function zeros is not (yet!) found,
the Riemann zeta function is, in many aspects, similar to dynamical zeta
functions and the investigation of such relations already mutually enriched
both quantum chaos and number theory (see e.g. the calculation by Keating and
Snaith moments of the Riemann zeta function using the random matrix theory 
\cite{KeatingSnaith}). 

The topics of these lectures were chosen specially to emphasize the
interplay between physics and mathematics which is typical for quantum chaos.

In Chap.~\ref{Trace} different types of trace formulas are discussed. The
main attention is given to the derivation of the Selberg trace formula which
relates the spectral density of automorphic Laplacian on hyperbolic surfaces
generated by discrete groups with classical periodic orbits for the free
motion on these surfaces. This question is rarely discussed in the physical
literature but is of general interest because it is the only case where the
trace formula is exact and not only a leading semiclassical contribution as
for general dynamical systems. Short derivations of trace formulas for dynamical
systems and for the Riemann zeta function zeros are also presented in this Chapter.

According to the well-known conjecture \cite{BohigasGiannoni} statistical 
properties of eigenvalues of energies of quantum chaotic systems are described 
by standard random matrix ensembles depending only on system symmetries. In
Chap.~\ref{Statistics} we discuss  analytical methods of
confirmation of this conjecture. The largest part of this Chapter is devoted
to a heuristic derivation of the `exact' two-point correlation function for
the Riemann zeros. The derivation is based on the Hardy--Littlewood
conjecture about the distribution of prime pairs which is also reviewed. The
resulting formula agrees very well with numerical calculations of Odlyzko.

In Chap.~\ref{Arithmetic} a special class of dynamical systems is considered,
namely, hyperbolic surfaces generated by arithmetic groups.
Though from viewpoint of classical mechanics these models are the best known
examples of classical chaos, their spectral statistics are close to the
Poisson statistics typical for integrable models. The reason for this
unexpected behavior is found to be related with exponential degeneracies of
periodic orbit lengths characteristic for arithmetical systems. The case of
the modular group is considered in details and the exact
expression for the two-point correlation function for this problem is derived. 

To be accessible for physics students the lectures are written in a
non-formal manner. In many cases analogies are used instead of theorems and
complicated mathematical notions are illustrated by simple examples.

\chapter{Trace Formulas}\label{Trace}

Different types of trace formula are the cornerstone of quantum chaos. Trace
formulas relate quantum properties of a system with its classical counterparts. 
In the simplest and widely used case the trace formula expresses the quantum 
density of states through a sum over periodic orbits and each term in this  
sum  can be calculated from pure classical mechanics.  

In general, dynamical trace formulas represent only the leading term of the
semiclassical expansion in  powers of $\hbar$. The computation of other
terms is possible though quite tedious \cite{Gaspard}. The noticeable exception
is the free motion on constant negative curvature surfaces generated by discrete 
groups where the trace formula (called the Selberg trace formula) is exact. The
derivation of this formula is the main goal of this Section. 

For clarity, in Sect.~\ref{Rectangular} the simplest case of the rectangular
billiard is briefly considered and the trace formula for this system is
derived. The derivation is presented in a manner which permits to generalize
it to the Selberg case of constant negative curvature surfaces generated by 
discrete groups which is considered in details in Sect.~\ref{Negative}. 
In Sects.~\ref{Integrable} and \ref{Chaotic} the derivations of the trace 
formula for, respectively,  classically integrable and chaotic systems are  
presented.  
In Sect.~\ref{Riemann} it is demonstrated that the density of Riemann zeta 
function zeros can be written as  a sort of  trace formula where the role
of  periodic orbits is played by prime numbers. Section~\ref{SummaryTrace} 
is the summary of this Chapter.

\section{Plane Rectangular Billiard}\label{Rectangular}

To clarify the derivation of trace formulas let us consider in details a
very simple example, namely, the computation of the
energy spectrum for the plane rectangular billiard with periodic boundary
conditions. 

This problem consists of solving the equation
\begin{equation}
(\varDelta +E_{\vec{n}})\Psi_{\vec{n}} (x,y)=0
\label{1}
\end{equation}
where $\varDelta=\partial^2/\partial x^2+\partial^2/\partial y^2$ is the usual
two-dimensional Laplacian with periodic boundary conditions  
\begin{equation}
\Psi_{\vec{n}} (x+a,y)=\Psi_{\vec{n}} (x,y+b)=\Psi_{\vec{n}} (x,y)
\label{2}
\end{equation}
where  $a$ and $b$ are sizes of the rectangle.

The plane wave 
$$
\Psi_{\vec{n}} (x,y)=e^{\imag k_1x+\imag k_2y}
$$
is an admissible solution of (\ref{1}). Boundary conditions (\ref{2}) determine
the allowed values of the momentum $\vec{k}$
$$
k_1=\frac{2\pi}{a}n_1,\;\;\;\;k_2=\frac{2\pi}{b}n_2\;,
$$
with $n_1,n_2=0,\pm 1,\pm 2,\ldots $, and, consequently,  energy eigenvalues are
\begin{equation}
E_{n_1n_2}=\left (\frac{2\pi}{a}n_1\right )^2+
\left (\frac{2\pi}{b}n_2\right )^2\;.
\label{energy}
\end{equation}
The first step of construction of trace formulas is to consider instead of
individual eigenvalues their density defined as the sum over all eigenvalues 
which explains the word `trace'
\begin{equation}
d(E)\equiv \sum_{n_1,n_2=-\infty}^{+\infty} \delta (E-E_{n_1 n_2})\;.
\label{density}
\end{equation}
To transforms this and similar expressions into a convenient form one  
often uses the Poisson summation formula 
\begin{equation}
\sum_{n=-\infty}^{+\infty}f(n)=\sum_{m=-\infty}^{+\infty}
\int_{-\infty}^{+\infty} \E^{2\pi \imag m n}f(n)\D n\;.
\label{Poisson}
\end{equation}
An informal proof of this identity can, for example, be done as follows.

First
$$
\sum_{n=-\infty}^{+\infty}f(n)=
\int_{-\infty}^{+\infty}f(x)g(x)\D x
$$
where $g(x)$ is the periodic $\delta$-function
$$
g(x)= \sum_{m=-\infty}^{+\infty}\delta(x-n)\;.
$$
As any periodic function with period 1, $g(x)$ can be expanded into the
Fourier series
$$
g(x)=\sum_{m=-\infty}^{+\infty}\E^{2\pi \imag mx}c_m\;.
$$
Coefficients $c_m$ are obtained by the integration of $g(x)$ over one period
$$
c_m=\int_{-1/2}^{+1/2}g(y)\E^{-2\pi \imag m y}\D y=1
$$
which gives (\ref{Poisson}).

By applying the Poisson summation formula (\ref{Poisson}) to the density of
states (\ref{density}) one gets
\begin{eqnarray*}
 d(E)&=&\sum_{m_1,m_2=-\infty}^{+\infty} \int \int  
\E^{2\pi \imag (m_1n_1+m_2n_2)}\times \\
&\times&\delta \left (E-\left (\frac{2\pi}{a}n_1\right )^2-
\left (\frac{2\pi}{b}n_2\right )^2\right )\D n_1\D n_2\;.
\nonumber
\end{eqnarray*}
Perform the following substitutions:  $E=k^2$,
$n_1=a r\cos \varphi/2\pi$, and $n_2=br\sin \varphi/2\pi$.
Then $\D n_1 \D n_2=ab r\D r \D \varphi/(2\pi)^2$ and
\begin{eqnarray*}
d(E)&=&\frac{\mu(D)}{(2\pi)^2}\sum_{m_1,m_2=-\infty}^{+\infty} 
\int \int \E^{\imag (m_1a\cos \varphi+m_2b\sin \varphi)r}\delta (k^2-r^2)r\D r
\D \varphi
\nonumber\\
&=&\frac{\mu(D)}{2(2\pi)^2}\sum_{m_1,m_2=-\infty}^{+\infty}
\int_0^{2\pi} \E^{\imag k\sqrt{(m_1a)^2+(m_2b)^2}\cos \varphi}\D \varphi
\nonumber\\
&=&\frac{\mu(D)}{4\pi}\sum_{m_1,m_2=-\infty}^{+\infty} J_0(k L_p)\;,
\end{eqnarray*}
where $\mu(D)=ab$ is the area of the rectangle,
$$
J_0(x)=\frac{1}{2\pi}\int_0^{2\pi} \E^{\imag x\cos \varphi}\D \varphi
$$
is the Bessel function of  order  zero (see e.g. \cite{Bateman}, 
Vol. 2, Sect. 7), and
$$
L_p=\sqrt{(m_1a)^2+(m_2b)^2}
$$
is (as it is easy to check) the length of a periodic orbit in the rectangle
with periodic boundary conditions.

Separating the term with $m_1=m_2=0$ one concludes that the eigenvalue density
of the rectangle with periodic boundary conditions can be written as the
sum of two terms
$$
d(E)=\bar{d}(E)+d^{(osc)}(E)\;,
$$
where
\begin{equation}
\bar{d}(E)=\frac{\mu(D)}{4\pi}
\label{smooth}
\end{equation}
is the smooth part of the density and
\begin{equation}
d^{(osc)}(E)=\frac{\mu(D)}{4\pi}\sum_{\mbox{p.o.}} J_0(k L_p)\;,
\label{osc}
\end{equation}
is the oscillating part equal to a sum over all periodic orbits in the
rectangle.

As
$$
J_0(z)\stackrel{z\to\infty}{\longrightarrow}\sqrt{\frac{2}{\pi z}}
\cos \left (z-\frac{\pi}{4}\right )
$$
the oscillating part of the level density in the semiclassical limit
$k\to\infty$  takes the form
\begin{equation}
d^{(osc)}(E)=\frac{\mu(D)}{\sqrt{8\pi k}}\sum_{\mbox{p.o.}}
\frac{1}{\sqrt{L_p}}\cos \left (kL_p-\frac{\pi}{4}\right )\;.
\label{asymptotic}
\end{equation}
Let us repeat the main steps which lead to this trace formula. One starts
with an explicit formula (like (\ref{energy})) which expresses eigenvalues
as a function of integers. Using the Poisson summation formula
(\ref{Poisson}) the density of states (\ref{density}) is transformed into a
sum over periodic orbits. In Sect.~\ref{Integrable} it will be demonstrated
that exactly this method can be applied for any integrable systems in the
semiclassical limit where eigenvalues can be approximated by the WKB formulas.

\paragraph{More Refined Approach}

The above method of deriving the trace formula for the rectangular billiard
can be applied only if one knows an explicit expression for eigenvalues. For
chaotic systems this is not possible and another method has to be used. 

Assume that one has to solve the equation
$$
(E_{\vec{n}}-\hat{H})\Psi_{\vec{n}}(\vec{x})=0
$$
for a certain problem with a Hamiltonian $\hat{H}$. Under quite general 
conditions eigenfunctions $\Psi_{\vec{n}}(\vec{x})$ can be chosen orthogonal 
$$
\int \Psi_{\vec{n}}(\vec{x})\Psi_{\vec{m}}^*(\vec{x})\D \vec{x}=\delta_{nm}
$$
and they form a complete system of functions
$$
  \sum_{\vec{n}}\Psi_{\vec{n}}(\vec{x})\Psi_{\vec{n}}^*(\vec{y})=
  \delta (\vec{x}-\vec{y})\;.
$$
The Green function of the problem, by definition, obeys the equation
$$
(E-\hat{H})G_E(\vec{x},\vec{y})=\delta(\vec{x}-\vec{y})
$$
and the same boundary conditions as the original eigenfunctions. Its
explicit form  can formally be written through exact eigenfunctions and
eigenvalues as follows
\begin{equation}
G_E(\vec{x},\vec{y})=\sum_{\vec{n}}\frac{\Psi_{\vec{n}}(\vec{x})
\Psi_{\vec{n}}^*(\vec{y})}{E-E_n+\imag \epsilon}\;.
\label{exactGreen}
\end{equation}
The $+\imag \epsilon$ prescription determines the so-called retarded Green
function.

\paragraph{Example}

To get used to Green functions let us consider in details the calculation of
the Green function for the free motion in $f$-dimensional Euclidean space. 
This Green function obeys the free equation 
\begin{equation}
(E+\hbar^2\varDelta)G_E^{(0)}(\vec{x},\vec{y})=\delta (\vec{x}-\vec{y})\;.
\label{FreeEquation}
\end{equation}
Let us look for the solution of the above equation in the form
$G_E^{(0)}(\vec{x},\vec{y})=G(r)$ where $r=|\vec{x}-\vec{y}|$ is the 
distance between two points.

Simple calculations shows that for $r\neq 0$ $G(r)$ obeys the equation
$$
\frac{\D^2 G}{\D r^2}+\frac{f-1}{r}\frac{\D G}{\D r}+\frac{k^2}{\hbar^2} G=0
$$
where $E=k^2$.

After the substitution 
$$
G(r)=r^{1-f/2}g \left ( \frac{k}{\hbar} r \right )
$$
one gets for $g(z)$ the Bessel equation (see e.g. \cite{Bateman}, Vol. 2, Sect. 7)
\begin{equation}
\frac{\D^2 g}{\D z^2}+\frac{1}{z}\frac{\D g}{\D z}+
\left (1-\frac{\nu^2}{z^2}\right )g=0
\label{BesselEquation}
\end{equation}
with $\nu=|f/2-1|$.

There are many solutions of this equation. The above $+\imag \epsilon$ prescription
means that when $k\to k+i\epsilon$ with a positive $\epsilon$ the Green function has
to decrease at large distances. It is easy to see that   $G(r)$ is
proportional to $\E^{\pm \imag kr/\hbar}$ at large $r$. The $+i\epsilon$ prescription
selects a solution which behaves at infinity like $\E^{+\imag k r/\hbar}$ with 
positive $k$. The required solution of (\ref{BesselEquation}) is the first
Hankel function (see \cite{Bateman}, Vol. 2, Sect. 7)
\begin{equation}
g(z)=C_f H_{\nu}^{(1)}(z)
\label{gz}
\end{equation}
where $C_f$ is a constant and $H_{\nu}^{(1)}(z)$ has the following 
asymptotics for large and small $z$
$$
H_{\nu}^{(1)}(z)\stackrel{z\to \infty}{\longrightarrow} 
\sqrt{\frac{2}{\pi z}}\E^{\imag (z-\pi\nu/2-\pi/4)}
$$
and 
$$
H_{\nu}^{(1)}(z)\stackrel{z\to 0}{\longrightarrow} \left \{
\begin{array}{cl}-\imag 2^{\nu}\Gamma (\nu )  z^{-\nu}/\pi\;,&\;\nu\neq 2\\ 
2\imag \ln z/\pi \;,&\;\nu=2 \end{array}\right .\;.
$$ 
The overall factor in (\ref{gz}) has to be computed from the requirement 
that the Green function will give the correct $\delta$-function contribution
in the right hand side of (\ref{FreeEquation}).  This term can appear only
in the result of differentiation of the Green function at small $r$ where it
has the following behaviour
$$
G(r)\stackrel{r\to 0}{\longrightarrow}G_0(r)= A_f r^{2-f}
$$
with
$$
A_f=C_f \frac{2^{\nu}\hbar^{\nu}\Gamma (\nu)}{\imag \pi k^{\nu}}\;.
$$
One should have 
\begin{equation}
\hbar^2\varDelta G_0(r)=\delta (\vec{r})\;.
\label{Gzero}
\end{equation}
Multiplying this equality by a suitable test function $f(r)$ quickly decreasing
at infinity one has
$$
\hbar^2\int f(r)\varDelta G_0(r)\D \vec{r}=f(0)\;.
$$
Integrating by parts one obtains
$$
\hbar^2 \int \frac{\partial}{\partial x_\mu} f(r)
\frac{\partial}{\partial x_\mu}G_0(r)\D \vec{r}=-f(0)\;.
$$
As both functions $f(r)$ and $G_0(r)$ depend only on the modulus of
$\vec{r}$ one finally finds
$$
\hbar^2\int_0^{\infty} \frac{\D f(r)}{\D r} 
\frac{\D G_0(r)}{\D r}r^{f-1}\D r S_{f-1}=-f(0)
$$
where $S_{f-1}$ is the volume of the $(f-1)$-dimensional sphere
$x_1^2+\ldots+x_f^2=1$. Using (\ref{Gzero}) one concludes that in order
to give the $\delta$-function term $A_f$ has to obey
$$
\hbar^2 A_f(f-2)S_{f-1}=-1\;.
$$
One of the simplest method of calculation of $S_{f-1}$ is the
following identity 
$$
\int_{-\infty}^{\infty}\E^{-x_1^2}\D x_1 
\int_{-\infty}^{\infty}\E^{-x_2^2}\D x_2\ldots  
\int_{-\infty}^{\infty}\E^{-x_f^2}\D x_f=\pi^{f/2}\;.
$$
By changing  Cartesian coordinates in the left hand side to
hyper-spherical ones  we obtain
$$
\int_0^{\infty}\E^{-r^2}r^{f-1}\D r S_{f-1}=\pi^{f/2}
$$
which gives
$$
S_{f-1}=\frac{2\pi^{f/2}}{\Gamma (f/2)}
$$
where $\Gamma(x)$ is the usual gamma-function 
(see e.g. \cite{Bateman}, Vol.~1, Sect.~1).

Combining together all terms and using the relation 
$x\Gamma (x)=\Gamma(x+1)$ one gets the explicit expression for the free
Green function in $f$ dimensions
\begin{equation}
G^{(0)}_E(\vec{x},\vec{y})=
\frac{k^{\nu}}{4i\hbar^2 (2\pi \hbar r)^{\nu}} 
H_{\nu}^{(1)}\left (\frac{k}{\hbar}|\vec{x}-\vec{y}|\right )
\label{GreenFdimensions}
\end{equation}
where $\nu=|f/2-1|$. In particular, in the two-dimensional Euclidean space 
\begin{equation}
G^{(0)}_E(\vec{x},\vec{y})=\frac{1}{4i\hbar^2} 
H_0^{(1)}\left (\frac{k}{\hbar}|\vec{x}-\vec{y}|\right )\;.
\label{Gfree}
\end{equation}
Another method of calculation of the free Green function is based on
(\ref{exactGreen}) which for the free motion is equivalent to the Fourier
expansion 
\begin{equation}
G_E^{(0)}(\vec{x},\vec{y})=\int \frac{\D \vec{p}}{(2\pi\hbar)^f}
\frac{\E^{\imag \vec{p}(\vec{x}-\vec{y})/\hbar }}{E-p^2+\imag \epsilon}\;.
\label{GreenFourier}
\end{equation}
Performing angular integration one obtains the same formulas as above.

\vspace{3em}
 
The knowledge of the Green function permits to calculate practically all
quantum mechanical quantities. In particular, using 
$$
\mbox{Im }\frac{1}{x+\imag \varepsilon}
\stackrel{\varepsilon \to 0}{\longrightarrow} - \pi \delta (x)
$$
one gets that the eigenvalue density is expressed through the exact Green
function as follows
\begin{equation}
d(E)=-\frac{1}{\pi} \mbox{Im} \int_DG_E(\vec{x},\vec{x})\D \vec{x}\;.
\label{ImGreen}
\end{equation}
This general expression is the starting point of all trace formulas.

For the above model of the rectangle with periodic boundary
conditions the exact Green function has to obey
\begin{equation}
(\frac{\partial^2}{\partial x^2}+\frac{\partial^2}{\partial y^2} +E)
G_E(x,y;x',y')=\delta(x-x')\delta(y-y')
\label{GreenEquation}
\end{equation}
and the periodic boundary conditions
\begin{equation}
G_E(x+na,y+mb;x',y')=G_E(x,y;x',y')
\label{GreenPeriodic}
\end{equation}
for all integer $m$ and $n$.

The fact important for us later is that the rectangular billiard with
periodic boundary conditions can be considered as the result of the
factorization of the whole plane $(x,y)$ with respect to the group of integer
translations
\begin{equation}
x\to x+na,\;\;y\to y+mb
\label{translations}
\end{equation}
with integer $m$ and $n$.

The factorization of the plan $(x,y)$ with respect to these transformations
means two things. First, any two points connected by a group
transformation is considered as one point. Hence (\ref{GreenPeriodic})
fulfilled. Second, inside the rectangle there is no points which are
connected by these transformations. In mathematical language the rectangle
with sizes $(a,b)$ is the fundamental domain of the group (\ref{translations}).

Correspondingly, the exact Green function for the rectangular billiard with
periodic boundary conditions  equals the sum of the free Green 
function over all elements of the group of integer translations
(\ref{translations})
$$
G_E(x,y;x',y')=\sum_{n,m=-\infty}^{\infty} G^{(0)}_E(x+na,y+mb;x',y')\;.
$$
Here $G^{(0)}_E(\vec{x},\vec{x}')$ is the Green function corresponding to
the free motion without periodic boundary conditions. To prove formally
that it is really the exact Green function one has to note that (i) it obeys
(\ref{GreenEquation}) because each term in the sum obeys it, (ii) it obeys
boundary conditions (\ref{GreenPeriodic}) by construction (provided the
sum converges), and (iii) inside the initial rectangle only identity term
can produce a $\delta$-function contribution required in
(\ref{GreenEquation}) because all other terms will give  $\delta$-functions
outside the rectangle.   

The next steps are straightforward. The free Green function for the
two-dimensional Euclidean plane has the form (\ref{Gfree}). 
From (\ref{ImGreen}) it follows that  the eigenvalue density for
the rectangular billiard is
\begin{eqnarray}
d(E)&=&-\frac{1}{\pi}\mbox{Im } \int_DG_E(\vec{x},\vec{x})\D \vec{x}=\nonumber\\
&=&\frac{1}{4\pi} \sum_{mn} \int_D \mbox{Im } 
H_0^{(1)}\left (k\sqrt{(ma)^2+(nb)^2}\right )\D \vec{x}=\nonumber\\
&=&\frac{\mu(D)}{4\pi}+\frac{\mu(D)}{4\pi}
\sum_{\mbox{\scriptsize{p.o.}}}\nolimits^\prime J_0(kL_p)
\label{traceRectangle}
\end{eqnarray}
which coincides  exactly with (\ref{smooth}) and (\ref{osc}) obtained 
directly from the knowledge of the eigenvalues.

\begin{figure}
\center
\includegraphics[height=8cm, angle=-90]{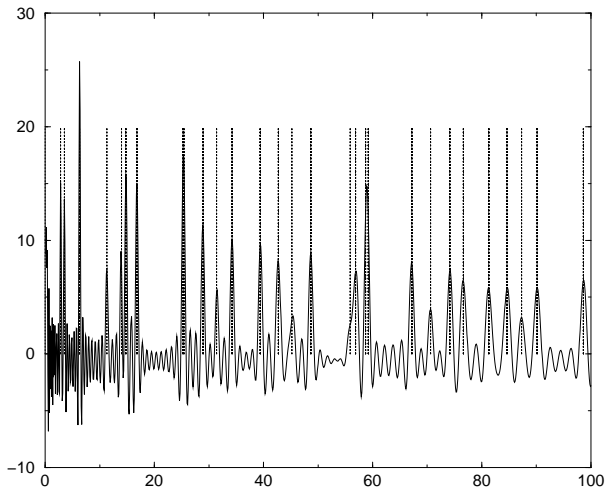}
\caption{
The trace formula for the rectangular billiard with periodic boundary
conditions calculated by taking into account 250 different periodic orbits. 
Dotted lines indicate the position of exact energy levels.}
\label{example}
\end{figure}

The principal drawback of all trace formulas is that the sum over periodic
orbits does not converge. Even the sum of the squares diverges. The simplest
way to treat this problem is to multiply both sides of
(\ref{traceRectangle}) by a suitable test function $h(E)$ and integrate them 
over $E$. In this manner one obtains
$$
\sum_n h(E_n)=\frac{\mu(D)}{4\pi} \int_0^{\infty} h(E)dE+
\frac{\mu(D)}{4\pi} \sum_{\mbox{\scriptsize{p.o.}}}\int_0^{\infty}
h(E)J_0(\sqrt{E}L_p) \D E\;.
$$
When the Fourier harmonics of $h(E)$ decrease quickly the sum over periodic
orbits converges and this expression constitutes  a mathematically well 
defined trace formula. Nevertheless for approximate calculations of
eigenvalues of energies one can still use `naive' trace formulas by
introducing a cut-off on periodic orbit sum. For example, in 
Fig.~\ref{example} the result
of numerical application of the above trace formula is presented.    
In performing this calculation one uses the asymptotic form of the
oscillating part of the density of state (\ref{asymptotic}) with only 250
first periodic orbits. Though additional oscillations are clearly seen, one
can read off this figure the positions of first energy levels for the
problem considered. In the literature many different methods of resummation
of trace formulas were discussed (see e.g. \cite{resummation} and references therein).

\section{Billiards on Constant Negative Curvature Surfaces}\label{Negative}

The crucial point in the second method of derivation of the trace 
formula for the rectangular billiard with periodic boundary conditions 
was a representation of the exact Green function as a
sum of a free Green function over all images of the initial point. 
This method of images can be applied for any problem which corresponds to a
factorization of a space over the action of a discrete group. In the
Euclidean plane (i.e. the space of zero curvature) there exist only a few 
discrete groups. Much more different discrete groups are possible in the 
constant negative curvature (hyperbolic) space. Correspondingly, one can 
derive the trace formula (called the Selberg trace formula) for all
hyperbolic surfaces generated by discrete groups. 

The exposition of this Section follows closely \cite{Bogomolny1}. In
Sect.~\ref{Hyperbolic} hyperbolic geometry is non-formally discussed. The
important fact is that on hyperbolic plane there exist an infinite
number of discrete groups (see e.g. \cite{Katok}). Their properties are
mentioned in Sect.~\ref{Discrete}. In Sect.~\ref{Classical} the classical 
mechanics on hyperbolic surfaces is considered and in Sect.~\ref{Quantum}
the notion of quantum problems on such surfaces is introduced. The 
construction of the Selberg  trace formula for hyperbolic surfaces generated
by discrete groups
consists of two steps. The first is the explicit calculation of the free
hyperbolic Green function performed in Sect.~\ref{Construction}. The second step
includes the summation over all group transformations. In Sect.~\ref{Density}
it is demonstrated that the identity group element gives  the mean density of
states. Other group elements contribute  to the oscillating part of the level density
and correspond to classical periodic orbits for the motion on systems
considered. The relation between group elements and periodic orbits is not
unique. All conjugated matrices correspond to one periodic orbit. The
summation over classes of conjugated elements is done in
Sect.~\ref{Conjugated}. Performing necessary integrations in
Sect.~\ref{Selberg} one gets the famous Selberg trace formula. Using this formula 
in Sect.~\ref{Huber} we compute the asymptotic density of periodic orbits
for discrete groups. In Sect.~\ref{SelbergZeta} the construction of the
Selberg zeta function is presented. The importance of this function follows 
from the fact that its non-trivial zeros coincide with eigenvalues of
the Laplace--Beltrami operator automorphic with respect to a discrete
group (see Sect.~\ref{SelbergZeros}). Though the Selberg zeta function is
defined formally only in a part of the complex plan, it obeys a functional
equation (Sect.~\ref{FunctionalSelberg}) which  permits the
analytical continuation to the whole complex plane.

\subsection{Hyperbolic Geometry}\label{Hyperbolic}

\begin{figure}
\center
\includegraphics[height=7cm, angle=-90]{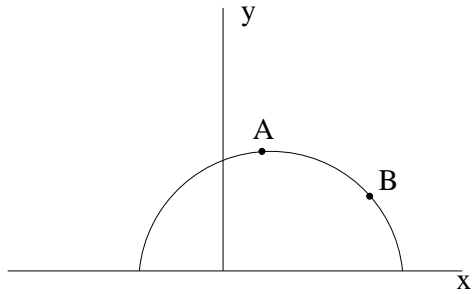}
\caption{The Poincar\'e model of constant negative curvature space. Solid 
line indicates the geodesic passing through points A and B.}
\label{poincare}
\end{figure}

The standard representation  of the constant  negative curvature space is the 
Poincar\'e upper half plane model $(x,y)$ with $y>0$ (see e.g. \cite{Balatz}
and \cite{Katok}) with the following metric form 
$$
\D s^2=\frac{1}{y^2}(\D x^2+\D y^2)\;.
$$
The geodesic in this space (= the straight line) connecting two points is
the arc of circle perpendicular to the abscissa axis which passes through
these  points (see Fig.~\ref{poincare}). 
The distance $d(\vec{x},\vec{y})$ between two points
$\vec{x}=(x_1,y_1)$ and $\vec{y}=(x_2,y_2)$ is defined as the length of the 
geodesic connecting these points. Explicitly
\begin{equation}
\cosh d(\vec{x},\vec{y})=1+\frac{(x_1-x_2)^2+(y_1-y_2)^2}{2y_1y_2}=
1+\frac{|z_1-z_2|^2}{2\mbox{Im } z_1 \mbox{Im }z_2}
\label{distance}
\end{equation}
where in the last equation one combined coordinates $(x,y)$ into a complex 
number $z=x+iy$. 

In the Euclidean plane the distance between two points remains invariant
under 3-parameter group of rotations and translations. For constant negative
curvature space the distance (\ref{distance}) is invariant under fractional 
transformations
\begin{equation}
z\to z'=g(z)\equiv \frac{az+b}{cz+d}
\label{fract}
\end{equation}
with real parameters $a,b,c,d$. This invariance follows from the following
relations
$$
z_1'-z_2'=\frac{az_1+b}{cz_1+d}-\frac{az_2+b}{cz_2+d}=
(ad-bc)\frac{z_1-z_2}{(cz_1+d)(cz_2+d)}\;,
$$
and
$$
y'=\frac{1}{2i}(z'-z'^{*})=(ad-bc)\frac{y}{|cz+d|^2}\;.
$$
Substituting these expressions to (\ref{distance}) one concludes that the
distance between two transformed points $z_1',z_2'$ is the same as between
initial points $z_1,z_2$.

As fractional transformations are not changed under the multiplication of
all elements $a,b,c,d$ by a real factor, one can normalize them by the
requirement
$$
ad-bc=1\;.
$$
In this case the distance preserving transformations are described by 
$2\times 2$ matrices with real elements and  unit determinant
$$
g=\left (\begin{array}{cc} a&b\\c&d\end{array}\right ) ,\;\;\mbox{and } \;
\det g\equiv ad-bc=1\;.
$$
It is easy to check that the result of two successive fractional
transformations (\ref{fract}) corresponds to the usual multiplication of the
corresponding matrices.

The collection of all such  matrices forms a group called
the projective special linear group over reals and it is denoted by
PSL(2,$\mbox{I\hspace{-.15em}R}$). `Linear' in the name means that it is a
matrix group, `special' indicates that the determinant equals 1, and
`projective' here has to remind that fractional transformations (\ref{fract})
are not changed when all elements are multiplied by $\pm 1$ which is
equivalent that two matrices $\pm {\bf 1}$ corresponds to the identity element
of the group.

The free classical motion on the constant negative curvature surface is defined 
as  the motion along geodesics (i.e. circles perpendicular to the abscissa axis). 
The measure invariant under fractional transformations is the following
differential  form
\begin{equation}
\D \mu=\frac{\D x\D y}{y^2}\;.
\label{muinvariant}
\end{equation}
This measure is invariant in the sense that if two regions, $D$ and $D'$,
are related by a transformation (\ref{fract}), $D'=g(D)$, measures of these
two regions are equal, $\mu(D')=\mu(D)$. 

The operator invariant with respect to distance preserving
transformations (\ref{fract}) is called the Laplace--Beltrami operator 
and it has the following form
\begin{equation}
\varDelta_{LB}=y^2
\left (\frac{\partial^2}{\partial x^2}+\frac{\partial^2}{\partial
y^2}\right )\;.
\label{LB}
\end{equation}
Its invariance means that
$$
\varDelta_{LB}f(g(z))=g(\varDelta_{LB}f(z))
$$
for any fractional transformation $g(z)$. 

Practically all notions used for the Euclidean space can be translated to 
the constant negative curvature case (see e.g. \cite{Balatz}). 

\subsection{Discrete groups}\label{Discrete}

A rectangle (a torus) considered in Sect.~\ref{Rectangular} was the 
result of the factorization of the free motion on the plane by
a discrete group of translations (\ref{translations}). Exactly in the same 
way one can construct
a finite constant negative surface by factorizing the upper half plane by
the action of a discrete group $\in PSL(2,$\mbox{I\hspace{-.15em}R}$)$. 

A group is discrete if (roughly speaking) there is a finite vicinity of
every point of our space such that the results of all the group
transformations (except the identity) lie outside this vicinity.
The images of a point cannot approach each other too close. 

\paragraph{Example} 
The group of transformation of the unit circle into itself. The
group consists of all transformations of the following type
$$
z\to g(n)z,\;\;\;\;\mbox{and}\;\;\;g(n)=\exp (2\pi \imag \alpha n)\;,
$$
where $\alpha $ is a constant and $n$ is an integer. If $\alpha$ is a
rational number $\alpha=M/N$, $g(n)$ can take only a finite number of values
$(g(n))^N=1$ and the corresponding group is discrete. But if $\alpha$ is an
irrational number, the images of any point cover the whole circle uniformly
and the group is not discrete. 

\vspace{3em}

\paragraph{Modular Group}

Mathematical fact: in the upper half plane there exists an infinite number
of discrete groups (see e.g. \cite{Katok}). As  an example let us consider
the group of $2\times 2$ integer matrices with  unit determinant
$$
\left (\begin{array}{cc} m&n\\k&l\end{array}\right ),\;\;\;m,n,k,l\;\mbox{are
    integers and }\;ml-nk=1\;.
$$
This is evidently a group. It is  called the modular group
PSL(2,$\mbox{Z\hspace{-.3em}Z}$) ($\mbox{Z\hspace{-.3em}Z}$ means integers) 
and it is one of the most investigated groups in mathematics.

This group is generated by the translation $T:\;z\to z+1$ and the inversion
$S:\;z\to -1/z$ (see e.g. \cite{Katok})  which are represented by the 
following matrices
$$
  T: \left (\begin{array}{cc} 1&1\\0&1\end{array}\right ),\;\;
S: \left (\begin{array}{cc} 0&1\\-1&0\end{array}\right )\;.
$$
These matrices obey defining relations
$$
S^2=-1\;,\;(ST)^3=1
$$
and  are generators in the sense that any modular group matrix can
be represented as a product of a certain sequence of matrices corresponding to 
$S$ and $T$.

\paragraph{Fundamental Region}
  
Similarly to the statement that the rectangular billiard is a fundamental
domain of integer translations, one can construct a fundamental domain for
any discrete group.

By definition the fundamental domain of a group is defined as a region on the 
upper half plane such that (i) for all points outside the fundamental domain 
there exists a group transformation that puts it to fundamental domain and 
(ii) no two points inside the fundamental domain are connected by group 
transformations.

\begin{figure}
\center
\includegraphics[height=7cm, angle=-90]{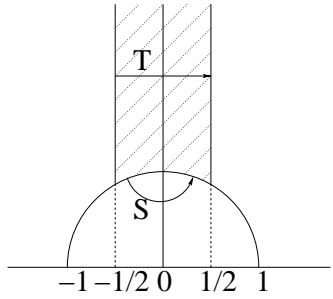}
\caption{Fundamental domain of the modular group. The indicated parts are
identified by the corresponding generators}
\label{modular}
\end{figure}

The fundamental domain for the modular group is presented in
Fig.~\ref{modular}. 
In general, the fundamental region of a discrete group has a shape of a
polygon built from segments of geodesics. Group generators  identify
corresponding sides of the polygon. 

\subsection{Classical Mechanics}\label{Classical}

Assume that we have a discrete group $G$ with corresponding matrices
$M\in G\; \in$ PSL(2,$\mbox{I\hspace{-.15em}R}$)
$$
M=\left (\begin{array}{cc} a&b\\c&d\end{array}\right )\;.
$$
The factorization over action of the group means that points $z$ and $z'$
where
\begin{equation}
z'=\frac{az+b}{cz+d}
\label{fractional}
\end{equation}
are identified i.e. they are considered as one point. The classical motion
on the resulting surface 
is the motion (with unit velocity) on geodesics (semi-circles perpendicular
to the real axis) inside the fundamental domain but when a trajectory hits a
boundary it reappears from the opposite side as prescribed by boundary
identifications. 

For each hyperbolic matrix $M\in G$ with $|\mbox{Tr }M|>2$  one can associate 
a periodic orbit defined as a geodesics
which remains invariant under the corresponding transformation. The equation
of such invariant geodesics has the form
\begin{equation}
c(x^2+y^2)+(d-a)x-b=0\;.
\label{hyperbolicPeriodic}
\end{equation}
This equation is  the only function which has the following property.
If $z=x+iy$ belongs to this curve then 
$$
z'=\frac{az+b}{cz+d}
$$  
also belongs to it.

The length of the periodic orbit is the distance along these geodesics
between a point and its image. Let $z'$ as above be the result of transformation
(\ref{fractional}) then the distance between $z$ and $z'$ is 
$$
\cosh l_p=1+\frac{|z-z'|^2}{2yy'}\;.
$$
But $y'=y/|cz+d|^2$ and
$$
z-\frac{az+b}{cz+d}=\frac{c(x+iy)^2-(d-a)(x+iy)-b}{cz+d}=
y\frac{-2cy+i(d-a+2cx)}{cz+d}\;.
$$
Here we  have used the fact that point $z$ belongs to the periodic orbit
(i.e. its coordinates obey (\ref{hyperbolicPeriodic})).
Therefore
\begin{eqnarray*}
\cosh l_p&=&1+\frac{1}{2}|-2cy+i(d-a+2acx)|^2=\\
&=&1+\frac{1}{2}[4bc+(d-a)^2]=\frac{1}{2}(a+d)^2-1\;.
\end{eqnarray*}
Notice that the length of periodic orbit does not depend on an initial point
and is a function only of the trace of the corresponding matrix. 

Finally one gets 
\begin{equation}
2\cosh \frac{l_p}{2}=|\mbox{Tr }M|\;.
\label{PeriodicLength}
\end{equation}
Periodic orbits are defined only for hyperbolic matrices with $|\mbox{Tr }M|>2$.
For discrete groups only a finite number of elliptic matrices with 
$|\mbox{Tr }M|<2$ can exist (see \cite{Katok}).
 
To each hyperbolic group matrix one can associate only one periodic orbit but 
each periodic orbit corresponds to infinite many group matrices. 
It is due to the
fact that $z$ and $g(z)$ for any group transformation has to be considered as
one point. Therefore all matrices in the form
$$
  SMS^{-1}
$$
for all $S\in G$ gives one periodic orbit. These matrices form a class of
conjugated matrices and periodic orbits of the classical motion are in
one-to-one correspondence with classes of conjugated matrices.

\subsection{Quantum Problem}\label{Quantum}

The natural `quantum' problem on hyperbolic plane consists in considering
the same equation as in (\ref{1}) but with the substitution of 
the invariant Laplace--Beltrami operator (\ref{LB}) instead of the usual
Laplace operator 
$$
\left ( y^2(\frac{\partial^2}{\partial x^2}+\frac{\partial^2}{\partial
  y^2})+E_n\right )\Psi_n(x,y)=0
$$
for the class of functions invariant (= automorphic) with respect to a given
discrete group $G$
$$
\Psi_n(x',y')=\Psi_n (x,y)
$$
where $z'=x'+iy'$ is connected with $z=x+iy$ by  group transformations
$$
z'=\frac{az+b}{cz+d}\;.
$$
It is easy to check that the Laplace--Beltrami operator (\ref{LB}) is 
self-adjoint with respect to the invariant measure (\ref{muinvariant}), i.e.
$$
\int \Psi^*(\varDelta \Psi) \D \mu=\int (\varDelta \Psi^*) \Psi \D \mu
$$
and all eigenvalues $E_n$ are real and $E_n\geq 0$.
 
\subsection{Construction of the Green Function}\label{Construction}

As in the case of plain rectangular billiards the construction of the Green 
function requires two main steps.
\begin{itemize}
  \item The computation of the exact Green function for the free motion on
      the whole upper half plane.      
  \item The summation of the free Green function over all images of the 
  initial point under group transformations.
\end{itemize}
The free hyperbolic Green function obeys the equation
$$
(\varDelta_{LB}+E)G_E^{(0)}(\vec{x},\vec{x}')=\delta (\vec{x}-\vec{x}')
$$
and should depend only on the (hyperbolic) distance between points 
$\vec{x},\vec{x}'$
$$
y=\cosh d(\vec{x},\vec{x}')=1+\frac{ (x-x')^2+(y-y')^2}{2yy'}\;.
$$
After simple calculations one gets that $G(y)$ with $y\neq 0$ obeys
the equation for the Legendre functions (see e.g. \cite{Bateman}, Vol.1, Sect. 3)
$$
(1-y^2)\frac{\D^2G}{\D y^2}-2y\frac{\D G}{\D y}+l(l+1)G=0
$$
where
$$
E=\frac{1}{4}+k^2=-l(l+1)
$$
and
$$
l=-\frac{1}{2}-\imag k\;.
$$
As for the plane case the required solution of the above equation should grow 
as $\E^{ikd}$ when $d\to \infty$ and should behave like $\ln d /2\pi$ 
when $d\to 0$. From  \cite{Bateman}, Vol.1, Sect. 3 it follows that 
$$
G_E^{(0)}(\vec{x},\vec{x}')=-\frac{1}{2\pi}
Q_{-\frac{1}{2}-\imag k}(\cosh d(\vec{x},\vec{x}'))\;.
$$
Here $Q_{-\frac{1}{2}-ik}(\cosh d)$  is the Legendre function of the second
kind with  the integral representation \cite{Bateman}, Vol.~1 (3.7.4)
$$
Q_{-\frac{1}{2}-\imag k}(\cosh d)=\frac{1}{\sqrt{2}}\int_d^{\infty}
\frac{\E^{\imag kr}\D r}{\sqrt{\cosh r-\cosh d}}
$$
and the following asymptotics
$$
Q_{-\frac{1}{2}-\imag k}(\cosh d)\stackrel{d\to 0}{\longrightarrow}-\log d
$$
and
$$
Q_{-\frac{1}{2}-\imag k}(\cosh d)\stackrel{d\to \infty}{\longrightarrow}
\sqrt{\frac{\pi}{2k\sinh d}}\E^{\imag(kd-\pi/4)}\;.
$$
The automorphic Green function is the sum over all images of one of the points
$$
G_E(\vec{x},\vec{x}')=\sum_{g}G_E^{(0)}(\vec{x},g(\vec{x}'))
$$ 
where the summation is performed over all group transformations.

\subsection{Density of State}\label{Density}

Using the standard formula (\ref{ImGreen})
$$
d(E)=-\frac{1}{\pi}\int_D\mbox{ Im }G_E(\vec{x},\vec{x})\D \mu 
$$
one gets the expression for the density of states as the sum over all group
elements
$$
d(E)=\frac{1}{2\sqrt{2}\pi^2}\sum_{g} \int_D\frac{\D x\D y}{y^2}
\left (\int_{d(z,g(z))}^{\infty}
\frac{\sin kr \D r}{\sqrt{\cosh r-\cosh d(z,g(z))}}\right ) \;.
$$

\paragraph{Mean Density of States}

The mean density of states corresponds to the identity element of our group.
In this case $g(z)=z$ and $d(z,g(z))=0$. Therefore
\begin{eqnarray*}
\bar{d}(E)&=&\frac{1}{2\sqrt{2}\pi^2} \int_D\frac{\D x \D y}{y^2}
\int_0^{\infty} \frac{\sin kr}{\sqrt{\cosh r-1}}\D r\nonumber\\
&=&\frac{\mu(D)}{(2\pi)^2}\int_0^{\infty} 
\frac{\sin kr}{\sinh (r/2)}\D r  
\end{eqnarray*}
where
$$
\mu(D)=\int_D\frac{\D x \D y}{y^2}
$$
is the (hyperbolic) area of the fundamental domain.

The last integral is
$$
\int_0^{\infty} \frac{\sin kr}{\sinh (r/2)}\D r=\pi \tanh \pi k
$$
and the mean density of states takes the form
$$
  \bar{d}(E)=\frac{\mu(D)}{4\pi}\tanh \pi k\;.
$$
When $k\to \infty$ it tends to $\mu(D)/4\pi$ as for the plane case. 

\subsection{Conjugated Classes}\label{Conjugated}

The most tedious step is the computation of the contribution from
non-trivial fractional transformations.

Let us divide all group matrices into classes of conjugated elements. It
means that all matrices having the form
$$
g'=SgS^{-1} 
$$
where $S$ belong to the group are considered as forming one class. 

Two classes either have no common elements or coincide. This
statement is a consequence of the fact that if
$$   
S_1g_1S_1^{-1}=S_2g_2S_2^{-1}  
$$
then $g_2=S_3g_1S_3^{-1}$ where $S_3=S_1^{-1}S_2$. Therefore $g_2$ belongs
to the same class as $g_1$ and group matrices are splitted into classes of 
mutually non-conjugated elements.

The summation over group elements can be rewritten as the double sum over
classes of conjugated elements and the elements in each class. Let $g$ be a
representative of a class. Then the summation over elements in this class
is
$$
\sum_S\int_Df(z,SgS^{-1}(z))\D \mu
$$
and the summation is performed over all group matrices $S$ provided there
is no double counting in the sum. The latter means that matrices $S$ should
be such that they do not contain matrices for which
$$
S_1gS_1^{-1}=S_2gS_2^{-1}
$$
or the matrix $S_3=S_1^{-1}S_2$ commutes with  matrix $g$
$$ 
S_3g=gS_3\;.
$$
Denote the set of matrices commuting with $g$ by $S_g$. They form a
subgroup of the initial group $G$ as their products also commute with
$g$. To ensure the unique decomposition of group matrices into
non-overlapping classes of conjugated elements the summation should be
performed over matrices $S$ such that no two of them can be represented as
$$
S_2=sS_1
$$
and $s$ belongs to $S_g$. This is equivalent to the statement
that we sum over all matrices but the matrices $sS$ are considered as one
matrix. It means that we factorize the group over $S_g$ and consider the
group $G/S_g$.

As the distance is invariant under simultaneous transformations of both
coordinates
$$
d(z,z')=d(S(z),S(z'))
$$
one has
$$ 
d(z,g(z))=d(S(z),Sg(z))=d(y,SgS^{-1}(y))
$$
where $y=S(z)$.

These relations give
$$
\int_D f(d(y,SgS^{-1}(y)))\D \mu =\int_{S^{-1}(D)}f(z,g(z))\D \mu
$$
and the last integral is taken over the image of the fundamental domain
under the transformation $S^{-1}$. Therefore
$$
\sum_{S}\int_D f(d(y,SgS^{-1}(y)))\D \mu =\sum_{S}\int_{S^{-1}(D)}
f(d(z,g(z)))\D \mu\;.
$$
For different $S$ images $S^{-1}(D)$ are different and do not overlap.
The integrand does not depend on $S$ and
$$
\sum_{S}\int_D f(d(y,SgS^{-1}(y)))\D \mu =\int_{D_g}
f(d(z,g(z)))\D \mu
$$
where
$$
D_g=\sum_S S^{-1}(D)\;.
$$
The sum of all images $S^{-1}(D)$ will cover the whole upper half plane but
we have to sum not over all $S$ but only over $S$ factorized by the action
the group of matrices commuting with a fixed matrix $g$. Therefore the sum
will be a smaller region.

Any matrix $g$ can be written as a power of a primitive element
$$
g=g_0^n
$$
and it is (almost) evident that matrices commuting with $g$ are precisely the
group of matrices generated by $g_0$. This is a cyclic abelian group
consisting of all (positive, negative, and zero) powers of $g_0$
$$
S_g=g_0^m,\;\;\;m=0,\pm 1,\pm 2,\ldots 
$$
and as a discrete group it has a fundamental domain $FD_g$.

Therefore
$$
\sum_{S\in G/S_g}\int_D f(d(y,SgS^{-1}(y)))\D \mu =\int_{FD_g}
f(d(z,g(z))) \D \mu\;.
$$
In the left hand side the integration is taken over the fundamental domain
of the whole group $G$ and the summation is done over matrices from $G$
factorized by the subgroup $S_g$ of matrices which commutes with a fixed
matrix $g$. In the right hand side there is no summation but the integration
is performed over the (large) fundamental domain of the subgroup $S_g$.

\subsection{Selberg Trace Formula}\label{Selberg}

We have demonstrated that the density of states of the hyperbolic 
Laplace--Beltrami operator automorphic over a discrete group can be represented as 
$$
d(E)=\bar{d}(E)+\sum_g d_g(E)
$$
where
$$
d_g(E)=\frac{1}{2\sqrt{2}\pi^2}\int_{FD_g}\D \mu \int_{d(z,g(z))}^{\infty}
\frac{\sin kr}{\sqrt{\cosh r-\cosh d(z,g(z))}}\D r
$$
and the summation is performed over classes of conjugated matrices.

Let us consider the case of hyperbolic matrices $g=g_0^m$ (i.e.
$|\mbox{Tr }g_0|>2$). By a suitable 
matrix $B$ such matrix can be transform to the diagonal form
$$
Bg_0B^{-1}=\left (\begin{array}{cc}\lambda_0&0\\0&\lambda_0^{-1}\end{array}
\right )\;.
$$
For hyperbolic matrices $\lambda_0$ is real and $|\lambda_0|>1$.
By the same transformation the matrix $g$ will be transformed to 
$$
BgB^{-1}=\left (\begin{array}{cc}\lambda&0\\0&\lambda^{-1}\end{array}
\right )
$$
and $\lambda=\lambda_0^m$. 

\begin{figure}
\center
\includegraphics[height=6cm, angle=-90]{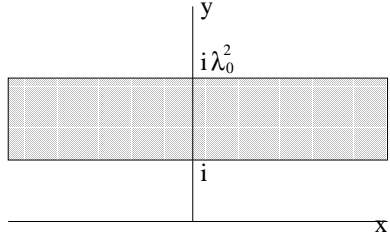}
\caption{Fundamental domain of multiplication group}
\label{fundamental}
\end{figure}

Assume that $g$ is in the diagonal form. Then $g(z)=\lambda^2 z$ and
$$
\cosh d(z,g(z))=1+\frac{(\lambda^2-1)^2(x^2+y^2)}{2\lambda^2 y^2}\;.
$$
Because $\lambda_0$ is real the transformation $z'=\lambda_0^2z$ gives
$y'=\lambda_0^2y$ and the fundamental domain of $S_g=\lambda_0^{2m}z$ 
has the form of a horizontal strip $1<y<\lambda_0^2$ indicated in Fig.~\ref{fundamental}.
Now
$$
d_g(E)=\int_{-\infty}^{\infty} \D x
\int_1^{\lambda_0^2}F\left (\frac{(\lambda^2-1)^2(x^2+y^2)}{\lambda^2 y^2}
\right )\frac{\D y}{y^2}\;.
$$
Introducing a new variable $\xi=xy$ one gets
\begin{eqnarray*}
d_g(E)&=&\int_1^{\lambda_0^2}\frac{\D y}{y}\int_{-\infty}^{\infty}
F\left ((1+\xi^2)\frac{(\lambda^2-1)^2}{\lambda^2}\right )\D \xi =\\
&=&\ln \lambda_0^2 \int_{-\infty}^{\infty}
F\left ((1+\xi^2)\frac{(\lambda^2-1)^2}{\lambda^2}\right )\D \xi\;.
\end{eqnarray*}
After the substitution
$$
u=(1+\xi^2)\frac{(\lambda^2-1)^2}{\lambda^2}
$$
one obtains
$$
d_g(E)=\frac{\ln
  \lambda_0^2}{\sqrt{u_0}}\int_{u_0}^{\infty}\frac{F(u)}{\sqrt{u-u_0}}\D u
$$
where
$$u_0=\frac{(\lambda^2-1)^2}{\lambda^2}=\lambda^2+\frac{1}{\lambda^2}-2\;.$$

The variable $u$ is connected with the distance by $\cosh d=1+u/2$ and the
function $F(\cosh d)$ has the form
$$
F(\cosh d)=\frac{1}{2\sqrt{2}\pi^2}\int_d^{\infty} \frac{\sin
  kr}{\sqrt{\cosh r-\cosh d}}\D r\;.
$$
Introduce a variable $\tau$ connected with $r$ as $u$ is connected with $d$
$$
\cosh \tau=1+\frac{r}{2}\;,\;\;\;
\frac{\D r}{\D \tau}=\frac{1}{\sqrt{\tau^2+4\tau}}\;.
$$
It gives
$$
F(u)=\frac{1}{2\pi^2}\int_{u}^{\infty} \frac{\sin kr(\tau)}{\sqrt{(\tau
    -u)(\tau^2+4\tau)}}\D \tau
$$
and
$$
d_g(E)=\frac{\ln \lambda_0^2}{2\pi^2 \sqrt{u_0}}f(u_0)
$$
where
$$
f(w)=\int_{w}^{\infty} \frac{\D u}{\sqrt{u-w}}\int_{u}^{\infty}
  \frac{\sin kr(\tau)}{\sqrt{(\tau -u)(\tau^2+4\tau)}}\D \tau\;.
$$
Changing the order of integration one obtains
$$
f(w)=\int_{w}^{\infty}\frac{\sin kr(\tau)}{\sqrt{\tau^2+4\tau}}\D \tau
\int_w^{\tau }\frac{\D u}{\sqrt{(u-w)(\tau-u)}}\;.
$$
The last integral is a half of the residue at infinity
$$
\int_w^{\tau }\frac{\D u}{\sqrt{(u-w)(\tau-u)}}=\pi
$$
and
$$
f(w)=\pi \int_w^{\infty} \frac{\sin kr(\tau)}{\sqrt{\tau^2+4\tau}}\D \tau=
\pi \int_{l_p}^{\infty}\sin (kr)\D r=\frac{\pi}{k}\cos kl_p\;.
$$
Here $l_p$ is the minimal value of $r$ corresponding to $u_0$
$$
\cosh l_p =1+\frac{u_0}{2}=1+\frac{1}{2}(\lambda^2+\frac{1}{\lambda^2}-2)=
\frac{1}{2}(\lambda+\frac{1}{\lambda})^2-1
$$
or
$$
2\cosh l_p =\lambda+\frac{1}{\lambda}\equiv \mbox{ Tr }g
$$
i.e. $l_p$ is the length of periodic orbit associated with the matrix $g$.

Therefore
$$
d_g(E)=\frac{\ln \lambda_0^2}{2\pi k \sqrt{\lambda+\lambda^{-1}-2}}
\cos kl_p=\frac{l_p^{(0)}}{4\pi k \sinh l_p/2}\cos kl_p
$$
where $l_p^{(0)}$ is the length of the primitive periodic orbit associated
with $g_0$.

Combining all terms together one finds that the eigenvalues 
density of the Laplace--Beltrami operator automorphic with respect to a discrete 
group with only hyperbolic matrices has the form  
$$
d(E)=\frac{\mu(D)}{4\pi}\tanh \pi k+\sum_{\mbox{\scriptsize{p.p.o.}}}\frac{l_p}{4\pi k} 
\sum_{n=1}^{\infty} \frac{\cos (knl_p)}{\sinh (nl_p/2)}\;.
$$
The oscillating part of the density is given by the double sum. 
The first summation is done over all primitive periodic orbits (p.p.o.) 
and the second
sum is performed over all repetitions of these orbits. Here $k$ is the
momentum related with the energy by $E=k^2+1/4$.

To obtain mathematically sound formula and to avoid problems with convergence it
is common to multiply  both parts of the above equality by a test
function $h(k)$ and to integrate over $\D E=2k \D k$. To assume the
convergence the test function $h(r)$ should have the following properties
\begin{itemize}
\item The function $h(r)$  is a function analytical in the region 
$|\mbox{Im }r|\leq 1/2+\delta$ with  certain $\delta >0$.
\item $h(-r)=h(r)$.
\item $|h(r)|\leq A(1+|r|)^{-2-\delta}$.
\end{itemize}
The left hand side of the above equation is
$$
\int d(E)h(k)\D E=\sum_n \delta (E-E_n)h(k)\D E=\sum_n h(k_n)\;.
$$
In the right hand side one obtains
$$
\int h(k)\frac{\cos kl}{2\pi k}k\D k=\frac{1}{2\pi} \int_{-\infty}^{\infty}
h(k)\E^{-\imag kl}\D k\;.
$$
The final formula takes the form
\begin{eqnarray}
\sum_n h(k_n)&=&\frac{\mu (D)}{2\pi}\int_{-\infty}^{\infty}k h(k)\tanh (\pi k)
\D k+\nonumber\\
&+& \sum_{\mbox{\scriptsize{p.p.o.}}}l_p\sum_{n=1}^{\infty} 
\frac{1}{2\sinh (nl_p/2)}g(nl_p)
\label{SelbergTrace}
\end{eqnarray}
where $k_n$ is related with eigenvalue $E_n$ as follows 
$$
E_n=k_n^2+\frac{1}{4}
$$
and $g(l)$ is the Fourier transform of $h(k)$
$$
g(l)=\frac{1}{2\pi}\int_{-\infty}^{\infty} h(k)\E^{-\imag kl}\D k\;.
$$
This is the famous Selberg trace formula. It connects eigenvalues of the
Laplace--Beltrami operator for functions automorphic with respect to a
discrete group having only hyperbolic elements with classical periodic
orbits.

\subsection{Density of Periodic Orbits}\label{Huber}
To find the density of periodic orbits for a discrete group let us choose
the test function $h(r)$ in (\ref{SelbergTrace}) as
$$
h(r)=\E^{-(r^2+1/4)T}\equiv \E^{-ET}
$$ 
with a parameter $T>0$. Its Fourier transforms is
$$
g(u)=\frac{1}{2\pi}\int_{-\infty}^{\infty} h(k)
\E^{-\imag k u}\D k=\frac{\E^{-T/4}}{2\sqrt{\pi T}}\E^{-u^2/4T}\;.
$$
In the left hand side of the Selberg trace formula one obtains
$$
\sum_n \E^{-E_nT}=1+\sum_{E_n>0} \E^{-E_nT}
$$
where we take into account that for any discrete group there is one zero
eigenvalue corresponding to a constant eigenfunction. Therefore when 
$T\to\infty$ the above sum tends to one
$$
\sum_n \E^{-E_nT}\stackrel{T\to\infty}{\longrightarrow} 1\;.
$$
One can easily check that in the right hand side of (\ref{SelbergTrace})
the contribution of the smooth part of the density goes to zero at large 
$T$ and the contribution of periodic orbits is important only for primitive
periodic orbits with $n=1$. The latter is 
$$
\frac{\E^{-T/4}}{2\sqrt{\pi T}}\sum_{\mbox{p}}l_p\E^{-l_p^2/4T-l_p/2}=
\frac{\E^{-T/4}}{2\sqrt{\pi T}}\int_0^{\infty}l\E^{-l^2/4T-l/2}\rho(l)\D l
$$
where $\rho(l)$ is the density of periodic orbits. Hence the Selberg
trace formula states that
$$
\lim_{T\to\infty}\frac{\E^{-T/4}}{2\sqrt{\pi T}}\int_0^{\infty}l
\E^{-l^2/4T-l/2}\rho(l)\D l=1\;.
$$
Assume that $\rho(l)=b\E^{al}/l$ with certain constants $a$ and $b$.  
Then from the above limit it follows that $a=b=1$ which demonstrates 
that the density of periodic orbits for a discrete group increases exponentially 
with the length
$$
\rho(l)=\frac{\E^{l}}{l}\;. 
$$

\subsection{Selberg Zeta Function}\label{SelbergZeta}

Amongst many applications of the Selberg trace formula let us consider
the construction of the Selberg zeta function.

Choose as  test function $h(k)$ the function
$$
h(k)=\frac{1}{k^2+\alpha^2} -\frac{1}{k^2+\beta^2}\;.
$$
Its Fourier transform is
$$
g(l)=\frac{1}{2\alpha}\E^{-\alpha |l|} -\frac{1}{2\beta}\E^{-\beta|l|}\;.
$$
The Selberg trace formula gives
\begin{eqnarray*}
&&
\sum_n \left (\frac{1}{k_n^2+\alpha^2} -\frac{1}{k_n^2+\beta^2}\right )=
\nonumber\\
&&=\frac{\mu(D)}{2\pi} \int_{-\infty}^{\infty} k\tanh \pi k 
\left (\frac{1}{k^2+\alpha^2} -\frac{1}{k^2+\beta^2}\right )\D k +\\
&&+\sum_{\mbox{\scriptsize{p.p.o.}}}\sum_{n=1}^{\infty} \frac{l_p}{2\sinh  nl_p/2}
\left (\frac{\E^{-\alpha l_p}}{2\alpha}-\frac{\E^{-\beta l_p}}{2\beta}\right )\;.
\end{eqnarray*}
The Selberg zeta function is defined as the following formal product
\begin{equation}
Z(s)=\prod_{\mbox{\scriptsize{p.p.o}}}\prod_{m=0}^{\infty} (1-\E^{-l_p(s+m)})\;.
\label{ZetaProduct}
\end{equation}
One has
\begin{eqnarray*}
\frac{1}{Z}\frac{dZ}{ds}&=&\sum_{\mbox{\scriptsize{p.p.o.}}} \sum_{m=0}^{\infty}
\frac{l_p\E^{-l_p(s+m)}}{1-\E^{-l_p(s+m)}}=
\sum_{\mbox{\scriptsize{p.p.o.}}} l_p \sum_{n=1}^{\infty} 
\sum_{m=0}^{\infty}\E^{-l_p(s+m)n}=\\
&=&\sum_{\mbox{p.p.o.}} l_p \sum_{n=1}^{\infty} 
\frac{\E^{-l_pns}}{1-\E^{-l_p n}}=
\sum_{\mbox{\scriptsize{p.p.o.}}} l_p \sum_{n=1}^{\infty} 
\frac{1}{2\sinh nl_p/2}\E^{-l_pn(s-1/2)}\;.
\end{eqnarray*}
Choose $\alpha=s-1/2$ and $\beta=s'-1/2$  then
\begin{eqnarray*}
&&\sum_n \left (\frac{1}{k_n^2+(s-1/2)^2} -\frac{1}{k_n^2+(s'-1/2)^2} \right )=\\
&&=
\frac{\mu(D)}{4\pi} \int_{-\infty}^{\infty} k\tanh \pi k
\left (\frac{1}{k^2+(s-1/2)^2} -\frac{1}{k^2+(s'-1/2)^2}\right )\D k+\\
&&+\frac{1}{2s-1}\frac{Z(s)'}{Z(s)}-\frac{1}{2s'-1}\frac{Z(s')'}{Z(s')}\;.
\end{eqnarray*}
The integral in the right hand side can be computed by the residues
$$
\int_{-\infty}^{\infty} k\tanh \pi k
\left (\frac{1}{k^2+(s-1/2)^2} -\frac{1}{k^2+(s-1/2)^2}\right )\D k=f(s)-f(s')
$$
where $f(s)$ is the sum over residues from one pole $k=i(s-1/2)$ and from
poles $k_n=i(n+1/2)$ of $\tanh \pi k$
\begin{eqnarray*}
& &f(s)=2\pi \imag \left [\frac{1}{2} \tanh [\imag \pi(s-1/2)]+\frac{\imag}{\pi}
\sum_{n=0}^{\infty} \frac{n+1/2}{(s-1/2)^2-(n+1/2)^2} \right ]=\\
& &= \pi \cot \pi s-
\sum_{n=1}^{\infty} \frac{1}{s-n}+\sum_{n=1}^{\infty}\frac{1}{s+n}\;.
\end{eqnarray*}
But
$$
\pi\cot \pi s=\sum_{n=1}^{\infty} \frac{1}{s-n}+\sum_{n=1}^{\infty}
\frac{1}{s+n}\;,
$$
therefore
$$f(s)=2\sum_{n=1}^{\infty}\frac{1}{s+n}\;.$$
Using these relations one gets the identity valid for all values of $s$ and $s'$
\begin{eqnarray}
\frac{1}{2s-1}\frac{Z'(s)}{Z(s)}&=&\frac{1}{2s'-1}\frac{Z'(s')}{Z(s')}-
\frac{\mu(D)}{2\pi}\sum_{n=0}^{\infty}\left (\frac{1}{s+n}-\frac{1}{s'+n}\right )+
\nonumber\\
&+&\sum_{n}\left (\frac{1}{k_n^2+(s-1/2)^2}-\frac{1}{k_n^2+(s'-1/2)^2}\right )\;.
\label{ZetaContinuation}
\end{eqnarray}
The right hand side of this identity has poles at $s=1/2+\imag k_n$ and $s=-n$.
The same poles have to be present in the left hand side. If
$$
\frac{Z'(s)}{Z(s)}\to \frac{\nu_k}{s-s_k}
$$
then
$$
Z(s)\to (s-s_k)^{\nu_k}\;\;\;\mbox{when }s\to s_k\;.
$$
When $\nu_k>0$ (resp. $\nu_k<0$)  point $s_k$ is a zero (resp. a pole) of
the Selberg zeta function $Z(s)$.

\subsection{Zeros of the Selberg Zeta Function}\label{SelbergZeros}
  
Combining all poles one concludes that the Selberg zeta function for
a group with only hyperbolic elements have two different sets of zero. 
The first consists of non-trivial zeros $$s=1/2\pm \imag  k_n,$$ coming
from eigenvalues of the Laplace--Beltrami operator for automorphic functions. 
The second set includes a zero from $E=0$ eigenvalue and zeros from the smooth
term. These zeros are called
trivial zeros and they are located at points
$$s=-m \;\;(m=1,2,\ldots $$
with multiplicity $\nu_m=(2m+1)\mu(D)/2\pi$, at point $s=0$ with
multiplicity $\nu_0=\mu(D)/2\pi$ and a single zero at $s=1$. These
multiplicities are integers because the area of a compact fundamental domain
$\mu(D)=4\pi (g-1)$ where $g$ is the genus of the surface.

\begin{figure}
\center
\includegraphics[height=7cm, angle=-90]{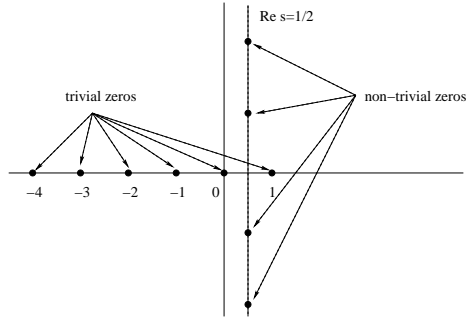}
\caption{Zeros of the Selberg zeta function}
\label{selberg}
\end{figure}

The structure of these zeros is presented schematically at Fig.~\ref{selberg}.
 
\subsection{Functional Equation}\label{FunctionalSelberg}

The infinite product defining the Selberg zeta function (\ref{ZetaProduct}) converges 
only when Re $s>1/2$. Nevertheless the Selberg zeta function 
can be analytically continued to the whole complex plane $s$ with the aid 
of (\ref{ZetaContinuation}).

Put $s'=1-s$ in (\ref{ZetaContinuation}). The sum over eigenvalues cancels and
$f(s)-f(1-s)=2\pi \cot \pi s$. Therefore
$$
\frac{1}{2s-1}\left (\frac{Z'(s)}{Z(s)}+\frac{Z'(1-s)}{Z(1-s)}\right )=
-\frac{\mu(D)}{2}\cot \pi s
$$
which is equivalent to the following  relation (called functional equation) 
\begin{equation}
Z(s)=\varphi(s)Z(1-s)
\label{FunctionalEquation}
\end{equation}
where
$$\frac{\varphi'(s)}{\varphi(s)}=-\mu(D)(s-\frac{1}{2})\cot \pi s$$
and $\varphi(1/2)=1$. 

Explicitly
$$
\varphi(s)=\exp \left (\mu(D)\int_0^{s-1/2}u\tan \pi u \D u \right )\;.
$$
Therefore if one knows the Selberg zeta function when Re $s>1$ 
(\ref{FunctionalEquation}) gives its continuation to the mirror region
Re $s<0$.

\section{Trace Formulas for Integrable Dynamical Systems}\label{Integrable}

A $f$-dimensional system is called integrable if its classical Hamiltonian
can be written as a function of  action variables only
$$
H(\vec{I})=H(I_1,\ldots ,I_f)\;.
$$
In this representation the classical equations of motion take especially
simple form
$$
\dot{\vec{I}}=-\frac{\partial H}{\partial \vec{\varphi}}=0\;,\;\;
\dot{\vec{\varphi}}=\frac{\partial H}{\partial \vec{I}}=\vec{\omega}\;.
$$
The semiclassical quantization consists of fixing the values of the action
variables
$$
I_j=\hbar (n_j+\frac{\mu_j}{4})
$$
where $n_j$ are integers and $\mu_j$ are called the Maslov indices.

In this approximation eigenvalues of energy of the system are a function 
of these integers
$$
E(\vec{n})=H\left (\hbar(n_1+\frac{\mu_1}{4}),\ldots ,
\hbar(n_f+\frac{\mu_f}{4})\right )\;.
$$
The eigenvalue density is the sum over all integers $n_j$
$$
d(E)=\sum_{\vec{n}} \delta (E-H(\hbar (\vec{n}+\frac{1}{4}\vec{\mu}))\;.
$$
Using the Poisson summation formula (\ref{Poisson}) one transforms this
expression as follows
\begin{eqnarray}
d(E)&=&\sum_{\vec{N}}\int e^{2\pi i\vec{N}\vec{n}} 
\delta (E-H(\hbar (\vec{n}+\frac{1}{4}\vec{\mu}))\D \vec{n}=
\nonumber\\
&=&\frac{1}{\hbar^f}\sum_{\vec{N}}\E^{-\imag \pi\vec{N}\vec{\mu}/2}\int 
\E^{2\pi i\vec{N}\vec{I}/\hbar}\int  \delta
(E-H(\vec{I}\,))\D \vec{I}
\label{DensityIntegrable}
\end{eqnarray}
where the summation is taken over $f$ integers $N_j$. 

\subsection{Smooth Part of the Density}\label{Smooth}

The term with $\vec{N}=0$ in (\ref{DensityIntegrable}) corresponds to 
the smooth part of the density
$$
\bar{d}(E)=\frac{1}{\hbar^f}\int  \delta (E-H(\vec{I}\,))\D \vec{I}\;.
$$
As $\D \vec{I}\D \vec{\varphi}$ is the canonical invariant,
$\D \vec{I}\D \vec{\varphi}=\D \vec{p}\D \vec{q}$
where $\vec{p}$ and $\vec{q}$ are the momenta and coordinates and, because
$\int \D \vec{\varphi}=(2\pi )^f$,  the formula
for the smooth part of the level density can be rewritten in the
Thomas-Fermi form
\begin{equation}
\bar{d}(E)=\int \delta(E-H(\vec{p},\vec{q}\,))
\frac{\D \vec{p}\D \vec{q}}{(2\pi \hbar)^f}\;. 
\label{meanDensity}
\end{equation}
The usual interpretation of this formula is that each quantum state
occupies $(2\pi \hbar)^f$ volume on the constant energy surface. 
For general systems (\ref{meanDensity}) represents the leading term
of the expansion of the smooth part of the level density when 
$\hbar\to 0$. Other terms can be found e.g. in \cite{BaltesHill}. 
See also \cite{BerryHowls} for the resummation of such series for  certain models.

\subsection{Oscillating Part of the Density}\label{Oscillating}

In the semiclassical approximation $\hbar\to 0$  terms with
$\vec{N}\neq 0$ in (\ref{DensityIntegrable}) can be calculated by the 
saddle point method. Our derivation differs slightly from the one 
given in \cite{TaborBerry}. First it is convenient to represent
$\delta$-function as follows 
$$
\delta (x)=\frac{1}{2\pi \hbar}\int_{-\infty}^{\infty} 
\E^{\imag \alpha x/\hbar }\D\alpha\;.
$$
Then
$$
d^{(osc)}(E)=\frac{1}{2\pi \hbar^{f+1}}\sum_{\vec{N}}\E^{-\imag \pi \vec{N}\vec{\mu}/2}
\int_{-\infty}^{\infty}
\D \alpha \int \E^{\imag S(\vec{I},\alpha)/\hbar }\D \vec{I}
$$
where the effective action, $S(\vec{I},\alpha)$, is
$$
S(\vec{I},\alpha)=2\pi\vec{N}\vec{I}+\alpha(E-H(\vec{I}))\;.
$$
The integration over $\vec{I}$ and $\alpha$ can be performed by the saddle      
point method. The saddle point values, $\vec{I}_{sp}$ and $\alpha_{sp}$, are 
determined from equations 
$$
\frac{\partial S}{\partial \alpha}=E-H(\vec{I}_{sp} )=0\;,\;\;\;
\frac{\partial S}{\partial \vec{I}}=2\pi \vec{N}-
\alpha_{sp} \vec{\omega}_{sp}=0\;.
$$
The first equation shows that in the leading approximation $\vec{I}_{sp}$ 
belongs to  the constant energy surface and the second equation
selects special values of $\vec{I}_{sp}$ for which frequencies $\omega_j$
are commensurable
$$
\vec{\omega}_{sp}=\frac{2\pi}{\alpha_{sp}}\vec{N}\;.
$$
Together the saddle point conditions demonstrate that in the limit
$\hbar \to 0$ the dominant contribution to the term with fixed integer vector
$\vec{N}$ comes from the classical periodic orbit with period
$$
T_p=\alpha_{s.p}
$$
and the saddle point action coincides with the classical action along this
trajectory
$$
S_{sp}=2\pi \vec{N}\vec{I}_{sp}\;.
$$
To compute remaining integrals it is necessary to expand the full action
up to quadratic terms on deviations from the saddle point values. One has
$$
S(\vec{I}_{sp}+\delta \vec{I},\alpha_{s.p}+\delta \alpha)=S_{sp}+
\frac{T_p}{2}(\delta I_i H_{ij}\delta I_j) -\delta \alpha (\omega_j\delta
I_j)
$$
where the summation over repeating indexes is assumed.
$H_{ij}$ is the matrix of the second derivatives of the Hamiltonian
computed at the saddle point
$$
H_{ij}\equiv \frac{\partial^2 H}{\partial I_i \partial I_j}
\left |_{\vec{I}=\vec{I}_{sp}}\right .\;.
$$
The following steps are straightforward
\begin{eqnarray*}
& &\int \D \delta \vec{I} \D \delta \alpha 
\exp \left (\frac{\imag}{\hbar}S(\vec{I},\alpha)\right )=
\nonumber\\
&=&\E^{\imag S_{sp}/\hbar}\int \D \delta \alpha \int \D \delta \vec{I}
\exp \left (\frac{\imag}{2\hbar} T_p(\delta I_i H_{ij}\delta I_j) -
  \frac{\delta \alpha}{\hbar} (\omega_j\delta I_j)\right )=
  \nonumber\\
&=&\left (\frac{2\pi \hbar}{T_p}\right )^{f/2}
\frac{\E^{\imag S_{sp}/\hbar}}{\sqrt{|\det H_{ij}|}}
\int \delta \alpha \exp \left (-\frac{\imag}{2\hbar T_{p}}(\delta \alpha )^2
(\omega_i H_{ij}^{-1}\omega_j) +\frac{\imag}{4}\pi \beta' \right )=
\nonumber\\
&=&\left (\frac{2\pi \hbar}{T_p}\right )^{f/2}
\frac{\sqrt{2\pi \hbar T_p}}
{\sqrt{|\det H_{ij}|(\omega_k H_{kl}^{-1}\omega_l)}}
\exp \left (\frac{\imag}{\hbar} S_{sp}+\frac{\imag}{4} \pi \beta \right )=
\nonumber\\
&=&\frac{(2\pi)^{(f-1)/2}\hbar^{(f+1)/2}}
{T_p^{(f-3)/2}|(N_i Q_{ij} N_j)|^{1/2}}
\exp \left (\frac{\imag}{\hbar} S_{sp} +\frac{\imag}{4} \pi \beta \right )
\end{eqnarray*}
where $Q_{ij}=H_{ij}^{-1}\det H$ called the co-matrix of $H_{ij}$ is
the determinant obtained from $H_{ij}$ by omitting the $i$-th row and
the $j$-th column. The phase $\beta$ is the signature of $H_{ij}$ minus the
sign of $(\omega H^{-1} \omega)$.

The final expression for the oscillating part of the level density of an
integrable system with a Hamiltonian $H(\vec{I})$ is
$$
d^{(osc)}(E)=\sum_{\vec{N}}P_{\vec{N}}
\exp \left (\imag \frac{S_p}{\hbar}-\imag \frac{\pi}{4} \vec{N}\vec{\mu}+
\imag \frac{\pi}{4}\beta\right )
$$
where $S_p=2\pi \vec{N}\vec{I}$ is the action over a classical periodic
orbit with fixed winding numbers and
$$
P_{\vec{N}}=\left (\frac{2\pi }{\hbar T_p}\right )^{(f-3)/2}
\frac{1}{\hbar^2|(N_i Q_{ij} N_j)|^{1/2}}\;.
$$
The summation over integer vectors $\vec{N}$ is equivalent to the summation over all
classical periodic orbit families of the system.

\section{Trace Formula for Chaotic Systems}\label{Chaotic}

To compute the eigenvalue density for a chaotic system one has to start
with general expression (\ref{ImGreen})  
$$
d(E)=-\frac{1}{\pi}\int \mbox{Im }G_{E}(\vec{x},\vec{x})\D \vec{x}
$$
which relates the quantum density  with  the Green function of the system,
$G_E(\vec{x},\vec{y})$, obeying  the Schroedinger
 equation with a $\delta$-function term in the right hand side
$$
(E-\hat{H})G_E(\vec{x},\vec{y})=\delta (\vec{x}-\vec{y})\;.
$$
For concreteness let us consider the usual case
$$
\hat{H} =-\hbar^2\varDelta+V(\vec{x})\;.
$$
The exact Green function can be computed exactly only in very limited cases.
For generic systems the best which can be achieved is the calculation of the
Green function in the semiclassical limit $\hbar \to 0$. 

\subsection{Semiclassical Green Function}\label{Green}

Let us try to obey the Schroedinger equation in the following form 
(see \cite{Gutzwiller})
\begin{equation}
G_E(\vec{x},\vec{y})=A(\vec{x},\vec{y})\E^{\imag S(\vec{x},\vec{y})/\hbar }
\label{Wave}
\end{equation}
where the prefactor $A(\vec{x},\vec{y})$ can be expanded into a power series
of $\hbar$.

Separating the real and imaginary parts of the Schroedinger equation one gets
two equations
$$
\left (E-(\nabla S)^2-V(\vec{x})\right )+ \hbar^2\varDelta A=0
$$
and
$$
2\nabla S\nabla A+\varDelta S A=0\;.
$$
In the leading order in $\hbar$ the first equation reduces to the
Hamilton-Jacobi equation for the classical action $S(\vec{x},\vec{y})$
$$
E=(\nabla S)^2 +V(\vec{x})\;.
$$
It is well known that the solution of this equation can be obtained in the
following way. 

Find the solution of the usual classical equations of motion
$$
\ddot{\vec{x}}=-\frac{\partial V}{\partial \vec{x}}
$$
with energy $E$ which starts at a fixed point $\vec{y}$ and ends at a
point $\vec{x}$. Then
$$
S(\vec{x},\vec{y})=\int_{\vec{y}}^{\vec{x}}\vec{p}d\vec{x}
$$
where $\vec{p}$ is the momentum and the integral is taken over this trajectory.

Instead of proving this fact we illustrate it on an example of the free
motion. The free motion equations $\ddot{\vec{x}}=0$ have a general solution
$$
\vec{x}=\vec{k}t+\vec{y}
$$
with a fixed vector $\vec{k}$. One has
$$
\vec{k}=\frac{\vec{x}-\vec{y}}{t}
$$
and the conservation of energy $|\vec{k}|^2=E$ determines the time of motion
$$
t=\frac{|\vec{x}-\vec{y}|}{\sqrt{E}}\;.
$$
Therefore
$$
S(\vec{x},\vec{y})=\sqrt{E}|\vec{x}-\vec{y}|
$$
which, evidently, is the solution of the free Hamilton--Jacobi equation.

The next order equation
$$
2\nabla S\nabla A+\varDelta S A=0
$$
is equivalent to the conservation of current. Indeed, for the semiclassical 
wave function (\ref{Wave})
$$
\vec{J}=\frac{1}{2\imag }(\Psi^*\nabla \Psi-\Psi \nabla \Psi^*)=
A^2\nabla S
$$
and
$$
\nabla \vec{J}=A(2\nabla A\nabla S+A\varDelta S)=0\;.
$$
The solution of the above transport equation has the form
$$
A(\vec{x},\vec{y})=\frac{\pi}{(2\pi \hbar)^{(f+1)/2}}\left |\frac{1}{k_ik_f}
\det \left (-\frac{\partial^2 S}{\partial t_{i\bot}\partial t_{f\bot}}
\right ) \right |^{1/2}
$$
where $t_{i\bot}$ and $t_{f\bot}$ are coordinates perpendicular to the
trajectory in the initial, $\vec{y}$, and final, $\vec{x}$, points respectively
and $k_i$, $k_f$ are the initial and final momenta. The derivation of this
formula can be found e.g. in \cite{Gutzwiller}. The overall prefactor
in this formula can be fixed by comparing with the asymptotics of the
free Green function (\ref{GreenFdimensions}) at large distances. 

The final formula for the semiclassical Green function takes the form
\begin{eqnarray*}
G_E(\vec{x},\vec{y})&=&
   \sum_{\begin{array}{c}\mbox{\scriptsize{classical}}\\
                         \mbox{\scriptsize{trajectories}}\end{array}}
\frac{\pi}{(2\pi \hbar)^{(f+1)/2}}\left |\frac{1}{k_ik_f}
\det \left (-\frac{\partial^2 S}{\partial t_{i\bot}\partial t_{f\bot}}\right ) 
\right |^{1/2}
\times\nonumber\\
&\times&
\exp \left (\frac{\imag}{\hbar}S_{cl}(\vec{x},\vec{y}) -
\frac{\imag}{4}\pi \mu \right )
\end{eqnarray*}
where the sum is taken over all classical trajectories with energy $E$
which connect points $\vec{y}$ and $\vec{x}$. $\mu$ is the Maslov index
which, roughly speaking, counts the number of points along the trajectory 
where semiclassical
approximation cannot be applied.

\subsection{Gutzwiller Trace Formula}\label{Gutzwiller}

\begin{figure}
\center
\includegraphics[height=5cm, angle=-90]{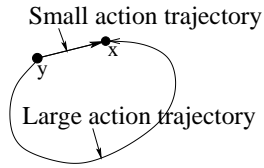}
\caption{
Small and large action contributions to the Green function for nearby points}
\label{action}
\end{figure}

The knowledge of the Green function permits the calculation of the density
of eigenstates by the usual formula (\ref{ImGreen})
$$
d(E)=-\frac{1}{\pi}\int \mbox{ Im }G_E(\vec{x},\vec{x})\D \vec{x}\;.
$$
The Green function $G_E(\vec{x},\vec{y})$ at points $\vec{x}$ and $\vec{y}$ 
very close to each other has two different contributions (see Fig.~\ref{action}). 
The first comes from very short trajectories  where semiclassical approximation 
cannot, in general, be applied. The second is related with long trajectories. 
The first contribution can be computed by using the
Thomas--Fermi (local) approximation for the Green function. In this 
approximation one uses the local formula (cf. (\ref{GreenFourier})) 
$$
G_E(\vec{x},\vec{y})\stackrel{\vec{y}\to\vec{x}}{\longrightarrow}
\int\frac{\D \vec{p}}{(2\pi \hbar)^f}\frac{\E^{i\vec{p}(\vec{x}-\vec{y})/\hbar}}
{(E-H(\vec{p},\vec{x})+i\epsilon)}\;.
$$
Therefore
$$
\mbox{ Im }G_E(\vec{x},\vec{x})=-\pi\int\frac{\D \vec{p}}{(2\pi \hbar)^f}
\delta \left (E-H(\vec{p},\vec{x})\right )
$$
and the smooth part of the level density in the leading approximation equals
the phase-space volume of the constant energy surface divided by
$(2\pi \hbar)^f$
$$
\bar{d}(E)=\int\frac{\D \vec{p}\D \vec{x}}{(2\pi \hbar)^f}
\delta \left (E-H(\vec{p},\vec{x})\right )\;.
$$
The contributions from long classical trajectories with finite actions
corresponds to the oscillating part of the density and  
can be calculated using the semiclassical approximation of the Green 
function (\ref{Wave}).

One has
$$
d^{(osc)}(E)=-\frac{1}{\pi}\mbox{ Im }\sum_{\begin{array}{c}
  \mbox{\scriptsize{classical}}\\
  \mbox{\scriptsize{trajectories}}\end{array}}
\int A(\vec{x},\vec{x})\E^{\imag S(\vec{x},\vec{x})/\hbar }\D \vec{x}\;.
$$
When $\hbar\to 0$ the integration can be performed in the saddle point
approximation. The saddles are solutions of the equation
$$
\left [\frac{\partial S(\vec{x},\vec{y})}{\partial \vec{x}}+
\frac{\partial S(\vec{x},\vec{y})}{\partial \vec{y}}\right ]_{\vec{y}=\vec{x}}=0\;.
$$
But
$$
\frac{\partial S(\vec{x},\vec{y})}{\partial
  \vec{x}}=\vec{k}_f\;,\;\;\;
\frac{\partial S(\vec{x},\vec{y})}{\partial \vec{y}}=-\vec{k}_i
$$
where $\vec{k}_f$ and $\vec{k}_i$ are the momenta in the final and initial
points respectively.

Hence the saddle point equations select special classical orbits which
start and end in the same point with the same momentum. It means that the
saddles are classical periodic orbits of the system and
$$ 
S_{sp}=S_p\;.
$$
To calculate the integral around one particular periodic orbit it is
convenient to split the integration over the whole space to one
integration along the orbit and $(f-1)$ integrations in directions 
perpendicular to the orbit. For simplicity we consider the two-dimensional case. 

\begin{figure}
\center
\includegraphics[height=4cm, angle=-90]{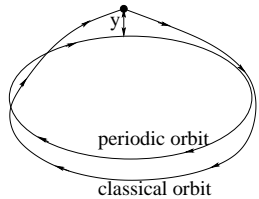}
\caption{
A periodic orbit and a closed classical orbit in its vicinity}
\label{deviation}
\end{figure}

The change of the action when a point is at the distance $y$ from the
periodic orbit is
$$
\delta S=\frac{1}{2} y^2\frac{\partial^2 S(y,y)}{\partial y^2}|_{y=0}
$$
where $S(y,y)$ is the classical action for a classical orbit in a vicinity
of the periodic orbit (see Fig.~\ref{deviation}). 
To compute such derivatives it is useful to use the
monodromy matrix, $m_{ij}$,  which relates initial and final coordinates 
and momenta in a vicinity of periodic orbit in the linear approximation
$$
\left (\begin{array}{c}\delta y_f\\  \delta p_f\end{array}\right )=
\left (\begin{array}{cc}m_{11}&m_{12}\\m_{21}&m_{22}\end{array}\right )
  \left (\begin{array}{c}\delta y_i\\  \delta p_i\end{array}\right )\;.
$$
As the classical motion preserves the canonical invariant $\D p\D q$ it follows
that $\det M=1$.

One has
\begin{eqnarray*}
  \delta y_f&=&m_{11}\delta y_i+m_{12}\delta p_i\;,\nonumber \\
  \delta p_f&=&m_{21}\delta y_i+m_{22}\delta p_i\;.
\end{eqnarray*}
But
$$
  p_i=-\frac{\partial S}{\partial y_i}\;,\;\;\;
  p_f=\frac{\partial S}{\partial y_f}\;.
$$  
Therefore
$$
  \delta p_i= -\frac{\partial^2 S}{\partial y_i^2} \delta y_i
  -\frac{\partial^2 S}{\partial y_i\partial y_f}\delta y_f\;,\;\;
  \delta p_f=\frac{\partial^2 S}{\partial y_i\partial y_f}\delta y_i
  +\frac{\partial^2 S}{\partial y_f^2}\delta y_f\;.
$$
From comparison of these two expression one obtains the expressions of the
second derivatives of the action through monodromy matrix elements
$$
\frac{\partial^2 S}{\partial y_i\partial y_f}=-\frac{1}{m_{12}}\;,\;\;
\frac{\partial^2 S}{\partial y_i^2}=\frac{m_{11}}{m_{12}}\;,\;\;
\frac{\partial^2 S}{\partial y_f^2}=\frac{m_{22}}{m_{12}}\;.
$$
Substituting these expressions to the contribution to the trace formula from
one periodic orbit one gets (in two dimensions)
$$
d^{(osc)}_p(E)=\frac{1}{\imag (2\pi \imag \hbar)^{3/2}}
\int |m_{12}|^{-1/2}\exp (\frac{\imag }{\hbar}S_p+\imag
\frac{m_{11}+m_{22}-2}{2\hbar \;m_{12}}y^2)\D y\frac{\D x}{k(x)}
$$ 
where $x$ and $y$ are respectively coordinates parallel and  perpendicular to the trajectory.

Computing the resulting integrals one obtains
$$
d^{(osc)}_p(E)=\frac{T_p}{\pi \hbar }
\frac{\E^{\imag S_p/\hbar-\imag \pi \mu_p/2}}{\sqrt{|m_{11}+m_{22}-2|}}
$$
where $T_p=\int \D x/k(x)$ is the geometrical period of the trajectory.

Finally the Gutzwiller trace formula takes the form (valid in arbitrary
dimensions)
$$
d^{(osc)}(E)=\sum_{\begin{array}{c}
    \mbox{\scriptsize{primitive}}\\
    \mbox{\scriptsize{periodic}}\\
    \mbox{\scriptsize{orbits}}
                   \end{array}}
\frac{T_p}{\pi \hbar}\sum_{n=1}^{\infty} \frac{1}{|\det (M_p^n-1)|^{1/2}}
\cos \left [n(\frac{S_p}{\hbar}-\frac{\pi}{2}\mu_p)\right ]\;.
$$
In the derivation of this formula we assumed that all periodic orbits are
unstable and $M_p$ is the monodromy matrix for a primitive periodic orbit.

\section{Riemann Zeta Function}\label{Riemann}
  
The trace-like formulas exist not only for dynamical systems but also for 
the Riemann zeta function  (and others number-theoretical zeta functions as 
well). 

The Riemann zeta function is a function of complex variable $s$ 
defined as follows
\begin{equation}
\zeta (s)=\sum_{n=1}^{\infty}\frac{1}{n^s}=\prod_p (1-p^{-s})^{-1}
\label{ZetaDefinition}
\end{equation}
where the product is taken over prime numbers. 
The second equality (called the Euler product) is a consequence of the unique 
factorization of integers into a product of prime numbers.

This function converges only when Re$s>1$ but can analytically
be continued in the whole complex $s$-plane.

\subsection{Functional Equation}\label{FunctionalRiemman}

The possibility of this continuation is connected with the fact that 
the Riemann zeta function satisfies the important functional equation
\begin{equation}
\zeta(s)=\varphi(s)\zeta(1-s)
\label{functionalZeta}
\end{equation}
where
\begin{equation}
\varphi(s)=2^s\pi^{s-1}\sin \left (\frac{\pi s}{2} \right ) 
\Gamma (1-s)\;.
\label{phis}
\end{equation}
We present one of numerous method of proving this relation 
(see e.g. \cite{Titchmarsh}). 

When Re $s>0$ one has the equality
$$\int_0^{\infty}x^{s/2-1}\E^{-\pi n^2 x}\D x= \frac{\Gamma( s/2)}{n^s \pi^{s/2}}
$$
where $\Gamma (x)$ is the Gamma function (see e.g. \cite{Bateman}, Vol. 1, Sect.~1).
Therefore if Re $s>1$
$$
\frac{\Gamma (s/2)\zeta (s)}{\pi^{s/2}}= \int_0^{\infty} x^{s/2-1}\Psi(x)\D x
$$
where $\Psi(x)$ is given by the following series
$$
\Psi(x)=\sum_{n=1}^{\infty} \E^{-\pi n^2 x}\;.
$$
Using the Poisson summation formula (\ref{Poisson}) one obtains
$$
\sum_{n=-\infty}^{\infty} \E^{-\pi n^2 x}=\frac{1}{\sqrt{x}}
\sum_{n=-\infty}^{\infty} \E^{-\pi n^2/x}
$$
which leads to the identity
$$
2\Psi(x)+1=\frac{1}{\sqrt{x}}\left (2\Psi(\frac{1}{x})+1\right )\;.
$$
Hence
\begin{eqnarray*}
\xi(s)&\equiv & \pi^{-s/2}\Gamma(\frac{1}{2} s)\zeta(s)=\int_0^1
x^{s/2}\Psi(x)\D x +\int_1^{\infty} x^{s/2}\Psi(x)\D x=\nonumber \\
&=&
\int_0^1 x^{s/2}\left (\frac{1}{\sqrt{x}}\Psi(\frac{1}{x})
+\frac{1}{2\sqrt{x}}-\frac{1}{2}\right )\D x+
\int_1^{\infty} x^{s/2}\Psi(x)\D x=\\
&=&\frac{1}{s-1}-\frac{1}{s}+\int_0^1 x^{s/2-3/2}\Psi(\frac{1}{x})\D x
+\int_1^{\infty} x^{s/2}\Psi(x)\D x=
\nonumber \\
&=&\frac{1}{s(s-1)}+
\int_1^{\infty}\left (x^{-s/2-1/2}+ x^{s/2-1}  \right )\Psi(x)\D x\;.
\end{eqnarray*}
The last integral is convergent for all values of $s$ and gives the
analytical continuation of the Riemann zeta function to the whole
complex $s$-plane, the only singularity being the pole at $s=1$ with unit  residue 
$$
\zeta(s)\stackrel{s\to 1}{\longrightarrow} \frac{1}{s-1}\;.
$$
(The pole at $s=0$ is canceled by the pole of $\Gamma(s/2)$ giving 
$\zeta(0)=-1/2$.)
 
One of important consequences of the above formula of analytical
continuation  is that it does not change under the substitution 
$s\to 1-s$. Therefore for all values of $s$
$$
\xi (s)=\xi (1-s)
$$
or
$$
\zeta(s)=\varphi(s)\zeta(1-s)
$$
where
\begin{equation}
\varphi(s)=\pi^{s-1/2}\frac{\Gamma(1/2-s/2)}{\Gamma(s/2)}
\label{newphis}
\end{equation}
By standard formulas (see e.g. \cite{Bateman}, Vol. 1, 1.2.5, 1.2.15)
$$
\Gamma (x)\Gamma (1-x)=\frac{\pi}{\sin \pi x}\;,\;\;\;
\Gamma(2x)=2^{2x-1}\pi^{-1/2} \Gamma (x)\Gamma (x+\frac{1}{2})
$$
the last expression can be transformed to (\ref{phis}) which proves
the functional equation (\ref{functionalZeta}).

From the functional equation  (\ref{functionalZeta}) is follows
that $\zeta(s)$ has 'trivial' zeros at negative even integers (except zero) 
$s=-2,-4,\ldots$ which appear from $\sin (\pi s/2)$ in (\ref{phis}). 
All other non-trivial zeros, $\zeta(s_n)=0$, are situated in the 
so-called critical strip 
$0<\mbox{ Re }s<1$. If one denotes these zeros as 
$s_n=1/2+\imag \gamma_n$ then functional equation together with the
fact that $\zeta(s)^*=\zeta(s^*)$ state that in general there exit
4 sets of zeros: $\gamma_n\;,\;-\gamma_n \;,\;\gamma_n^*\;,\;-\gamma_n^*$. 

According to the famous {\em Riemann conjecture} (see e.g. \cite{Titchmarsh}) 
all nontrivial zeros of $\zeta(s)$ lie at the symmetry line Re $s=1/2$  or
$\gamma_n$ are all  real quantities. Numerical calculations confirms this 
conjecture for exceptionally large number of zeros (see e.g. \cite{OdlyzkoPaper}
and the web site of Odlyzko \cite{Web}) but a mathematical proof is still absent. 

\subsection{Trace Formula for the Riemann Zeros}\label{TraceRiemann}

Let us fix a test function $h(r)$ exactly as it was done for the Selberg
trace formula in Sect.~\ref{Selberg} i.e.
\begin{itemize}
\item $h(r)$  is a function analytical in the region $|\mbox{Im }r|$ $\leq 1/2+\delta$,
\item $h(-r)=h(r)$,
\item $|h(r)|\leq A(1+|r|)^{-2-\delta}$.
\end{itemize}
Denote as in that Section
$$
g(u)=\frac{1}{2\pi}\int_{-\infty}^{+\infty}h(r)\E^{-iru}\D r
$$
and define
$$
H(s)=\int_{-\infty}^{+\infty}g(u)\E^{(s-1/2)u}\D u\;.
$$
Now let us compute the integral
$$
\frac{1}{2\pi \imag}\oint \D s H(s)\frac{\zeta'(s)}{\zeta(s)}
$$
where the contour of integration is taken over the rectangle
$-\eta \leq \mbox{ Re }s\leq 1+\eta$ and 
$-T \leq \mbox{ Im }s\leq T$ with  $0<\eta<\delta$ and $T\to +\infty$. Inside
this rectangle there are poles of $\zeta'(s)/\zeta(s)$ coming from
non-trivial zeros of the Riemann zeta function, $s_n=1/2+\imag \gamma_n$, and
the one from the pole of $\zeta(s)$ at $s=1$. The total contribution from these poles is
$$
\sum_{n}h(\gamma_n)-h(-\frac{\imag}{2})\;.  
$$
One can check that the limit $T\to\infty$ exists and, consequently, one has the identity
$$
\sum_{n}h(\gamma_n)-h(-\frac{\imag}{2})=
\frac{1}{2\pi \imag}\int_{1+\eta-\imag \infty}^{1+\eta+\imag \infty}
\D s H(s)\frac{\zeta'(s)}{\zeta(s)}-
\frac{1}{2\pi \imag}\int_{-\eta-\imag \infty}^{-\eta+\imag \infty}
\D s H(s)\frac{\zeta'(s)}{\zeta(s)}\;.
$$  
Let us substitute in the second integral the functional equation 
(\ref{functionalZeta}) with $\varphi(s)$ from (\ref{newphis}). One has
$$
\frac{\zeta'(s)}{\zeta(s)}=\ln \pi -\frac{\zeta'(1-s)}{\zeta(1-s)}-
\frac{1}{2}\left [\frac{\Gamma'}{\Gamma}\left (\frac{s}{2}\right)+
\frac{\Gamma'}{\Gamma}\left (\frac{1-s}{2}\right ) \right ]\;.
$$
Now all integrals converge and one can move the integration contour
till $s=1/2+\imag r$ with real $r$. In this manner one obtains
\begin{eqnarray*}
&&\frac{1}{4\pi \imag}\int_{-\eta-\imag \infty}^{-\eta+\imag \infty}
\D s H(s)\left [\frac{\Gamma'}{\Gamma}\left (\frac{s}{2}\right)+
\frac{\Gamma'}{\Gamma}\left (\frac{1-s}{2}\right ) \right ]
=\nonumber\\
&&=h(\frac{\imag}{2})+
\frac{1}{2\pi}\int_{-\infty}^{+\infty}h(r)\frac{\Gamma'}{\Gamma}
\left (\frac{1}{4}+\frac{\imag}{2} r \right )\D r \;.
\end{eqnarray*}
The first term in the right hand side of this equality is due to the
appearance of the pole of $\Gamma(s/2)$ at $s=0$ when the integration
contour shifted till $s=1/2+\imag r$. Also we have used that $h(-r)=h(r)$.

For terms with the Riemann zeta function one can use the expansion which
follows from (\ref{ZetaDefinition})
$$
\frac{\zeta'(s)}{\zeta(s)}=-\sum_{p}\ln p \sum_{n=1}^{\infty}p^{-ns}\;. 
$$
Shifting the integration contour as above (i.e. till $s=1/2+\imag r$), using
that $g(-u)=g(u)$, and combining all terms together one gets the
following Weil explicit formula for the Riemann zeros
\begin{eqnarray*}
\sum_{\begin{array}{c}\mbox{\scriptsize{non-trivial}}\\
\mbox{\scriptsize{zeros}}\end{array}}h(\gamma_n)&=&
\frac{1}{2\pi}\int_{-\infty}^{\infty}
h(r)\frac{\Gamma'}{\Gamma}\left (\frac{1}{4}+\frac{\imag }{2}r\right )\D r+
h(\frac{\imag }{2})+h(-\frac{\imag }{2})-\\
&-&g(0)\ln \pi-2\sum_{\mbox{\scriptsize{primes}}}\ln p\sum_{n=1}^{\infty}
  \frac{1}{p^{n/2}}g(n\ln p)\;.
\end{eqnarray*}
Here $\gamma_n$ are related with non-trivial zeros of the Riemann zeta 
function, $s_n$, as follows
$$
s_n=\frac{1}{2}+\imag \gamma_n\;.
$$
This formula is an analog of usual trace formulas as it relates zeros of
the Riemann zeta function defined in a quite complicated manner with prime
numbers which are a common notion. 

The similarity with dynamical trace formulas is more striking if one
assumes the validity of the Riemann conjecture which states that $\gamma_n$
are real quantities (which in a certain sense can be considered as
energy levels of a quantum system). In  'semiclassical' limit $r\to \infty$ using
the Stirling formula (see e.g. \cite{Bateman}, Vol. 1, 1.9.4)
$$
\ln \Gamma (z)\stackrel{|z|\to \infty }{\longrightarrow}
(z-\frac{1}{2})\ln z -z +\frac{1}{2}\ln 2\pi
$$
one obtains that the density of Riemann zeros 
$$
d(E)=\sum_n\delta (E-\gamma_n)
$$ 
can be expressed by the following `physical' trace formula valid at large $E$
$$
d(E)=\bar{d}(E)+d^{(osc)}(E)
$$
where
$$
\bar{d}(E)=\frac{1}{2\pi}\ln \frac{E}{2\pi}+\mbox{ corrections}\;,
$$
and
$$
d^{(osc)}(E)=-\frac{1}{\pi}\sum_p\sum_{n=1}^{\infty} \frac{\ln p}{p^{n/2}}
\cos(En\ln p)
$$
where the summation is performed over all prime numbers.

\subsection{Chaotic Systems and the Riemann Zeta Function}\label{ChaoticRiemann}

By comparing the above equations with the trace formulas of chaotic systems 
one observes (see e.g. \cite{Hejhal3}, \cite{Berry3}, \cite{Berry2}) a remarkable
correspondence between different quantities in  these trace formulas
\begin{itemize}
\item periodic orbits of chaotic systems $\leftrightarrow $ primes,
\item periodic orbit period $T_p \;\;\leftrightarrow \;\;\ln p$,
\item convergence properties of both formulas are also quite similar.
\end{itemize}
The number of periodic orbits with period less than $T$ for chaotic systems is
asymptotically
$$
N(T_p<T)=\frac{\E^{hT}}{hT},
$$
where a constant $h$ is called the topological entropy. 

The number of prime numbers less than $x$ is given by the prime number
theorem (see e.g. \cite{Titchmarsh})
$$
N(p<x)=\frac{x}{\ln x}\;.
$$
As $\ln p \equiv T_p$ this expression has the form similar to number of
periodic orbits of chaotic systems with $h=1$
$$
N(T_p<T)=\frac{\E^{T}}{T}\;.
$$
Due to these similarities number-theoretical  zeta functions play the role
of a simple (but by far non-trivial) model of quantum chaos.

Notice that the overall signs of the oscillating part of trace formulas for
the Riemann zeta function  and  dynamical systems are different. According 
to Connes \cite{Connes} it may be interpreted as Riemann zeros belong not to
a spectrum of a certain self-adjoint operator but to an 'absorption' 
spectrum. Roughly speaking it means the following. Let us assume that
the spectrum of a 'Riemann Hamiltonian' is continuous and it covers the
whole axis. But exactly when eigenvalues equal Riemann zeros 
corresponding eigenfunctions of this Hamiltonian vanish. Therefore these 
eigenvalues do not belong to the spectrum and Riemann zeros correspond
to such missing points similarly to black lines (forming absorption 
spectra) which are visible when light passes through an absorption 
media. In Connes' approach the 'Riemann Hamiltonian' may be very
simple (see also \cite{BerryKeating}) but the functional space where
it has to be defined is extremely intricate.

\section{Summary}\label{SummaryTrace}

Trace formulas can be constructed for all `reasonable' systems. They express
the quantum density of states (and other quantity as well) as a sum over
classical periodic orbits. All quantities which enter trace formulas can be
computed within pure classical mechanics. 

Trace formulas consist of two terms
$$
d(E)=\bar{d}(E)+d^{(osc)}(E).
$$
The smooth part of the density, $\bar{d}(E)$, for all dynamical systems 
is given by the Thomas--Fermi formula (plus corrections if necessary)
$$
\bar{d}(E)=\int \frac{\D \vec{p}\D \vec{x}}{(2\pi \hbar)^f}\delta
\left (E-H(\vec{p},\vec{x})\right )\;.
$$
For integrable systems the oscillating part of the density, $d^{(osc)}(E)$, is 
$$
d^{(osc)}(E)=\sum_{\vec{N}}\left (\frac{2\pi }{\hbar T_p}\right )^{(f-3)/2}
\frac{1}{\hbar^2\sqrt{|(N_i Q_{ij} N_j)|}}
\exp \left (\imag \frac{S_p}{\hbar}-\imag \frac{\pi}{4}\vec{N}\vec{\mu}
+\imag \frac{\pi}{4}\beta\right )
$$
where $S_p=2\pi \vec{N}\vec{I}$ is the action over a classical periodic
orbit with fixed winding numbers $\vec{N}$ and $Q_{ij}$ is the
co-matrix of the matrix of the second derivatives of the Hamiltonian.

For chaotic systems $d^{(osc)}(E)$ is represented as a sum over all
classical periodic orbits
$$
d^{(osc)}(E)=\sum_{\mbox{\scriptsize p.p.o.}}
\frac{T_p}{\pi \hbar}\sum_{n=1}^{\infty} \frac{1}{|\det (M_p^n-1)|^{1/2}}
\cos \left (n\frac{S_p}{\hbar}-n\frac{\pi}{2} \mu_p\right )
$$
where $S_p$ is the classical action along a primitive periodic
trajectory and $M_p$ is its monodromy matrix.

Usually trace formulas represent the dominant contribution when 
$\hbar \to 0$. They are exact only in very special cases as for constant
negative curvature surfaces generated by discrete groups where they coincide
with the Selberg trace formula. For a group with only hyperbolic elements 
$$
\bar{d}(E)=\frac{\mu(D)}{4\pi}\tanh \pi k
$$
where $\mu(D)$ is the area of the fundamental domain of the group and
$$
d^{(osc)}(E)=\sum_{\mbox{\scriptsize{p.p.o.}}}\frac{l_p}{4\pi k} 
\sum_{n=1}^{\infty} \frac{\cos (knl_p)}{\sinh (nl_p/2)}
$$
where $l_p$ are lengths of  periodic orbits. 

The formulas similar to trace formulas exist also for number-theoretical zeta 
functions (assuming the generalized Riemann conjecture). In
particular, for the Riemann zeta function
$$
\bar{d}(E)=\frac{1}{2\pi}\ln \frac{E}{2\pi}
$$
and
$$
d^{(osc)}(E)=-\frac{1}{\pi}\sum_{\mbox{\scriptsize prime}}
\sum_{n=1}^{\infty} \frac{\ln p}{p^{n/2}}
\cos(En\ln p)\;.
$$
The principal difficulty of all trace formulas is the divergence of the sums 
over periodic orbits. To obtain a mathematically meaningful formula one 
considers instead of the singular density of states its smoothed version
defined as a sum over all eigenvalues of a suitable chosen smooth 
test-function. When its Fourier harmonics decrease quickly the resulting formula 
represent a well defined object. 

\paragraph{Suggestions for Further Readings}
\begin{itemize}

\item A very detailed account of trace formulas derived by multiple 
scattering method can be found in a series of papers by Balian and 
Bloch \cite{BalianBloch}.

\item A short but concise mathematical review of hyperbolic geometry
is given in \cite{Katok}.

\item Explicit forms of the Selberg trace formula for general discrete groups
with elliptic and parabolic elements are presented in two volumes
of Hejhal's monumental work  \cite{Hejhal2} which contains practically
all known information about the Selberg trace formula.

\item In \cite{Hejhal3} one can find a mathematical discussion about different
relations between number-theoretical zeta functions and dynamical systems.

\end{itemize} 
    
\chapter{Statistical Distribution of Quantum Eigenvalues}\label{Statistics}
\setcounter{section}{0}

Wigner and Dyson in the fifties had proposed to describe complicated 
(and mostly unknown) Hamiltonian of heavy nuclei by a member of an ensemble of
random matrices and they argued that  the type of this
ensemble  depends only on the symmetry of the Hamiltonian. 
For systems without time-reversal invariance the relevant ensemble is the
Gaussian Unitary Ensemble (GUE), for systems invariant with respect to 
time-reversal the  ensemble is the Gaussian Orthogonal Ensemble (GOE) 
and for systems with time-reversal invariance but with half-integer spin
energy levels have to be described according to the Gaussian Symplectic
Ensemble (GSE) of random matrices. 

For these classical ensembles all correlation functions which determines
statistical properties of eigenvalues $E_n$ can be written explicitly
(see e.g. \cite{Mehta}, \cite{Bohigas}). The simplest of them is the
one-point correlation function or the mean level density, $\bar{d}(E)$, which
is the probability density of finding a level in the interval
$(E, \;E+\D E)$. When $\bar{d}(E)$ is known one can construct a new sequence
of levels, $e_n$, called unfolded spectrum  as follows
$$
e_n=\int^{E_n}\bar{d}(E)\D E\;.
$$
This artificially constructed sequence has automatically unit local 
mean density which signifies  that   the mean level density (provided it is a
smooth function of $E$) plays a minor role in describing statistical properties
of a spectrum at small intervals.

The two-point correlation function, $R_2(\epsilon )$, is the probability density
of finding two levels separated by a distance in the interval 
$(\epsilon,\;\epsilon+\D \epsilon)$.
The characteristic properties of the above ensembles is the phenomenon 
of level repulsion which manifest itself in the vanishing of the two-point 
correlation function at small values of argument
$$
R_2(\epsilon)\stackrel{\epsilon \to 0}{\longrightarrow} \epsilon^{\beta}
$$
where the parameter $\beta=1,2,$ and $4$ for, respectively, GOE, GUE, and GSE.
This behaviour is in contrast with the case of the Poisson statistics
of independent random variables where
$$
R_2(\epsilon)\stackrel{\epsilon \to 0}{\longrightarrow} \bar{d}(E)\neq 0\;.
$$
For  later use we present the explicit form of the two-point
correlation function for GUE with mean density $\bar{d}$
\begin{equation}
\tilde{R}_2(\epsilon)=\bar{d}^2+\bar{d}\delta (\epsilon)+
\bar{R}_2(\epsilon)+R_2^{(osc)}(\epsilon)
\label{GUEd}
\end{equation}
where the smooth part of the connected two-point correlation function
is given by
\begin{equation}
\bar{R}_2(\epsilon)=-\frac{1}{2\pi^2 \epsilon^2}
\label{smoothGUE}
\end{equation}
and its oscillating part is
\begin{equation}
R_2^{(osc)}(\epsilon)=
\frac{\E^{2\pi \imag  \bar{d}\epsilon}+\E^{-2\pi \imag \bar{d}\epsilon}}
{4\pi^2 \epsilon^2}\;.
\label{oscGUE}
\end{equation}
The term $\bar{d}\delta (\epsilon)$ in (\ref{GUEd}) corresponds to
taking into account two identical levels and it is universal for
all systems without spectral degeneracy. It is a matter of convenience to
include it to $R_2(\epsilon)$ or not. When one adopts the
definition (\ref{CorrFunct}) the appearance of such terms is inevitable. 

Another useful quantity is the two-point correlation form factor defined
as the Fourier transform of the two-point correlation function 
(unfolded to the unit density)
\begin{equation}
K(t)=\int_{-\infty}^{\infty}R_2(x)e^{2\pi i  t x}dx\;.
\label{Ktwo}
\end{equation}
For convenience one introduces a factor $2\pi$ in the definition of time. 

\begin{figure}
\center
\includegraphics[height=7cm, angle=-90]{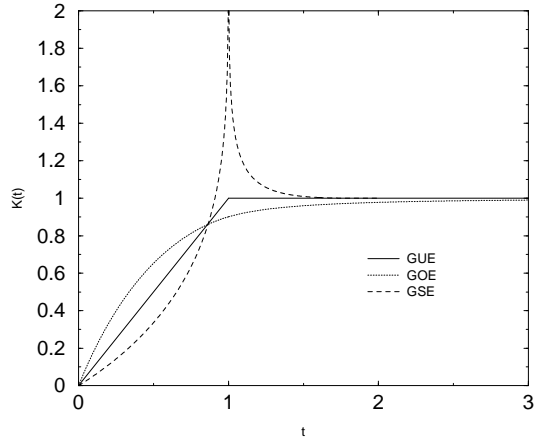}
\caption{Two point correlation form factor of classical random matrix
ensembles.}
\label{kt}
\end{figure}

In Fig.~\ref{kt} the two-point correlation form factors for usual
random matrix ensembles are presented.
Their explicit formulas can be found in \cite{Mehta}, \cite{Bohigas}.
For these classical ensembles small-$t$ behaviour of the form factors is
\begin{equation}
K(t)\stackrel{t\to 0}{\longrightarrow} \frac{2}{\beta}t
\label{smallt}
\end{equation}
with the same $\beta$ as above.

\begin{figure}
\center
\includegraphics[height=7cm, angle=-90]{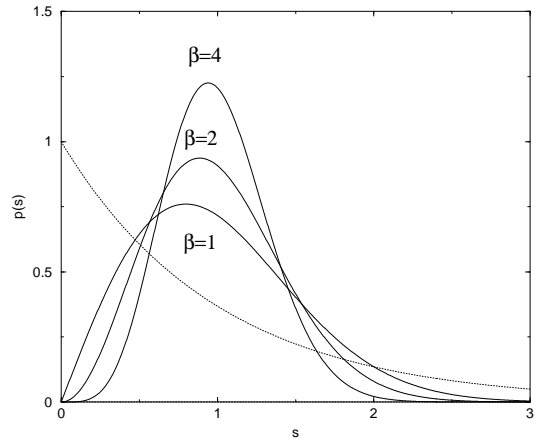}
\caption{Nearest-neighbor distribution for the standard random matrix
ensembles. Dotted line -- the Poisson prediction}
\label{psrmt}
\end{figure}

The nearest-neighbor distribution, $p(s)$, is defined as the probability
density of finding two levels separated by distance $s$ but, contrary
to the two-point correlation function, no levels inside this interval
are allowed. For classical ensembles the nearest-neighbor distributions
can be expressed through solutions of  certain integral equations and
numerically they are close to the Wigner surmise (see e.g. \cite{Bohigas})
$$
p(s)=as^{\beta}\E^{-bs^2}
$$
where $\beta$ is the same as above and constants $a$ and $b$ are
determined from normalization conditions
$$
\int_0^{\infty}p(s)\D s=\int_0^{\infty}sp(s)\D s=1\;.
$$
These functions are presented at Fig.~\ref{psrmt} together with the
Poisson prediction for this quantity  $p(s)=\E^{-s}$.

Though random matrix ensembles were first introduced to describe spectral
statistics of heavy nuclei later it was understood that the same
conjectures can be applied also for simple dynamical systems and
to-day standard accepted conjectures are the following

\begin{itemize}
\item The energy levels of classically integrable systems on the scale of
  the mean level density behave as independent random variables and their
  distribution is close to the Poisson distribution \cite{BerryTabor}. 

\item The energy levels of classically chaotic systems are not independent
  but on the scale of the mean level density they are distributed as 
  eigenvalues of random matrix ensembles depending
  only on symmetry properties of the system considered \cite{BohigasGiannoni}. 
\begin{itemize}
\item  
  For systems without time-reversal invariance the distribution of energy
  levels should be close to the distribution of the Gaussian Unitary
  Ensemble (GUE) characterized by quadratic level repulsion. 
\item
  For systems with time-reversal invariance the corresponding distribution
  should be close to that of the Gaussian Orthogonal Ensemble (GOE) with
  linear level repulsion.
\item
  For systems with time-reversal invariance but with half-integer spin
  energy levels should be described according to the Gaussian Symplectic
  Ensemble (GSE) of random matrices with quartic level repulsion.
\end{itemize}  
\end{itemize}
These conjectures are well confirmed by numerical calculations. 

The purpose of this Chapter is to investigate methods which permit to obtain 
spectral statistics analytically. For a large part of the Section we follow
\cite{Bogomolny2}. In Sect.~\ref{Correlation} a formal expression is obtained
which relates correlation functions with products of trace formulas. In 
Sect.~\ref{Diagonal} the simplest approximation to compute such products is 
discussed. It is called the diagonal approximation and it consists of taking 
into account only terms with exactly the same actions. Unfortunately, for 
chaotic systems this approximation can be used, strictly speaking, only for very 
small time estimated in Sect.~\ref{Criterion}. To understand the 
behaviour of the correlation functions for longer time  more complicated methods 
of calculation of non-diagonal terms have to be developed. In Sect.~\ref{Beyond} 
this goal is achieved for the Riemann zeta function. To obtain the information about
correlations of prime pairs we use the Hardy--Littlewood conjecture which 
is reviewed in Sect.~\ref{HardyLittlewood}. The explicit form of the two-point
correlation function for the Riemann zeros is obtained in Sec.~\ref{CorrelationRiemann}.
In Sect.~\ref{SummaryStatistics} it is 
demonstrated that the obtained expression very well agrees with numerical
calculations of spectral statistics for Riemann zeros.      

\section{Correlation Functions}\label{Correlation}

Formally $n$-point correlation functions of energy levels are defined as the
probability density of having $n$ energy levels at given positions. Because
the density of states, $d(E)$, is the probability density of finding
one level at point $E$, correlation functions are connected to
the density of states as follows
\begin{equation}
R_n(\epsilon_1,\epsilon_2,\ldots,\epsilon_n)=
\left \langle d(E+\epsilon_1)d(E+\epsilon_2)\ldots d(E+\epsilon_n)\right \rangle \;.
\label{CorrFunct}
\end{equation}
The brackets $\left\langle \dots\right\rangle$ denote a smoothing
over an appropriate energy window
$$
\left\langle f(E)\right \rangle=\int f(E')\sigma (E-E')\D E'
$$
with a certain function $\sigma (E)$. Such smoothing means that one considers
eigenvalues of quantum dynamical systems at different intervals of energy as
forming a statistical ensemble.

\begin{figure}
\center
\includegraphics[height=5cm, angle=-90]{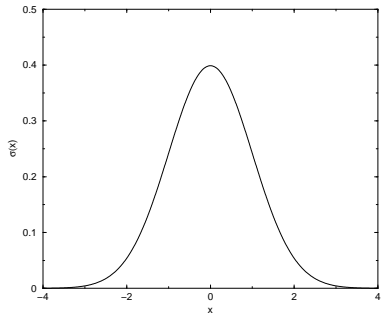}
\caption{Schematic form of smoothing function.}
\label{sigma}
\end{figure}

The function $\sigma (E)$ is assumed to fulfill the normalization condition 
$$
\int \sigma(E)dE=1
$$
and to be centered around zero with a width $\Delta E$  obeying inequalities
\begin{equation}
\Delta E_q \ll \Delta E \ll \Delta E_{cl} \ll E\;.
\label{inequalities}
\end{equation}
Here $\Delta E_q $ has to be of the order of the mean level spacing,
$\Delta E_q \approx 1/\bar{d}$, and
$\Delta E_{cl} $ denotes the energy scale at which classical dynamics
changes noticeably. A schematic picture of $\sigma(E)$ is represented 
at Fig.~\ref{sigma}. 

The trace formula for the density of states of chaotic systems was discussed in 
Chap.~\ref{Trace} and it has the form 
$$ 
d(E)=\bar{d}(E)+\sum_{p,n}A_{p,n}\E^{\imag n S_p(E)/\hbar} +\mbox{c.c.} 
$$ 
where the summation is performed over all primitive periodic orbits and 
its repetitions, and  
\begin{equation} 
A_{p,n}=\frac{T_p}{2\pi\hbar|\det (M_p^n-1)|^{1/2}}\E^{-\pi \imag n\mu_p/2}\;. 
\label{Apn} 
\end{equation} 
Substituting this expression in the formula for the 
two-point correlation function one gets 
\begin{eqnarray*} 
&&R_2(\epsilon_1, 
\epsilon_2)=\bar{d}^2+\\ &&+\sum_{p_i,n_i}A_{p_1,n_1}A_{p_2,n_2}^* \left \langle 
\exp \frac{i}{\hbar}(n_1S_{p_1}(E+\epsilon_1)-n_2S_{p_2}(E+\epsilon_2)) 
\right \rangle +\mbox{c.c.} \nonumber 
\end{eqnarray*} 
and the terms with the sum of actions are assumed to be washed out by the
smoothing procedure.

Expanding the actions and taking into account that 
$\partial S(E)/\partial E=T(E)$ where $T(E)$ is the classical period
of motion one finds
\begin{eqnarray*}
R_2^{(c)}(\epsilon_1, \epsilon_2)&=&
\sum_{p_i,n_i}A_{p_1,n_1}A_{p_2,n_2}^*
\left \langle \exp \frac{\imag }{\hbar}(n_1 S_{p_1}(E)-n_2 S_{p_2}(E))
\right \rangle \nonumber\\
&& \times 
\exp \frac{\imag}{\hbar}(n_1 T_{p_1}(E)\epsilon_1-n_2 T_{p_2}(E)\epsilon_2)+
\mbox{c.c.}\;.
\end{eqnarray*}
Here $R_2^{(c)}(\epsilon_1, \epsilon_2)$ is the connected part of the two-point 
correlation function $R_2(\epsilon_1, \epsilon_2)=\bar{d}^2+
R_2^{(c)}(\epsilon_1, \epsilon_2)$.

The most difficult part is the computation of the mean value of terms with 
the difference of actions
$$
\left \langle \exp \frac{\imag }{\hbar}(n_1 S_{p_1}(E)-n_2 S_{p_2}(E))
\right \rangle \;.
$$

\subsection{Diagonal Approximation}\label{Diagonal}

Berry \cite{Berry} proposed to estimate such  sums in an  approximation 
(called the diagonal approximation) by taking into account only terms with
{\em exactly} the same actions having in mind that terms with different
values of actions will be small after the smoothing. 

Let $g$ be the mean multiplicity of periodic orbit actions. 
Then the connected part of the two-point correlation function in the 
diagonal approximation is 
\begin{equation}
R_2^{(diag)}(\epsilon)=g\sum_{p,\;n\geq 1}|A_{p,n}|^2
\E^{\imag  n T_p(E)\epsilon/\hbar} +\mbox{c.c.}\;.
\label{R2diag}
\end{equation}
Here $\epsilon=\epsilon_1-\epsilon_2$ and the sum is taken over all
primitive periodic  orbits.

From (\ref{R2diag}) it follows that the two-point correlation form factor  
$$
K(t)=\int_{-\infty}^{+\infty}R_2(\epsilon) \E^{2\pi \imag t \epsilon}\D\epsilon.
$$ 
in the diagonal approximation equals the following sum over classical periodic orbits 
\begin{equation}
K^{(diag)}(t)= 2\pi g\sum_{p,n}|A_{p,n}|^2
\delta \left (2\pi t - \frac{nT_p(E)}{\hbar}\right ) +\mbox{c.c.}\;.
\label{Kdiag}
\end{equation}
According to the Hannay-Ozorio de Almeida sum rule \cite{HannayOzorio} sums
over periodic orbits of a chaotic systems can be calculated
by using the local density of periodic orbits related with the 
monodromy matrix, $M_p$,  as follows
$$
\D \rho_p=\frac{\D T_p}{T_p}|\det (M_p-1)|\;.
$$
Using (\ref{Apn}) one gets 
$$
K^{(diag)}(t)=\frac{g}{2\pi \hbar} \int T_p
\delta (2\pi t-\frac{T_p}{\hbar})dT_p=gt
$$
where $g$ is the mean multiplicity of periodic orbits (i.e. the mean proportion
of periodic orbits with exactly the same action). For generic systems
without time-reversal invariance there is no reasons for equality of actions
for different periodic orbits and $g=1$ but for systems with time-reversal
invariance each orbit can be traversed in two directions therefore  in general 
for such systems $g=2$. Comparing these expressions  one concludes
that the diagonal approximation reproduces the correct  
small-$t$ behavior of form-factors of classical ensembles 
(cf.~(\ref{smallt})). 

Unfortunately, $K^{(diag)}(t)$ grows with increasing of $t$ but the exact
form-factor for systems without spectral degeneracy should tends to $\bar{d}$
for large $t$. This is a consequence of the following arguments.
According to (\ref{CorrFunct}) 
\begin{eqnarray*}
R_2(\epsilon)&=&\left \langle 
\sum_{m,n}\delta (E-E_n)\delta (E+\epsilon -E_m)\right\rangle=
\nonumber\\
&=&\left \langle 
\sum_{n}\delta (E-E_n) \delta (\epsilon -E_m+E_n) \right\rangle\;.
\end{eqnarray*}
If there is no levels with exactly the same energy the
second $\delta$-function in the right hand side of this equation
tends to $\delta(\epsilon)$ when $\epsilon\to 0$ and the first
one gives $\bar{d}$. Therefore  
$$
R_2(\epsilon)\rightarrow \bar{d}\delta(\epsilon)\;, \;\;\mbox{ when }
\epsilon \rightarrow 0
$$
which is equivalent to the following asymptotics of the form factor 
$$
K(t)\rightarrow \bar{d}\;, \;\;\mbox{ when }t\rightarrow \infty\;.
$$
This evident contradiction clearly indicates that the diagonal approximation
for chaotic systems cannot be correct for all values of $t$ and
more complicated tools are needed to obtain the full form factor.

\subsection{Criterion of Applicability of Diagonal
Approximation}\label{Criterion}

One can give a (pessimistic) estimate till what time the diagonal
approximation can be valid by the following method. The main ingredient of the
diagonal approximation is the assumption that after smoothing all
off-diagonal terms give negligible contribution. This condition is almost the same
as the condition of the absence of quantum interference. But it is known that
the quantum interference is not important for times smaller than the
Ehrenfest time which is of the order of
$$
t_E\approx \frac{1}{\lambda_0}\ln (1/\hbar),
$$
where $\lambda_0$ is a (classical) constant of the order of the Lyapunov
exponent defined in such a way that the mean splitting of two nearby
trajectories at
time $t$ grows as $\exp (\lambda_0 t)$. For billiards $(ka)^{-1}$, 
where $a$ is of the order of system size, plays the role of $\hbar$   
and $\lambda_0=k \lambda$
where  $k$ is the momentum and $\lambda$ determines the deviation of two
trajectories with length $L=kt$. The constant $\lambda$ which we
also called the Lyapunov exponent  is independent on $k$ for billiards and
$$
t_E\approx \frac{1}{\lambda k}\ln(ka)\;.
$$
In the semiclassical limit $k\to \infty$ the Ehrenfest time and, 
consequently, the time during which one can use the diagonal
approximation tends to zero as $\ln k/k$.

More careful argumentation can be done as follows. The
off-diagonal terms can be neglected if
$$
\left | \left \langle \exp \frac{\imag}{\hbar}(S_{p_1}(E)-S_{p_2}(E)) 
\right \rangle \right |\ll 1\;.
$$
But this quantity is small provided the difference of periods of two orbits
$\Delta T=T_{p1}-T_{p_2}$ times the energy window $\Delta E$ used in the
definition of smoothing procedure is large
\begin{equation}
\frac{1}{\hbar}(T_{p1}-T_{p_2})\Delta E \gg 1\;.
\label{criter}
\end{equation}
For billiards $T_p=L_p/k$ and this condition means that one has to consider
all periodic orbits such that their difference of lengths is
$$
L_{p1}-L_{p_2}\gg \frac{\hbar k}{\Delta E}\;.
$$
But the number of periodic orbits with the length $L$ for chaotic systems 
grows exponentially
$$
N(L_p<L)=\frac{e^{\lambda L}}{\lambda L}
$$
where $\lambda$ is a constant of the order of the Lyapunov exponent.
Therefore in the interval $L, L+\delta l$ there is $e^{ \lambda L}\delta l/L$
orbits and  the mean difference of lengths between orbits with the lengths
less than $L$ is of the order of
$$
\Delta L =L\exp (-\lambda L)\;.
$$
To fulfilled the above condition one has to restrict the maximum length of
periodic orbits, $L_m$, by
$$
L_m\exp (-\lambda L_m)\approx \frac{k \hbar}{ \Delta E}\;.
$$
In the limit of large $L_m$ with logarithmic accuracy this relation  gives
\begin{equation}
L_m\approx \frac{1}{\lambda }\ln \frac{ \Delta E}{k\hbar \lambda }
\label{LMax}
\end{equation}
which corresponds to the maximal time till the diagonal approximation 
can be applied
$$
t_m=\frac{L_m}{k}\sim \frac{1}{\lambda k}\ln \frac{\Delta E}{\lambda k}\;.
$$
As $\Delta E \ll E=k^2$, $t_m<t_E$. 

Another important time scale for bounded quantum systems is called the
Heisenberg time, $t_H$. It is the time during which one can see the
discreteness of the spectrum 
$$
t_H=2\pi \bar{d}\;.
$$
As for billiards $\bar{d}$ is a constant 
$$
t_E\ll t_H\;.
$$
For the Riemann zeta function the situation is better because (i) in
this case `momentum'  plays the role of `energy' (the 'action' $E\ln p$ is 
linear in $E$ and not quadratic as for dynamical systems) and (ii) the density 
of states for the Riemann zeta function is  $(\ln (E/2\pi))/(2\pi)$. 

The analog of (\ref{criter}) in this case is
$$
(\ln p_1-\ln p_2) \Delta E\gg 1\;.
$$
It means that to apply the diagonal approximation prime numbers have to be such 
that the difference between any two of them  obeys
$$
\frac{\delta p}{p}\Delta E\gg 1.
$$
The difference between primes near $p$ is of the order of $\ln p$. Hence
from the above inequalities it follows that diagonal approximation can be
used till time $t_m=\ln p_m$ where $p_{m}$ is such that
$$
\frac{\ln p_m}{p_m}\geq \frac{1}{\Delta E}\;.
$$     
Or with logarithmic precision $p_m\leq \Delta E$. As $\Delta E\leq E$ 
(see (\ref{inequalities})),  $p_m\sim E$ and the maximum time 
$$
t_m\sim \ln E =2\pi \bar{d}(E)
$$ 
i.e. the diagonal approximation for the Riemann zeta function is valid till
the Heisenberg time which agrees with the Montgomery theorem \cite{Montgomery}.

This type of estimates clearly indicates that the diagonal approximation
for chaotic dynamical systems can not, strictly speaking,  be used to obtain an
information about the form-factor for large value of $t$. Only
the short-time behaviour of correlation functions can be calculated by
this method. (Notice that for GUE systems the diagonal approximation gives the
expected answer till the Heisenberg time but it just signifies that one has 
to find special reasons why all other terms cancel.) 

\section{Beyond the Diagonal Approximation}\label{Beyond}

The simplest and the most natural way of semi-classical computation of 
the two-point correlation functions is to find a method of calculating 
off-diagonal terms.
We shall discuss here this type of  computation on the example of the
Riemann zeta function where much more information then for dynamical systems 
is available (for the latter see \cite{KeatingBogomolny} and \cite{Bogomolny2}). 

The trace formula for the Riemann zeta function may be rewritten in the
form
$$
d^{(osc)}(E)=-\frac{1}{\pi}\sum_{n=1}^{\infty} \frac{1}{\sqrt{n}}
\Lambda(n)\cos(E\ln n)
$$
where
$$
\Lambda(n)=\left \{ \begin{array}{lr}
\ln p, & \mbox{if } n=p^k\\
0, & \mbox{otherwise}
\end{array} \right . \;.
$$
The connected two-point correlation function of the Riemann zeros,
$R_2^{(c)}=R_2-\bar{d}^2$, is
$$
R_2^{(c)}(\epsilon_1,\epsilon_2)=\frac{1}{4\pi^2}\sum_{n_1,n_2}
\frac{\Lambda(n_1)\Lambda(n_2)}{\sqrt{n_1n_2}}
\left \langle \E^{\imag (E+\epsilon_1)\ln n_1-\imag (E+\epsilon_2)\ln n_2}
\right \rangle  + \mbox{c.c.}\;.
$$
The diagonal approximation corresponds to taking into account terms
with $n_1=n_2$
\begin{eqnarray*}
R_2^{(diag)}(\epsilon_1,\epsilon_2)&=&\frac{1}{4\pi^2}\sum_{n}
\frac{\Lambda^2(n)}{n}
\E^{\imag (\epsilon_1-\epsilon_2)\ln n} +\mbox{c.c.}=
\nonumber \\
&=&
\frac{1}{4\pi^2}\sum_{p,m}
\frac{\ln^2 p}{p^{m}}
\E^{\imag (\epsilon_1-\epsilon_2)m\ln p} +\mbox{c.c.}\;.
\end{eqnarray*}
This expression may  be transform as follows (cf. \cite{AltshulerAndreev})
$$
R_2^{(diag)}(\epsilon)=-\frac{1}{4\pi^2}
\frac{\partial^2}{\partial \epsilon ^2}\ln \Delta(\epsilon)
$$
where
$$
\Delta(\epsilon)=|\zeta(1+\imag \epsilon)|^2\Phi^{(diag)}(\epsilon)\;,
$$
and function $\Phi^{(diag)}(\epsilon)$ is given by a convergent sum over
prime numbers
$$
\Phi^{(diag)}(\epsilon)=\exp \left ( 2\sum_p \sum_{m=1}^{\infty} 
\frac{1-m}{m^2p^m}\cos (m\epsilon \ln p )\right )\;.
$$
In the limit $\epsilon \rightarrow 0$,
$\zeta(1+\imag \epsilon)\rightarrow (\imag \epsilon)^{-1}$ and
$\Phi^{(diag)}(\epsilon)\rightarrow$ const. Therefore in this limit
$$
R_2^{(diag)}(\epsilon)\rightarrow -\frac{1}{2\pi^2\epsilon^2}
$$
which agrees with the smooth part of the GUE result (\ref{smoothGUE}).

The off-diagonal contribution  takes the form
$$
R_2^{(off)}(\epsilon_1,\epsilon_2)=\sum_{n_1\neq n_2}
\frac{\Lambda(n_1)\Lambda(n_2)}{4\pi^2\sqrt{n_1n_2}}
\left \langle \E^{\imag E\ln (n_1/n_2)+
\imag (\epsilon_1 \ln n_1-\epsilon_2\ln n_2)} \right \rangle  +
\mbox{c.c.}\;.
$$
The term $\exp (\imag E\ln (n_1/n_2))$ oscillates quickly if $n_1$ is not
close to $n_2$. Denoting
$$
n_1=n_2+r
$$
and expanding all smooth functions on $r$ one gets
$$
R_2^{(off)}(\epsilon)=\frac{1}{4\pi^2}\sum_{n,r}
\frac{\Lambda(n)\Lambda(n+r)}{n}
\left \langle \E^{\imag E r/n +\imag \epsilon \ln n}\right \rangle  +\mbox{c.c.}
$$ 
where $\epsilon=\epsilon_1-\epsilon_2$.

The main problem is clearly seen here. The function
$$
F(n,r)=\Lambda(n)\Lambda(n+r)
$$
is quite a wild function as it is nonzero only when both $n$ and $n+r$ are
powers of prime numbers. As we have assumed that $r\ll n$, the dominant
contribution to the two-point correlation function will come from the
mean value of this function over all $n$, i.e.
one has to substitute into $R_2^{(off)}(\epsilon )$ instead of $F(n,r)$ its mean value
$$
\alpha (r)=\lim_{N\rightarrow \infty}\frac{1}{N}\sum_{n=1}^N\Lambda(n)\Lambda(n+r)\;.
$$

\subsection{The Hardy-Littlewood Conjecture}\label{HardyLittlewood}

Fortunately the explicit expression for this function comes from the famous
Hardy--Littlewood conjecture. There are two different methods which permit
to `find' this conjecture. We start with the original Hardy-Littlewood
derivation \cite{HardyLittlewood}. 

First, let us remind two known facts. The number of prime numbers less that
a given number $N(p<x)$ is asymptotically (see e.g. \cite{Titchmarsh})
$$
N(p<x)=\frac{x}{\ln x}\;.
$$
Conveniently it can also be expressed in the following form
$$
\lim_{N\to \infty}\frac{1}{N}\sum_{n=1}^N\Lambda(n)=1\;.
$$
The number of prime number $N_{q,r}(p<x)$ in arithmetic progression of the
form $mq+r$ with $(r,q)=1$ and $r<q$ is given by the following asymptotic 
formula (see e.g. \cite{Dirichlet})
$$
N_{q,r}(p<x)=\frac{x}{\varphi(q)\ln x}
$$
where $\varphi(n)$ is the Euler function which counts integers less than
$n$ and co-prime with $n$
$$
\varphi(n)=n\prod_{p|n}\left (1-\frac{1}{p}\right )\;.
$$
As above, this relation can be rewritten in the equivalent form
\begin{equation}
\lim_{N\to \infty}\frac{1}{N}\sum_{m=1}^N\Lambda(mq+r)=\frac{1}{\varphi(q)}\;.
\label{Lqr}
\end{equation}
In the Hardy-Littlewood method \cite{HardyLittlewood} one introduces the function
$$
f(x)=\sum_{n=1}^{\infty} \Lambda(n)x^n
$$
which converges for all complex $x$ such that $|x|<1$. 

In the circle method of Hardy and Littlewood \cite{HardyLittlewood} one considers
the behaviour of
this function close to the unit circle when the phase of $x$ is near a
rational number $2\pi p/q$ with co-prime integers $p$ and $q$. One gets
$$
f(\E^{-u}\E^{2\pi \imag  p/q+\imag \delta})=\sum_{n=1}^{\infty} 
\Lambda(n)\E^{-n u}\E^{2\pi \imag n p/q+\imag n\delta}
$$
with $u,\delta \to 0$.

In the exponent  there is a quickly changing function $2\pi n p/q$. It is quite
natural to consider $n$ from the arithmetic progression
$$
n=mq+r
$$
with fixed $q$ and $r<q$. In this case
$$
f(\E^{-u}\E^{2\pi \imag p/q+\imag \delta})=\sum_{m,\ r} \Lambda(mq+r)
\E^{-(mq+r) ( u-\imag \delta)}\E^{2\pi \imag r p/q}\;.
$$
Substituting instead of $\Lambda(mq+r)$ its mean value (\ref{Lqr}) one gets
$$
f(\E^{-u}\E^{\imag 2\pi p/q+\imag \delta})\approx
\frac{1}{\varphi(q)}\sum_{(r,\ q)=1}\E^{2\pi \imag r p/q}
\int_{0}^{\infty}\E^{-n(u-\imag \delta)}\D n=
\frac{\mu(q)}{\varphi(q)(u-\imag \delta)}\;.
$$
In the last step we use that fact that \cite{Mobius}
$$
\sum_{(r,\ q)=1}\E^{2\pi \imag r/q}=\mu(q)
$$
where $\mu(q)$ is the M\"obius function defined through the
factorization of $q$ on prime factors
$$
\mu(q)=\left \{ \begin{array}{cl}
1& \mbox{if } q=1\\
(-1)^k& \mbox{if } q=p_1\ldots p_k\\
0&\mbox{if $q$ is divisible on $p^2$}
\end{array}\right .\;.
$$
The final expression means that function $f(x)$ has a pole singularity at
the unit circle at every rational point.

The knowledge of  $f(x)$ permits formally to compute the mean value of the
product of two $\Lambda$-functions.

Let
$$
J_r(R)=\frac{1}{2\pi}\int_0^{2\pi} f(R \E^{\imag \varphi})
f(R \E^{-\imag \varphi})\E^{-\imag r\varphi}\D\varphi
=R^r\sum_{m}\Lambda(m+r)\Lambda(m)R^{2m}\;.
$$
As the function $f(x)$ has a pole singularity at
the unit circle at every rational point one can try to approximate this
integral by the sum over singularities
\begin{eqnarray*}
J_r(\E^{-u})&=&\frac{1}{2\pi}\int_0^{2\pi} f(R\E^{\imag \varphi})
f(R \E^{-\imag \varphi})\E^{-\imag r\varphi}\D \varphi=\nonumber\\
&=&\frac{1}{2\pi} \sum_{(p,\ q)=1} \int f(\E^{-u+\imag 2\pi p/q +\imag \delta})
f(\E^{-u-2\pi \imag p/q -\imag \delta})\E^{-\imag r(2\pi p/q+\imag \delta)}\D \delta
=\nonumber\\
&=&\frac{1}{2\pi} \sum_{(p,\ q)=1}\E^{2\pi \imag r p/q}
\left (\frac{\mu(q)}{\varphi(q)}\right )^2\int \frac{\D \delta}{u^2+\delta^2}
=\nonumber\\
&=&\frac{1}{2u} \sum_{(p,\ q)=1}\E^{2\pi \imag pr/q}
\left (\frac{\mu(q)}{\varphi(q)}\right )^2\;.  
\end{eqnarray*}
Therefore
$$
\sum_{n=1}^{\infty} \Lambda(n)\Lambda (n+r)\E^{-2nu}\stackrel{u\to
  0}{\longrightarrow }
\frac{1}{2u} \sum_{(p,\ q)=1}\E^{2\pi \imag r p/q}
\left (\frac{\mu(q)}{\varphi(q)}\right )^2
$$
from which it follows that
$$
\lim_{N\to \infty} \frac{1}{N}\sum_{n=1}^{N} \Lambda(n)\Lambda (n+r)=\alpha(r)
$$
where
\begin{equation}
\alpha(r)=\sum_{q=1}^{\infty}
\left (\frac{\mu(q)}{\varphi(q)}\right )^2\sum_{(p,\ q)=1}e^{2\pi \imag r p/q}\;.
\label{HardySum}
\end{equation}
Using properties of such singular series one can prove \cite{HardyLittlewood} that
for even $r$ $\alpha(r)=0$ and for odd $r$ it can be
represented as the following product over prime numbers
\begin{equation}
\alpha (r)=C_2\prod_{p|r}\frac{p-1}{p-2}
\label{HardyProduct}
\end{equation}
where the product is taken over all prime divisors of $r$ bigger than 2 and
$C_2$ is the so-called twin prime constant
\begin{equation}
C_2=2\prod_{p>2}(1-\frac{1}{(p-1)^2})\approx 1.32032\ldots .
\label{TwinConstant}
\end{equation}
Instead of  demonstration the formal equivalence of (\ref{HardySum}) and
(\ref{HardyProduct})  we present another heuristic 'derivation' based on
the probabilistic interpretation of prime numbers which gives
directly (\ref{HardyProduct}) and (\ref{TwinConstant}). 

The argumentation consists on the following steps.
\begin{itemize}

\item Probability that a given number is divisible by a prime $p$ is
$$
\lim_{N\to \infty}\frac{1}{N}[\mbox{number of integers divisible by 
  $p\leq  N$}]
=\frac{1}{p}\;.
$$
In general to find such probabilities it is necessary to consider only the
residues modulo $p$ and find how many of them obey the requirement.

\item Probability that a given number is not divisible by a prime $p$ is
$$
1-\frac{1}{p}\;.
$$
\item Probability that a number is not divisible by primes
  $p_1,p_2,\ldots p_k$ is
\begin{equation}
\prod_{j=1}^k(1-\frac{1}{p_j})\;.
\label{prodpj}
\end{equation}
\end{itemize}
The above formula is correct for any finite collection of primes but
for  computations with infinite number of primes it may be wrong. 

For example, when used naively it gives that 
\begin{itemize}
\item probability that a number $x$ is a prime is
$$
\prod_{p<\sqrt{x}}(1-\frac{1}{p})\;.
$$
\end{itemize}
This prime number 'theorem' is false because from it it follows that the number
of primes less than $x$ is \cite{Titchmarsh}
$$
\Pi(x)=x\prod_{p<\sqrt{x}}(1-\frac{1}{p})\stackrel{x\to
  \infty}{\longrightarrow }
\frac{x}{\ln x}2\E^{-\gamma}
$$
which differs from the true prime number theorem by a factor 
$2\E^{-\gamma}\approx 1.123$ 
where $\gamma$ is the Euler constant. The origin of this discrepancy is
related with the approximation frequently used above: $[x/p]=x/p$ where
$[x]$ is the integer part of $x$. Instead of (\ref{prodpj}) one should have 
$\prod_p(1-[x/p]/x)$. For a finite number of primes and $x\to\infty$ it
tends to   (\ref{prodpj}). But when the number of primes considered
increases with $x$ errors are accumulated giving a constant factor.

Nevertheless one could try to use probabilistic arguments by forming
artificially  convergent quantities. One has 
\begin{itemize}
\item Probability that $x$ and $x+r$ are primes is 
$$
\lim_{N\to \infty} \frac{1}{N}\;[\mbox{ number of integers $x<N$ such that $x$
  and $x+r$ are primes }]\;.
$$
\end{itemize}
Let consider a prime $p$. Two cases are possible. Either $p\,|r$ or 
$p\, |\hspace{-.39em}/ r$.
In the first case the probability that  both number $x$ and $x+r$ are not
divisible by $p$ is the same as the probability that only number $x$ is not
divisible by $p$ which is
$$
\prod_{p|r}\left (1-\frac{1}{p}\right )\;.
$$
When $p\, |\hspace{-.39em}/  r$ one has to remove two numbers from the
set of residues as
$x=0,1,..,p-1$ (mod $p$) and $x+r=0,1,..,p-1$ (mod $p$). Therefore 
the probability that both
numbers $x$ and $x+r$ are not divisible by a prime $p$ is
$$
\prod_{p\, |\hspace{-.33em}/  r}\left (1-\frac{2}{p}\right )\;.
$$
Finally 
\begin{itemize}
\item Probability that both $x$ and $x+r$ are primes 
$= 
\prod_{p|r}\frac{p-1}{p}\prod_{p \, |\hspace{-.33em}/ r}\frac{p-2}{p}\;.
$
\end{itemize}
To find a convergent expression we divide  both sides by the probability
that numbers $x$ and $x+r$ are independently prime numbers computed also in
the probabilistic approximation. The latter quantity is
$$
[\mbox{ Probability that $x$ is prime and $x\leq N$ }]=\prod_p\frac{p-1}{p}\;.
$$
Therefore
\begin{eqnarray*}
&&
\frac{[\mbox{ Probability that both $x$ and $x+r$ are primes
with $x, x+r\leq N$ }]}{[\mbox{ Probability that $x$ is prime }]^2}\approx \\
&&\approx \prod_{p|r}\frac{p-1}{p}
\prod_{p \, |\hspace{-.33em}/ r}\frac{p-2}{p}
\prod_p(\frac{p}{p-1})^2=
2\prod_{p>2}\left (1-\frac{1}{(p-1)^2}\right )
\prod_{p|r}\frac{p-1}{p-2}\;.
\end{eqnarray*}
As the denominator in the above expression is $1/\ln^2 N$  it follows that
the probability that both $x$ and $x+r$ are primes with $x\leq N$, and 
$x+r\leq N$ is asymptotically
$$  
\frac{\alpha(r)}{\ln^2 N}
$$
with the same function $\alpha(r)$ as in (\ref{HardyProduct}). 

We stress that the Hardy--Littlewood conjecture is still not proved.
Even the existence of infinite number of twin primes
(primes separated by 2) is not yet proved while the Hardy--Littlewood conjecture 
states that their density is $C_2/\ln^2N$.
 
\subsection{Two-Point Correlation Function of Riemann Zeros} 
\label{CorrelationRiemann}

Taking  the above expression of the Hardy--Littlewood conjecture  as granted we get
$$
R_2^{(off)}(\epsilon)=\frac{1}{4\pi^2}\sum_{n\geq 1}
\frac{1}{n}\E^{\imag \epsilon \ln n}
\sum_r \alpha(r)\E^{\imag E r/n} +\mbox{c.c.}\;.
$$
After substitution the formula for $\alpha(r)$ and performing the sum over
all $r$ one obtains
$$
R_2^{(off)}(\epsilon)=\frac{1}{4\pi^2}\sum_n
\frac{1}{n}\E^{\imag \epsilon \ln n}
\sum_{(p,q)=1}\left (\frac{\mu(q)}{\varphi(q)}\right )^2
\delta \left (\frac{p}{q}-\frac{E}{2\pi n}\right ) +\mbox{c.c.}
$$
where the summation is taken over all pairs of mutually co-prime positive
integers $p$ and $q$ (without the restriction $p<q$).

Changing the summation over $n$ to the integration permits to transform
this expression to contributions  of values of $n$ where 
$$
\frac{p}{q}-\frac{E}{2\pi n}=0\;.
$$
In this approximation
$$
R_2^{(off)}(\epsilon)=\frac{1}{4\pi^2}\E^{\imag \epsilon \ln E/2\pi}
\sum_{(p,q)=1}\left (\frac{\mu(q)}{\varphi(q)}\right )^2
\left (\frac{q}{p}\right )^{1+\imag \epsilon} +\mbox{c.c.}\;.
$$
Using the formula (which is a mathematical expression of the 
inclusion--exclusion principle)
$$
\sum_{(p,q)=1}f(p)=\sum_{k=1}^{\infty}\sum_{\delta |q}f(k\delta)\mu(\delta)
$$
and taking into account that $2\pi \bar{d}=\ln (E/2\pi)$ one obtains
\begin{equation}
R_2^{(off)}(\epsilon)= \frac{1}{4\pi^2}|\zeta(1+\imag \epsilon)|^2
\E^{2\pi \imag \bar{d}\epsilon }\Phi^{(off)}(\epsilon)+\mbox{c.c.}
\label{R2offRiemann}
\end{equation}
where function $\Phi^{(off)}(\epsilon)$ is given by a convergent product over primes
$$
\Phi^{(off)}(\epsilon)= \prod_p\left (1-\frac{(1-p^{\imag \epsilon})^2}{(p-1)^2}\right )
$$
and $\Phi^{(off)}(0)=1$.

In the limit of small $\epsilon$
$$
R_2^{(off)}(\epsilon)= \frac{1}{(2 \pi \epsilon)^2}
\left (\E^{2\pi \imag \bar{d}\epsilon }+\E^{-2\pi \imag \bar{d}\epsilon }\right )
$$
which exactly corresponds to the GUE results for the oscillating part of the
two-point correlation function (\ref{oscGUE}).

The above calculations demonstrate how one can compute the two-point
correlation function through the knowledge of correlation function of
periodic orbit pairs. For the Riemann case one can prove under the same
conjectures that all $n$-point correlation functions of Riemann zeros tend
to corresponding GUE results \cite{BogomolnyKeating}.

\section{Summary}\label{SummaryStatistics}

Trace formulas can formally be used to calculate spectral correlation
functions for dynamical systems. In particular, the
two-point correlation function is the product of two densities of states
$$
R_2(\epsilon)\equiv \left \langle d(E+\epsilon)d(E) \right \rangle \;.
$$
The diagonal approximation consists of taking into account in such products
only terms with exactly the same action. For chaotic systems this approximation
is valid only for very small time. In particular, it permits to obtain
the short-time behaviour of correlation form factors which agrees
with predictions of standard random matrix ensembles.
  
The main difficulty in such approach to spectral statistics is the
necessity to compute  contributions from non-diagonal terms which
requires the knowledge of correlation functions of periodic orbits
with nearby actions. 

For the Riemann zeta function zeros it can be done using the
Hardy--Littlewood conjecture which claims that the number of
prime pairs $p$ and $p+r$ such that $p<N$ for large $N$ is asymptotically 
$$
\alpha (r)\frac{N}{\ln^2 N}
$$
where $\alpha(r)$ (with even $r$) is given by the product over 
all odd prime divisors of $r$
$$
\alpha(r)=C_2\prod_{p|r}\frac{p-1}{p-2} 
$$
and 
$$
C_2=2\prod_{p>2}\left( 1-\frac{1}{(p-1)^2}\right )\;.
$$
Using this formula one gets that the two-point correlation function of Riemann zeros is 
$$
R_2(\epsilon)=\bar{d}^2(E)+R_2^{(diag)}(\epsilon)+R_2^{(off)}(\epsilon)
$$
where the diagonal part
$$
R_2^{(diag)}(\epsilon)=-\frac{1}{4\pi^2}
\frac{\partial^2}{\partial \epsilon ^2}
\ln \left [|\zeta(1+\imag \epsilon)|^2\Phi^{(diag)}(\epsilon)\right ]
$$
and non-diagonal part 
$$
R_2^{(off)}(\epsilon)= \frac{1}{4\pi^2}|\zeta(1+\imag \epsilon)|^2
\E^{2\pi \imag \bar{d}\epsilon }\Phi^{(off)}(\epsilon)+\mbox{c.c.}\; .
$$
The functions $\Phi^{(diag)}(\epsilon)$ and  $\Phi^{(off)}(\epsilon)$ are 
given by convergent products over all primes
$$
\Phi^{(diag)}(\epsilon)=\exp \left ( 2\sum_p\sum_{m=1}^{\infty} 
\frac{1-m}{m^2p^m}\cos( m \epsilon \ln p) \right )
$$
and
$$
\Phi^{(off)}(\epsilon)= \prod_p
\left (1-\frac{(1-p^{\imag \epsilon})^2}{(p-1)^2}\right )\;.
$$
In \cite{Bogomolny2} a few other methods were developed  to 'obtain'
the two-point correlation function for Riemann zeros. These methods
were based on different ideas and  certain of them can be 
generalized for dynamical systems. Though neither of the methods
can be considered as a strict mathematical proof, 
all lead to the same expression (\ref{R2offRiemann}). 

\begin{figure}
\center
\includegraphics[height=8cm, angle=-90]{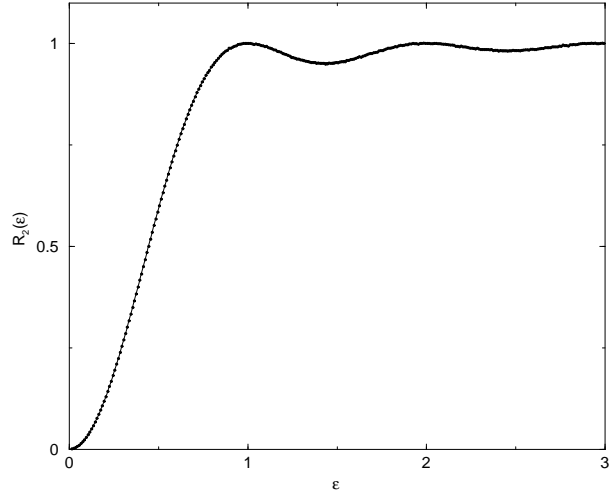}
\caption{Two point correlation function of the Riemann zeros near the
  $10^{23}$-th zero (dots) and the GUE prediction (solid line).}
\label{fig1}
\end{figure}

\begin{figure}
\center
\includegraphics[height=8cm, angle=-90]{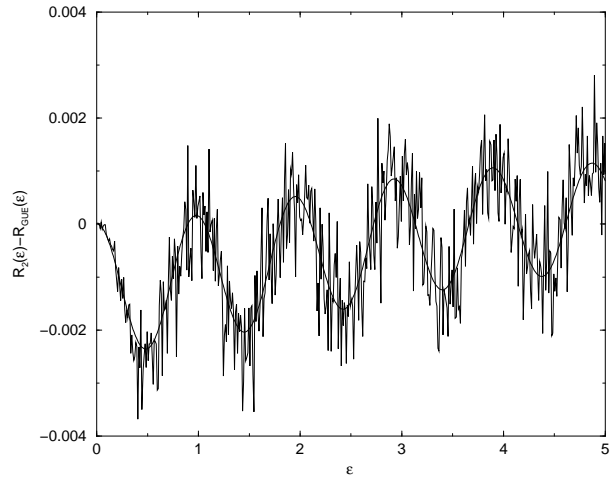}
\caption{The difference between the two point correlation function of the
  Riemann zeros and the GUE prediction in the interval $0<\epsilon<5$. The solid line is the difference
  between the `exact' correlation function and the GUE prediction in this
  interval. }
\label{fig2}
\end{figure}

\begin{figure}
\center
\includegraphics[height=8cm, angle=-90]{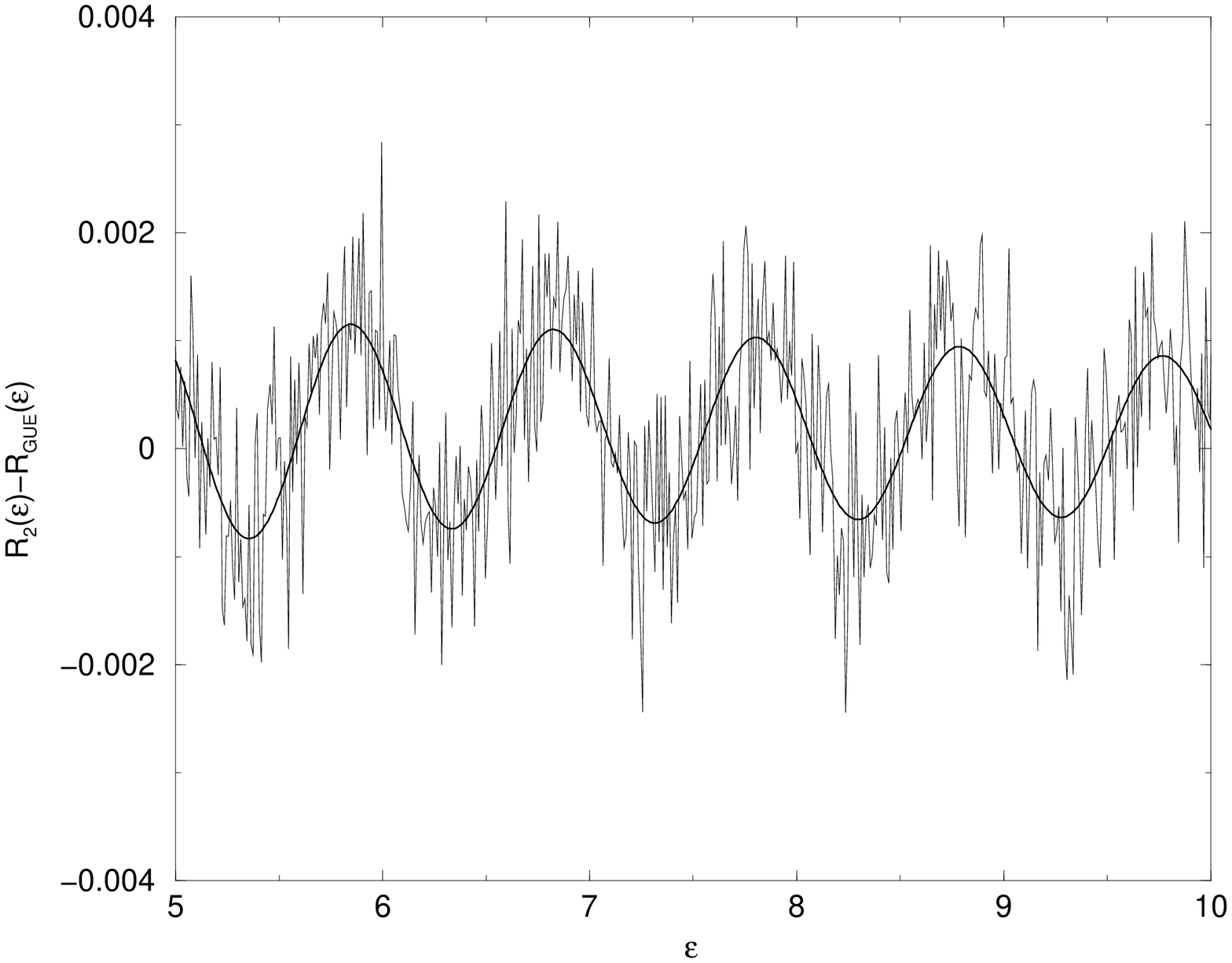}
\caption{The same as at Fig.~\ref{fig2} but in the interval $5<\epsilon<10$.}
\label{fig3}
\end{figure}

\begin{figure}
\center
\includegraphics[height=8cm, angle=-90]{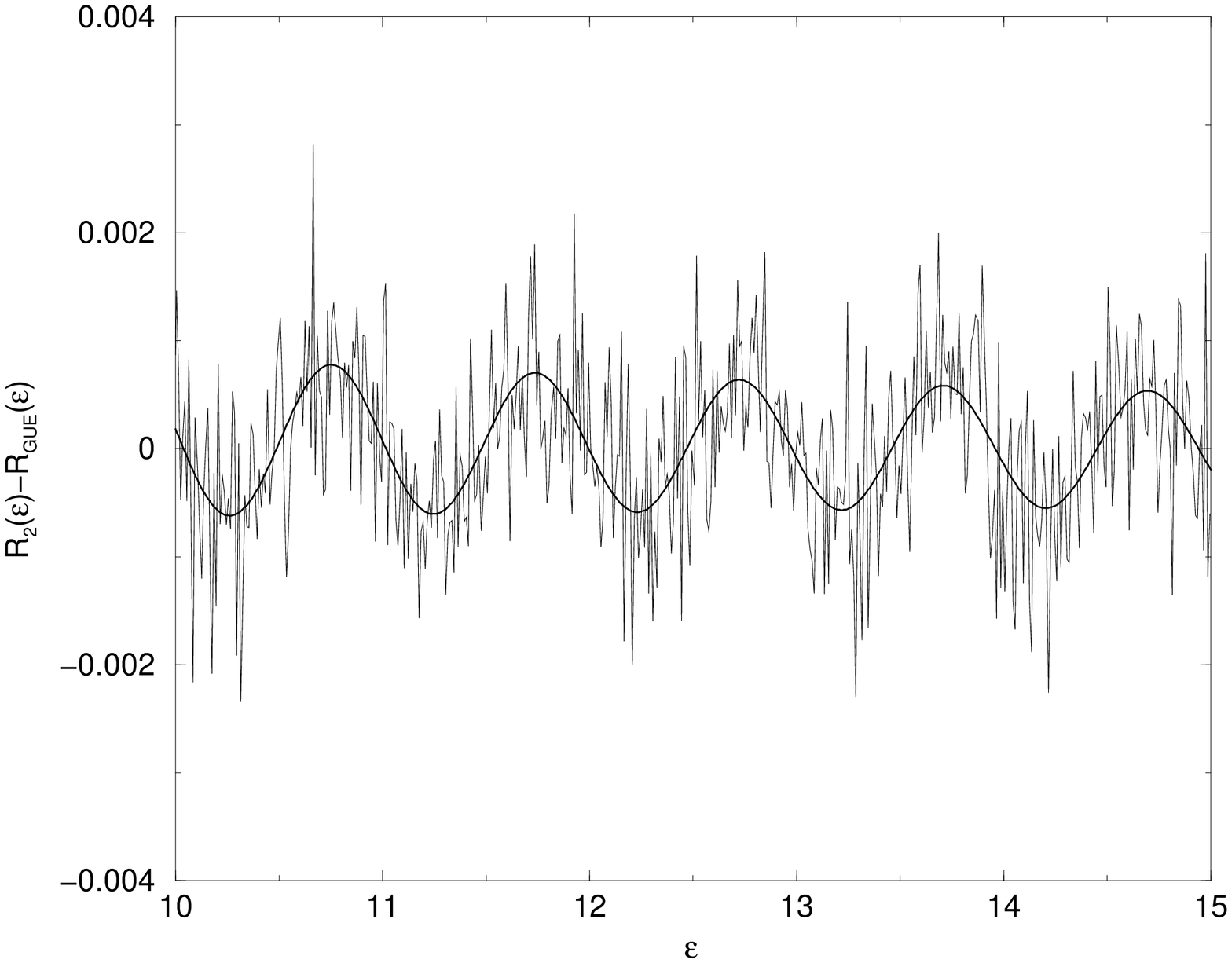}
\caption{The same as at Fig.~\ref{fig2} but in the interval $10<\epsilon<15$.}
\label{fig4}
\end{figure}

\begin{figure}
\center
\includegraphics[height=8cm, angle=-90]{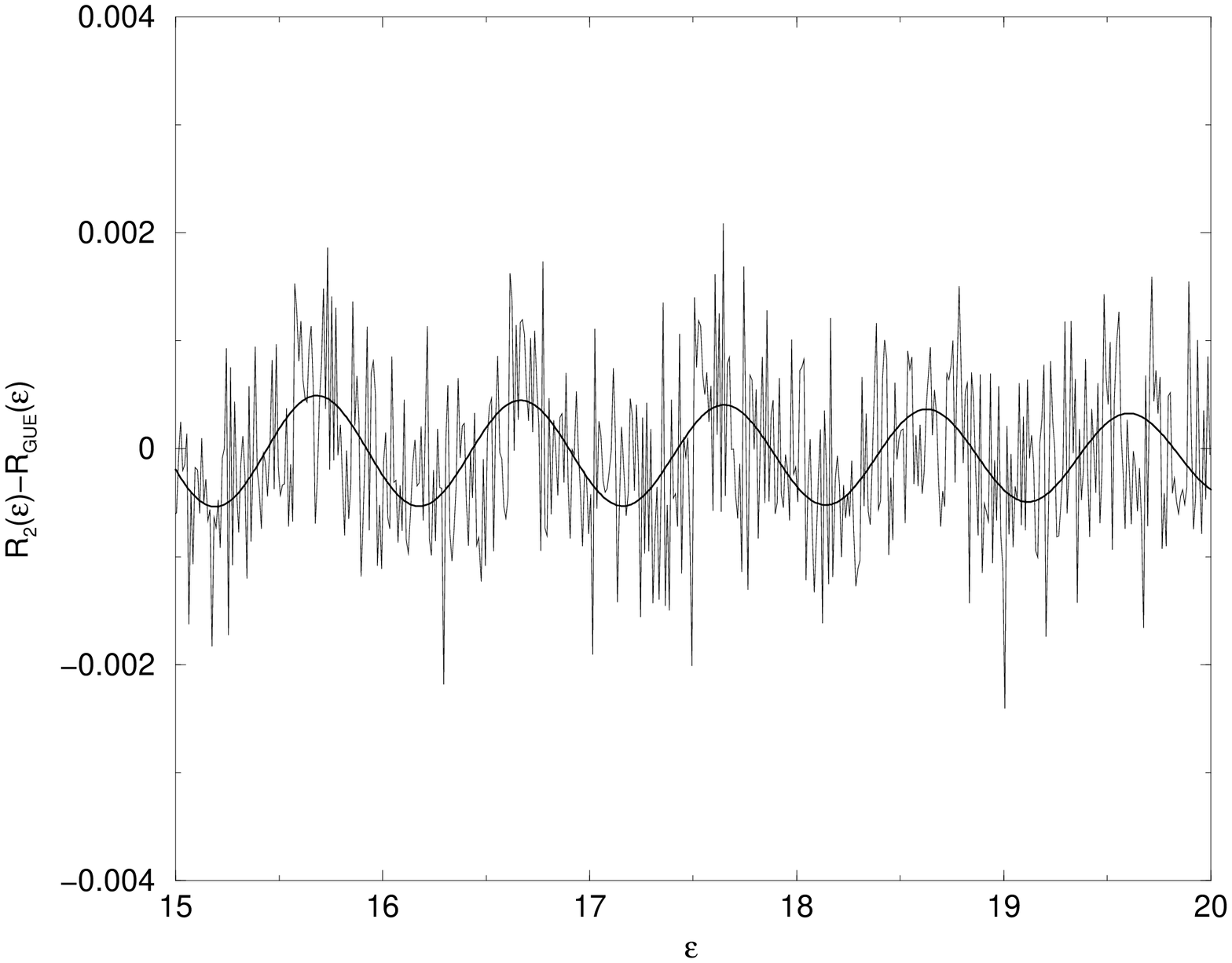}
\caption{The same as at Fig.~\ref{fig2} but in the interval $15<\epsilon<20$.}
\label{fig5}
\end{figure}

\begin{figure}
\center
\includegraphics[height=8cm, angle=-90]{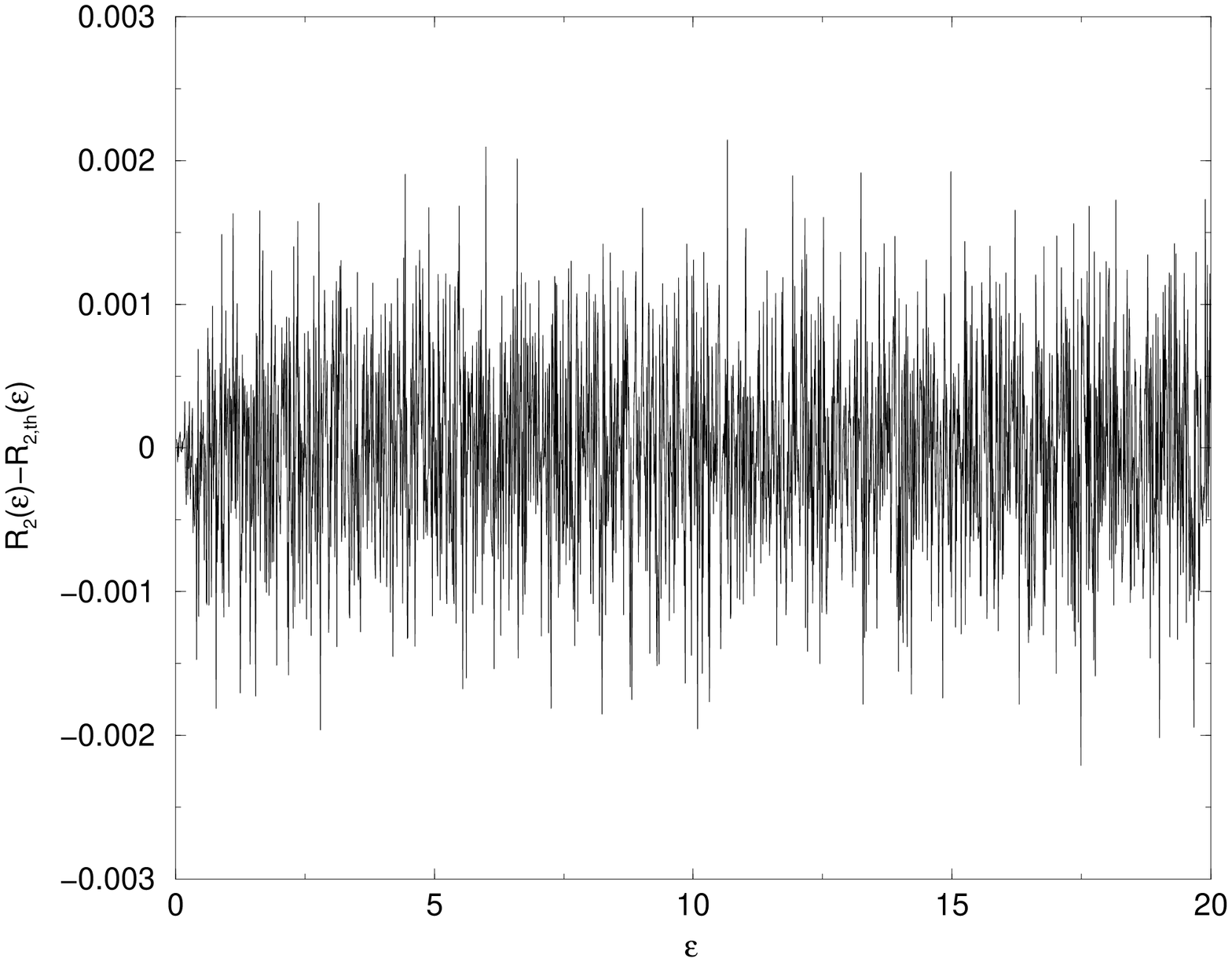}
\caption{The difference between numerically computed
two-point correlation function and the `exact' function}
\label{fig6}
\end{figure}

\begin{figure}
\center
\includegraphics[height=8cm, angle=-90]{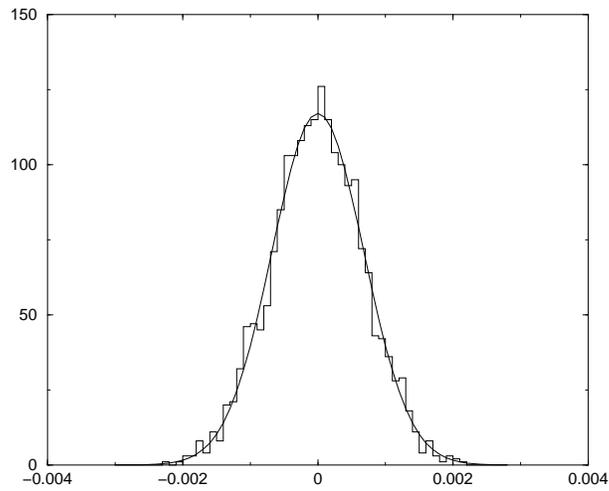}
\caption{The histogram of the deviations of the numerically computed
  two-point correlation function of Riemann zeros and the `exact' formula. 
  Solid line is the Gaussian fit to the histogram. }
\label{fig7}
\end{figure}

It is also of interest to check numerically the above formulas.  
When numerical calculations are performed one considers usually
correlation functions for the unfolded spectrum. For the two-point
correlation function this procedure corresponds to the following transformation
$$
R_2^{(\mbox{\scriptsize{unfolded}})}(\epsilon)=
\frac{1}{\bar{d}^2(E)}R_2\left (\frac{\epsilon}{\bar{d}(E)}\right )\;.
$$
At Fig~\ref{fig1} we present the two-point correlation function for
$2\cdot 10^{8}$ zeros near the $10^{23}$-th zero computed numerically by
Odlyzko \cite{Odlysko} together with the GUE prediction for this quantity
$$
R_2^{\mbox{\scriptsize{GUE}}}(\epsilon)=1-\left (\frac{\sin \pi \epsilon}{\pi \epsilon}
\right )^2\;.
$$
At Figs.~\ref{fig2}-\ref{fig5} we present the difference between the
two-point correlation function computed numerically and the GUE prediction. 
At Fig.~\ref{fig6}  we present the difference between numerically computed
two-point correlation function and the `exact' function and at Fig.~\ref{fig7} 
the histogram of  differences is given. Notice that these differences are structure less and the histogram corresponds practically exactly  to statistical errors inherent in the calculation of the two-point correlation functions which signifies that the obtained formula agrees very well with the numerics.  

\chapter{Arithmetic Systems}\label{Arithmetic}
\setcounter{section}{0}

As was discussed above it is well accepted that spectral statistics of 
classically chaotic systems in the universal limit coincides with 
spectral statistics of the usual random matrix ensembles.
But it is also known (see e.g. \cite{Balatz}, \cite{Colin} and references 
therein)  that the motion on constant negative curvature surfaces generated
by discrete groups (considered in Chap.~\ref{Trace}) is the best example of
classical chaos. 
Consequently,  models on constant negative curvature seem to
be ideal tools to check the conjecture on spectral fluctuations of 
classically chaotic systems. Their classical motion is extremely chaotic and 
time-reversal invariant and   {\em a priori} assumption was that all of them 
should have energy levels distributions close to predictions of the 
Gaussian orthogonal ensemble (GOE) of random matrices. 

Nevertheless when the first large scale numerical calculations  were performed
\cite{Aurich}, \cite{Schmit} they clearly indicated that for 
certain hyperbolic models the spectral statistics were quite close to 
Poisson statistics typical for integrable systems.

\begin{figure}
\center
\includegraphics[height=8cm, angle=-90]{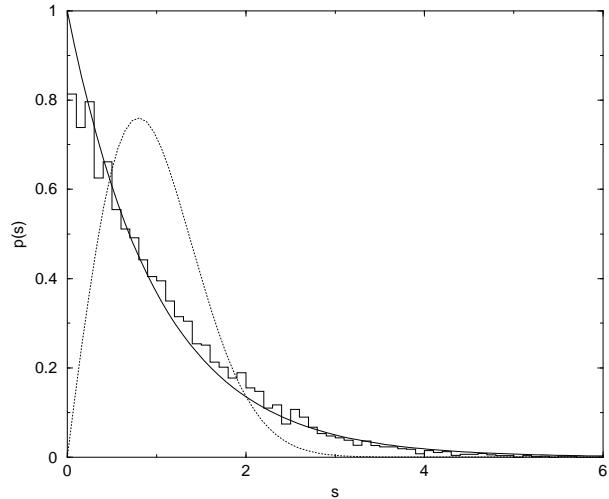}
\caption{The nearest neighbor distribution for 10000 first levels of the 
triangle $(\pi/2,\pi/3,0)$  (the modular triangle). Solid line - the Poisson
distribution. Dotted line - the GOE distribution.}
\label{modulardomain}
\end{figure}

As an example we present in Fig.~\ref{modulardomain} the nearest-neighbor distribution 
for the hyperbolic triangle with angles $(\pi/2, \pi/3,0)$
corresponding to the well-known modular triangle with  Dirichlet boundary
conditions. 
The agreement with Poisson prediction is striking though classical
motion for this system is perfectly chaotic.

The purpose  of this Chapter is to show that this strange behaviour is 
the consequence of exponentially large exact degeneracy of periodic orbit
lengths in systems considered \cite{BogomolnyGeorgeot}. 
In all hyperbolic surfaces where the Poisson-like statistics was observed
there is on average $\exp (l/2)$ classical periodic orbits with exactly
the same length $l$. It will be demonstrated that this is the characteristic 
property of models generated by the so-called arithmetic groups.  
As classical mechanics is not sensitive to lengths of periodic orbits
all these models remain completely chaotic. But the cumulative effect
of interference of degenerated periodic orbits changes drastically 
the quantum mechanical properties. 

This Chapter is based mostly on \cite{BGGS}. In Sect.~\ref{Peculiarities}
simple calculations prove exponential degeneracy
of periodic orbit lengths for the modular group. The main peculiarity of the
modular group matrices is that their traces are integers. Therefore if one
considers all matrices with $|\mbox{Tr }M|<X$ the number of different traces 
increases at most linearly with $X$. In Sect.~\ref{ArithmeticGroups} it is 
shown that this property is typical for all arithmetic groups. 
An informal mini-review of such groups is given in this Section and it is 
demonstrated that for all these groups exponentially many periodic orbits have
exactly the same length. From the results of \cite{Takeuchi} it follows 
that there is exactly 85 triangles generated by discrete arithmetic groups. 
All triangular models were the Poisson-like spectral statistics was 
numerically observed are in this list. 
In Sect.~\ref{DiagonalArithmetic} it is shown that in the diagonal approximation 
the two-point correlation form factor of arithmetic systems jumps very quickly
to the Poisson value thus confirming unusual nature of arithmetic systems.
In Sect.~\ref{ExactModular} the exact two-point correlation function for the 
modular domain is calculated. The correlations of multiplicities are obtained 
by a generalization of the Hardy--Littlewood method discussed in
Sect.~\ref{HardyLittlewood}. The resulting formula proves that in the universal
limit the two-point correlation function of eigenvalues of the 
Laplace--Beltrami operator automorphic with respect 
to the modular group tends to the Poisson prediction. 
Arithmetic groups have many other interesting properties. 
In particular, for all arithmetic groups it is possible to construct an 
infinite number of mutually commuting operators which commute also with the 
Laplace--Beltrami operator. Properties of these operators   
called the Hecke operators are discussed in Sect.~\ref{Hecke}. 
The Jacquet--Langlands correspondence between different arithmetic groups is
mentioned in Sect.~\ref{JL}.
In Sect.~\ref{Nonarithmetic} non-arithmetic models are briefly discussed.

\section{Modular group}\label{Peculiarities}

The modular group is the group of all $2\times2$ matrices
$$
M=\left (\begin{array}{cc}m&n\\k&l\end{array}\right )
$$
with integer $m$, $n$, $k$, $l$ and  the unit determinant
$ml-nk=1$.

The periodic orbits correspond in a unique way to the conjugacy classes of hyperbolic elements of the group (see Sect.~\ref{Classical}).
The length of periodic orbit $l_p$ is related with the
trace of a representative matrix of the conjugacy class $M$ as follows 
$$
2\cosh \frac{l_p}{2}=|\mbox{Tr } M|\;.
$$
As all elements of  modular group matrices are integers,  the 
trace is also an integer
\begin{equation}
|\mbox{Tr} M|=n\;.
\label{integer}
\end{equation}
Here the arithmetical nature of the group clearly comes into the play. This
simple property is very important. It signifies that for the modular group
there is just a quite restrictive set of all possible traces and,
consequently, of periodic orbit lengths. 
For modular group  the number of possible different lengths is the number
of different integers less than $2\cosh L/2$ (see (\ref{integer})), hence
$$
N_{\mbox{\scriptsize dif. lengths}}=2\cosh \frac{L}{2}\stackrel{L\to
  \infty}{\longrightarrow} \E^{L/2}\;.
$$
On the other hand, for any discrete group the number of periodic orbits of 
length less than $L$ grows asymptotically as
$$
N(l_p<L)=\frac{\E^L}{L}\;.
$$
Let $g(l)$ be the multiplicity of periodic orbits with length $l$.
One has obvious relations valid for large $L$
$$
\sum_{l<L}g(l)={e^L\over L}\;,\;\;\;\sum_{l<L}1=\E^{L/2}
$$
where the summation extends over different lengths of periodic orbits counted
without taking multiplicity into account. 

Let us define the  {\em mean} multiplicity $\left\langle g(l)\right\rangle $ 
as the following ratio
\begin{equation}
\left\langle {g(l)} \right\rangle=
{\hbox{Number of periodic orbits with $l<l_p<l+\Delta l$}
\over\hbox{Number of different lengths with $l<l_p<l+\Delta l$}}\:.
\label{MeanDefinition}
\end{equation}
Asymptotically for large $L$ the previous formulas gives
$$
\left\langle g(l) \right\rangle=2{e^{l/2}\over l}\
$$
which demonstrates that  periodic orbit lengths for the modular group are exponentially degenerated. 

\section{Arithmetic Groups}\label{ArithmeticGroups}

The crucial feature which led to the exponential degeneracy of 
periodic orbit lengths for the modular group was the fact 
that traces of modular group matrices were integers which was a direct 
consequence of the arithmetic  nature of modular group. 
But $2\times 2$ matrix groups with integer elements are exhausted by the 
modular group and  its subgroups. 

Nevertheless, one can construct a quite large class of discrete groups with
strong arithmetic properties by considering groups which are not equal to
$2\times 2$ integer matrices but which permit a {\em representation} by
$n\times n$ integer matrices ($n>2$). 

The existence of such representation means that for each $2\times 2$ 
group matrix, $g$, one can associate a $n\times n$ matrix with integer 
entries, $M(g)$, in such
a way that the matrix associated to the product of two groups matrices
equals the product of two matrices associated to the corresponding factors
$$
M(ab)=M(a)\times M(b)
$$
for all $a$ and $b$ from the group considered and $M(\vec{1})=\vec{1}$.

To define general arithmetic groups we need a few definitions. 
\begin{itemize}
\item A subset of a group $\Gamma$ is called a subgroup if it forms
itself a group.
\item A subgroup $g$ of a group $\Gamma$ is called a subgroup of finite 
index if $\Gamma$ can be represented as a finite union
$$
\Gamma=g+g\gamma_1+\ldots+g\gamma_k
$$
with $\gamma_k\in \Gamma$.
\item Two groups are called commensurable if they have a common
subgroup which is of finite index in both of them.
\end{itemize}

Groups which have a representation by integer matrices and all groups 
commensurable with them  are called  arithmetic groups. This Section is
devoted to the investigation of their properties.

In Sect.~\ref{AlgebraicFields} a non-formal review of algebraic fields is
given and in Sect.~\ref{QuaternionAlgebras} the construction of quaternion
algebras over algebraic fields is shortly discussed. 
It  appears that all arithmetic groups can be obtained from quaternion 
algebras with division and in Sect.~\ref{TakeuchiCriterion} the necessary
and sufficient conditions that a given group will be an arithmetic group is
presented. Using these conditions in Sect.~\ref{Multiplicities} it is proved
that  periodic orbit lengths for all arithmetic groups  have the same
exponential degeneracy (up to a constant factor) as for the modular group.

\subsection{Algebraic Fields}\label{AlgebraicFields}

Everybody is familiar with usual rational numbers
$$
u=\frac{p}{q}
$$
with integer $p$ and $q$. Their important properties  are (i)  that the sum 
and the product of any two rational numbers also have the same form and (ii)
all elements except $0$ have an 
inverse (i.e. the division is always possible).  From mathematical viewpoint
rational numbers form a field called  $\mbox{l\hspace{-.47em}Q}$. 

Algebraic fields of finite degree, $\mbox{I\hspace{-.15em}F}$, are 
a generalization of this reference field  obtaining  by adding to the set of 
rational numbers a root $\alpha$ of an irreducible polynomial
\begin{equation}
\sum_{k=0}^{n} c_k \alpha^k=0 
\label{defining}
\end{equation}
with integer coefficients $c_k$. This field is denoted
$\mbox{I\hspace{-.15em}F}=\mbox{ l\hspace{-.47em}Q}(\alpha)$.

Each element $u\in \mbox{l\hspace{-.47em}Q}(\alpha)$ can  be 
represented by the sum 
$$ 
u=\sum_{i=0}^{n-1} b_i \alpha^i
$$
where the $b_i$ are usual rationals. The summation and the multiplication of
these elements are done as with usual numbers except that all powers of
$\alpha$ larger than  $n-1$ have to be reduced using the defining equation 
(\ref{defining}). 

Integers of the field $\mbox{ l\hspace{-.47em}Q}(\alpha)$ are its elements
which obey a polynomial equation with integer coefficients with an additional condition that the highest power coefficients equals one (such polynomials are called monic polynomials).

In general, algebraic integers, $\omega$, of a field of degree $n$ are 
freely generated by $n$ linearly independent elements of the field $\beta_k$
with integer coefficients (in mathematical language it means that they form
a free $\mbox{Z\hspace{-.3em}Z}$-module of rank $n$). Explicitly
\begin{equation} 
\omega=\sum_{k=0}^{n-1} m_k \beta_k 
\label{AlgebraicIntegers}
\end{equation}
where all $m_k$ are usual integers. In simple cases $\beta_k=\alpha^k$ and
$$ 
\omega=\sum_{k=0}^{n-1} m_k \alpha^k
$$
with integer or demi-integer coefficients $m_k$. Algebraic integers like
usual integers form a ring (not a field) because the division
is not always possible. 

The polynomial equation  defining the field (\ref{defining}) has $n$ 
different roots $\alpha_i, i=0,1,...,n-1$ with $\alpha_0 = \alpha$.  Any
relation between elements of the field remains unchanged under the
transformations
$$ 
\phi_i:\:\:\alpha \rightarrow \alpha_i
$$
where one substitutes in all expressions instead of one root $\alpha$
another root $\alpha_i$. These transformations are called isomorphisms 
or embeddings of this field into $\mbox{l\hspace{-.47em}C}$ and   
they are the only transformations respecting the laws of the field.

\paragraph{Example} 
Add to the field of rational numbers $\mbox{ l\hspace{-.47em}Q}$ one root of 
the quadratic equation
$$ 
x^2=d 
$$
where $d$ is a square-free integer.
Elements of this field $\mbox{ l\hspace{-.47em}Q}(\sqrt{d})$ can be written
as
$$
u=p+q\sqrt{2}
$$ with $p$ and $q$ rationals.
Let
$$\omega=a+b\sqrt{2}
$$
be integers of this field. To find values of $a$ and $b$ one notes
that $\omega$ obeys the quadratic equation
$$
\omega^2-2a\omega+a^2-db^2=0\;.
$$
To describe integers $2a$ and $a^2-db^2$ have to be usual integers.
Depending on $d$ two types of solutions are possible.
\begin{itemize}
\item If $d \equiv 2$ or $d\equiv 3$ (mod $4$) then $a$ and $b$ have to be integers and
$$
\omega=m+n\sqrt{d}
$$
with integers $m$ and $n$.
\item If $d\equiv 1$ (mod $4$) then both $a$ and $b$ can be demi-integers and
$$
\omega=\frac{m}{2}+\frac{n}{2}\sqrt{d}
$$
with integers $m\equiv n$ (mod $2$).

To avoid the last restriction this expression can be rewritten in the form 
(\ref{AlgebraicIntegers})
\begin{equation}
\omega=m+n\frac{1+\sqrt{d}}{2}
\label{d1}
\end{equation}
with arbitrary integers $m$ and $n$.
\end{itemize}
As this field is defined by an equation of the second degree it has two 
isomorphisms 
\begin{eqnarray*}
\phi_0&:& p+q\sqrt{d}\longrightarrow p+q\sqrt{d}\;,\nonumber\\
\phi_1&:& p+q\sqrt{d}\longrightarrow p-q\sqrt{d}\;.
\end{eqnarray*}
Because the product of two algebraic integers is also an algebraic integer from 
(\ref{AlgebraicIntegers}) it follows that all algebraic integers permit 
a representation by matrices with integer elements in such a way that the 
matrix representing a product of two integers equals the product of matrices 
representing each factor.
   
For example, for the above considered case of 
$\mbox{ l\hspace{-.47em}Q}(\sqrt{d})$ with  $d\not\equiv 1$ (mod $4$)   one
can associate with an integer of this field, $\omega=m+n\sqrt{2}$, a 
$2\times 2$ matrix
\begin{equation}
M(\omega)=\left ( \begin{array}{cc}m & n\\d n& m\end{array}\right )\;.
\label{representation}
\end{equation}
It is easy to check that this is the true representation of field integers
because $M(\omega_1\omega_2)=M(\omega_1)M(\omega_2)$ and $M(1)={\bf 1}$.

When $d\equiv 1$ (mod $4$) the integers have the form (\ref{d1}) and one can 
check that the matrix representation can be chosen as follows
$$
M(\omega)=\left ( \begin{array}{cc}m & n\\\frac{d-1}{4} n& m+n\end{array}\right )\;.
$$
 
\subsection{Quaternion Algebras}\label{QuaternionAlgebras}

Algebras are   more general objects then fields. A (vector) algebra of finite
dimension $d$ is defined as formal sum
$$
\gamma=x_1\vec{i}_2+x_2\vec{i}_2+\ldots+x_d\vec{i}_d\;.
$$
Here $x_j$ belong to a basis field $\mbox{I\hspace{-.15em}F}$ and
$\vec{i}_j$ are formal objects (vectors) with a prescribed multiplication table 
$$
\vec{i}_j\vec{i}_k=\sum_{p=1}^dC_{jk}^p\vec{i}_p
$$
where $C_{jk}^p$ are from the basis field. The sum and the product of any
two elements of an algebra belong to it. General algebras should be neither commutative, nor associative.

An algebra is called a  normed algebras if there exists a function, $N(\gamma)$,
which associates to any element of the algebra a number from the basis field 
such that the norm of the product equals the product of the norms of both factors
$$
N(\gamma_1\gamma_2)=N(\gamma_1)N(\gamma_2)\;.
$$
An algebra is called a division algebra if the division is always possible (except a zero element).

Finite dimensional normed division algebras over real numbers  are exhausted 
by the following three possibilities  (the Frobenius theorem \cite{Frobenius}).
\begin{itemize} 
\item Commutative and associative division algebras are isomorphed either to the usual field of real numbers $\mbox{I\hspace{-.15em}R}$ or to the field of complex numbers $\mbox{l\hspace{-.47em}C}$. In the latter case the algebra is given by 
$$
\gamma = x_1+x_2 \imag 
$$ 
with $\imag^2=-1$. The norm in this case is
$$
N(\gamma)=x_1^2+x_2^2\;.
$$
\item Non-commutative but associative division algebras are isomorphed  to the
quaternion algebra
\begin{equation}
\gamma=x_1+x_2\vec{i}+x_3\vec{j}+x_4\vec{k}
\label{QuaternionAlgebra}
\end{equation}
where
\begin{equation}
\vec{i}^2=\vec{j}^2=\vec{k}^2=-1\;,\;\vec{k}=\vec{i}\vec{j}=-\vec{j}\vec{i}\;.
\label{Hamilton}
\end{equation}
The norm of the quaternion algebra is
$$
N(\gamma)=x_1^2+x_2^2+x_3^2+x_4^2\;.
$$
\item Non-associative normed division algebras are isomorphed  to the octonion algebra
$$
\gamma=\sum_{k=1}^8x_k\vec{i}_k
$$
with a complicated multiplication table and the norm given by the sum of
8 squares
$$
N(\gamma)=\sum_{k=1}^8x_k^2\;.
$$
\end{itemize}
Similarly, for an algebraic field $\mbox{I\hspace{-.15em}F}$ of finite degree
there exist quaternion normed algebras defined similarly to
Eqs.~(\ref{QuaternionAlgebra}) and (\ref{Hamilton}) \cite{Vigneras}. 
These  algebras are labeled by two elements $a,b\in \mbox{I\hspace{-.15em}F}$ 
and it is a four-dimensional non-commutative algebra with
basis $(\vec{1},\vec{i},\vec{j},\vec{k})$  as in (\ref{QuaternionAlgebra})
with the following multiplication table
\begin{equation}
\vec{i}^2=a\;,\;\;\vec{j}^2=b\;,\;\;\vec{k}=\vec{i}\vec{j}=-\vec{j}\vec{i}\;.
\label{relations}
\end{equation}
Such algebra is denoted by
$\left (\frac{a,b}{\mbox{I\hspace{-.15em}F}} \right )$ and its norm is
\begin{equation}
N(\gamma)=x_1^2-a x_2^2-b x_3^2+ab x_4^2\;.
\label{norm}
\end{equation}
The matrix representation of the quaternion algebra (\ref{relations}) is 
obtained by the isomorphism
$$
\vec{i}\rightarrow 
\left( \begin{array}{cc} \sqrt{a}& 0 \\ 0 & -\sqrt{a}  \end{array}
\right)\;,\;
\vec{j}\rightarrow 
\left( \begin{array}{cc} 0& 1 \\ b & 0  \end{array} \right)\;,\;
\vec{k}=\vec{i}\vec{j}\rightarrow 
\left( \begin{array}{cc}0& \sqrt{a} \\  -b\sqrt{a}&0  \end{array} \right)\;.
$$
Explicitly 
\begin{equation}  
\gamma =\left( \begin{array}{cc} x_1+x_2\sqrt{a}&x_3 +x_4\sqrt{a} \\
                   b (x_3 -x_4\sqrt{a}) & x_1-x_2\sqrt{a}  \end{array} \right)
\label{quaternion}
\end{equation}
with $x_1, x_2, x_3, x_4\in \mbox{I\hspace{-.15em}F}$. As it is a
representation of the quaternion algebra the product of these matrices also
has  the same form. In this representation the norm of the algebra
(\ref{norm}) equals the determinant of the matrix (\ref{quaternion})
$$
N(\gamma)=\det \gamma\;.
$$
From an algebraic field $\mbox{I\hspace{-.15em}F}$ of finite degree 
one can  build  also another simple set of matrices called 
M(2,$\mbox{I\hspace{-.15em}F}$)  
given by $2\times 2$ matrices  with entries from $\mbox{I\hspace{-.15em}F}$
\begin{equation}
\left ( \begin{array}{cc}x_1& x_2\\x_3& x_4\end{array}\right )\;.
\label{M2F}
\end{equation}
Are the two sets (\ref{quaternion}) and (\ref{M2F}) different or are they
isomorphic?  For example, if $a=u^2$ and 
$u\in \mbox{I\hspace{-.15em}F}$ the set (\ref{quaternion}) is, evidently, 
within M(2,$\mbox{I\hspace{-.15em}F}$). 

Let us show that if $\sqrt{a}\notin \mbox{I\hspace{-.15em}F}$ but there exist
certain elements
$q_1,q_2,q_3,q_4\in \mbox{I\hspace{-.15em}F}$ such that the 
determinant of the matrix (\ref{quaternion}) is zero
\begin{equation}
\det(\gamma)=q_1^2-q_2^2a-b(q_3^2-q_4^2a)=0
\label{determinant}
\end{equation}
then matrices (\ref{quaternion}) are isomorphic to 
M(2,$\mbox{I\hspace{-.15em}F}$) \cite{Katok}.

Indeed, from the above expression it follows that in this case $b$ has the form
$$
b=(q_1^2-q_2^2a)(q^3-q_4^2a)^{-1}=(u_1+u_2\sqrt{a})(u_1-u_2\sqrt{a})
$$
where
$$
u_1+u_2\sqrt{a}=(q_1+q_2\sqrt{a})(q_3+q_4\sqrt{a})^{-1}\;.
$$
As all fractions of field elements belong to the field 
$\mbox{I\hspace{-.15em}F}$, $u_1$ and $u_2$
are also elements of $\mbox{I\hspace{-.15em}F}$. Now one can check that
$$
\left( \begin{array}{cc} x_1+x_2\sqrt{a}&x_3 +x_4\sqrt{a} \\
b (x_3 -x_4\sqrt{a}) & x_1-x_2\sqrt{a}  \end{array} \right)=S^{-1}M S
$$
where
$$
M =\left( \begin{array}{cc} x_1+x_3u_1+x_4u_2a&x_2 -x_3u_2-x_4u_1\\
  a(x_2+x_3u_2+x_4u_1) & x_1-x_3u_1-x_4u_2a  \end{array} \right)
$$
and $S$ is a fixed (independent of $x_i$) matrix
$$
S=\left( \begin{array}{cc} u_1+u_2\sqrt{a}&1 \\
                   \sqrt{a} (u_1 +u_2\sqrt{a}) &-\sqrt{a}  \end{array}
		   \right)\;.
$$
The importance of such representation lies in the fact that the matrix $M$
contains only elements of our basis field $\mbox{I\hspace{-.15em}F}$  and 
does not contain $\sqrt{a}$. Therefore, it belongs to 
M(2,$\mbox{I\hspace{-.15em}F}$) and the set of matrices $\gamma$
(\ref{quaternion}) is the conjugation of matrices from 
M(2,$\mbox{I\hspace{-.15em}F}$) by a fixed matrix $S$. In other words,  
when the equation $\det (\gamma)=0$ has a solution 
$\gamma\in \mbox{I\hspace{-.15em}F}$ 
expression (\ref{quaternion}) is just a complicated way of
writing matrices from M(2,$\mbox{I\hspace{-.15em}F}$). These considerations 
demonstrate that in order to construct a group different from 
M(2,$\mbox{I\hspace{-.15em}F}$) it is necessary to require that
there exist {\em no} elements from the basis field such that the determinant
(\ref{determinant}) equals zero. Or, equivalently, any matrix
(\ref{quaternion}) should have an inverse element. In the language of
quaternion algebra this property corresponds to the division
algebra for which any element has an inverse.

As for  real fields this condition is quite restrictive and an explicit answer
can be obtained only in simple cases. Let us consider for example the case when 
$\mbox{I\hspace{-.15em}F}$ is the field of usual rational numbers 
$\mbox{I\hspace{-.15em}F}=\mbox{l\hspace{-.47em}Q}$. The following
theorem \cite{Katok} gives a series of division algebras over 
$\mbox{l\hspace{-.47em}Q}$. 

Let $b$ be a prime number and $a$ be an integer such that the equation
$$
x^2\equiv a\;(\mbox{mod }b)
$$
has no integer solution. Then the pair $(a,b)$ defines a division algebra
over $\mbox{l\hspace{-.47em}Q}$ or equivalently the equation
(\ref{determinant})
\begin{equation}
x_1^2-x_2^2a-b(x_3^2-x_4^2a)=0
\label{det}
\end{equation}
has only zero rational solutions. 

To prove the theorem note that due to homogeneity of this equation it is
sufficient to consider integer solutions $x_1,x_2,x_3,x_4$ without common
factors. From (\ref{det}) it follows that
$$
x_1^2\equiv x_2^2a \;(\mbox{mod }b)\;.
$$
Consider first the case when $b$ does not divide $x_2$,
$b\,|\hspace{-.38em}/ x_2$. As $b$ is assumed to be a prime, $x_2^{-1}$ (mod
$b$) exists and  $(x_1/x_2)^2\equiv a \;(\mbox{mod }b)$ which
contradicts our assumption. Hence $b\,|\,x_2$ but then $b\,|\,x_1$ and
$$
x_3^2\equiv x_4^2a \;(\mbox{mod }b)\;.
$$
The same arguments give that $b\,|\,x_4$ and $b\,|\,x_3$ which contradicts
the assumption about the absence of common factors of $x_i$. Therefore there is
no rational solution of (\ref{det}) and the quaternion algebra defined by $a$ and
$b$ is a division algebra.

Quaternion algebras with division are analogs of algebraic fields.
How one can define integers of a quaternion algebra? 

We have seen above that algebraic integers of a field of degree $n$ form a free
$\mbox{Z\hspace{-.3em}Z}$-module of rank $n$ i.e. they can be 
represented as  a sum of $n$ elements of the field with integer coefficients 
(see (\ref{AlgebraicIntegers}).  Similarly one
can define `integers' of a quaternion algebra over such field as a free
$\mbox{Z\hspace{-.3em}Z}$-module of rank $4n$ (which generates the whole
algebra). For technical reasons they are called not integers but
`the order' in the algebra. The word 'integers' in algebras is reserved for elements for which the trace and the determinant of matrix (\ref{quaternion}) are integers of the basis field. Different orders exist and the one which is  
not contained in  any other order is called the maximal order. 

The simplest case appears when $a$ and $b$ are integers of the basis field
$\mbox{I\hspace{-.15em}F}$. Then matrices of the form
$$
\left( \begin{array}{cc} x_1+x_2\sqrt{a}&x_3 +x_4\sqrt{a} \\
                   b (x_3 -x_4\sqrt{a}) & x_1-x_2\sqrt{a}  \end{array} \right)
$$
where all $x_k$ are integers of $\mbox{I\hspace{-.15em}F}$ form an order of 
the algebra $\left (\frac{a,b}{\mbox{I\hspace{-.15em}F}}\right )$ (but not necessarily the maximal order).

Matrices of the order in a division quaternion algebra with unit determinant
form a group. Each matrix of this group belongs to the order and, therefore,  is defined by $4n$ integers. The product of two group matrices have the same form and corresponds to a certain transformation of integers defining both matrices. It means that these groups can be represented by $4n\times 4n$ matrices with integer elements.  

All such groups, all their subgroups, and all groups commensurable
with them are {\em discrete arithmetic groups} with finite fundamental domain 
\cite{Katok}.

\paragraph{Example}

As $x^2 (\mbox{mod }5) $ takes only values $0,1,4$  the equation
$$
x^2\equiv 3 \;(\mbox{mod }5)
$$
has no integer solution. Hence the pair (3,5) defines a division algebra over
$\mbox{l\hspace{-.47em}Q}$. 

A simplest order of this algebra has the form
\begin{equation}
\left( \begin{array}{ll} m+n\sqrt{3}&k +l\sqrt{3} \\
5(k -l\sqrt{3}) & m-n\sqrt{3}  \end{array} \right)
\label{SimplestOrder}
\end{equation}
with integer $m,n,k,l$. When one considers these matrices 
with the unit determinant 
$$
m^2-3n^2-5k^2+15l^2=1
$$
they form a discrete arithmetic group $\Gamma_1$ with a finite fundamental area.

The order (\ref{SimplestOrder}) is not the maximal order. The latter can be 
chosen e.g. as follows
\begin{equation}
\left( \begin{array}{ll} \frac{1}{2}(m+n\sqrt{3})&\frac{1}{2}(k +l\sqrt{3}) \\
\frac{5}{2} (k -l\sqrt{3}) & \frac{1}{2}(m-n\sqrt{3})  \end{array} \right)
\label{MaximalOrder}
\end{equation}
with integer $m, n, k, l$ such that $m\equiv k\; (\mbox{mod } 2)$ and 
$n\equiv l\; (\mbox{mod } 2)$. Matrices (\ref{MaximalOrder})  
with the unit determinant
$$
m^2-3n^2-5k^2+15l^2=4
$$
constitute  another discrete arithmetic group $\Gamma_2$ whose fundamental domain 
is smaller than for the group (\ref{SimplestOrder}) as $\Gamma_1$ is a subgroup 
of  $\Gamma_2$.

Using the representation (\ref{representation}) one concludes that to each
$2\times 2$ matrix of the group  $\Gamma_1$ one can associate the $4\times 4$ matrix with integer elements
$$
M(\gamma)=\left (\begin{array}{rrrr}
m    & n  &  k  &  l \\
2n   & m  &  2l &  k \\
5k   &-5l &  m  & -n \\
-10k & 5k & -2n &  m
\end{array} \right )\;.
$$
It is straightforward to check that (i)
$M(\gamma_1\gamma_2)=M(\gamma_1)M(\gamma_2)$ for all $\gamma_1\;,\;\gamma_2 \in \Gamma_1$,
(ii) $M({\bf 1})={\bf 1}$, and (iii) $\det(M)=(\det(\gamma))^2=1$ . Together these expressions mean that this group is an arithmetic group. 

\subsection{Criterion of arithmeticity}\label{TakeuchiCriterion}

For general fields the situation is more complicated. 
To explain the  general criterion of arithmetic groups let us first
stress a difference between usual integers and algebraic integers.  

The usual integers correspond to a discrete set of points. But for 
general algebraic integers this is not the case. For example, in the field
$\mbox{l\hspace{-.47em}Q}(\sqrt{2})$ integers have the form $n+m\sqrt{2}$
with integer $n$ and $m$. But  it is evident that one can construct
sequences of these algebraic integers converging to zero, e.g.
$(\sqrt{2}-1)^k=M_k-N_k\sqrt{2}\to 0$ when $k\to \infty$. Therefore the set of algebraic
integers is not discrete as it have finite accumulation points. 

How can one deal with such problem? The main point is that these small terms
become large under the transformation
\begin{equation}
\sqrt{2}\to -\sqrt{2}
\label{sqrt2}
\end{equation}
Let consider in the above example not all algebraic integers $n+m\sqrt{2}$ but only those 
which after transformation (\ref{sqrt2}) remain bounded
$$
|n-m\sqrt{2}|<\mbox{constant}\;.
$$
It is clear that now arbitrary small integers are excluded and one gets a
discrete set of points. 

For more general fields the transformation (\ref{sqrt2}) is generalized to all
non-trivial isomorphisms of the field. To remove arbitrary small elements one has to
require that for all isomorphisms of the field (except the identity), $\phi_i$, 
transformed values of integers (\ref{AlgebraicIntegers}) are restricted 
\begin{equation}
\left |\sum_{k=0}^{n-1} m_k \phi_i(\beta_k)\right |<\mbox{constant}\;.
\label{bounded}
\end{equation}
In order to be sure that all small numbers are removed it is necessary that
all roots of defining equation (\ref{defining}) are real. Otherwise,
changing a root to its complex conjugate may not change modulus of integers.

These considerations make reasonable that in order to construct a 
{\em discrete} subset of algebraic integers (without finite accumulation
points) it is necessary that (i) the field be a totally real field
(i.e. all roots
of defining equation (\ref{defining}) are real) and (ii) for all non-trivial
isomorphisms of the field transformed integers remain bounded as in 
(\ref{bounded}). 

A precise criterion of arithmeticity obtained by Takeuchi \cite{Takeuchi} is
quite similar (see also \cite{BogomolnySteiner} and \cite{BGGS} for
particular examples).

Takeuchi proved that a group $\Gamma$ is an arithmetic group  if and 
only if the traces of group matrices  have the following properties
\begin{itemize}
\item All Tr($\gamma$) are integers of a totally real algebraic field of 
finite degree.
\item For any non-trivial isomorphism $\phi$ of this field
which changes some $|\mbox{Tr}(\gamma)|$  for certain $\gamma$,
the value of the transformed trace satisfies
$|\phi(\mbox{Tr}(\gamma))| \leq 2$.
\end{itemize}
There are two  types of arithmetic groups.  Non-compact groups, built from
$SL(2,\mbox{Z\hspace{-.3em}Z})$,
and compact ones built from quaternion algebra different from
$M(2,\mbox{l\hspace{-.47em}Q})$.

The above criterion is quite effective, in particular,  it permits 
to find all possible arithmetic groups with triangular fundamental domains
\cite{Takeuchi}. There are 85 triangular hyperbolic surfaces generated by
discrete arithmetic groups. All of them are given in Table~\ref{table1}.

\begin{table}
\caption{The list of arithmetic triangles from \cite{Takeuchi}. 
($n,m,p$) in the first column corresponds to the three angles
($\pi/n,\pi/m,\pi/p$).  The second column indicates the algebraic
field from which is built the corresponding arithmetic group.}
\begin{center}
\begin{tabular}{|lllll|c|} 
\hline\noalign{\smallskip}
 & & (m,n,p)& & & \mbox{I\hspace{-.15em}F} \\ 
\noalign{\smallskip}\hline\noalign{\smallskip}
(2,3,$\infty$)&(2,4,$\infty$)&(2,6,$\infty$)&(2,$\infty$,$\infty$)&(3,3,$\infty$)&
$\mbox{ l\hspace{-.47em}Q}$ \\
(3,$\infty$,$\infty$)&(4,4,$\infty$)&(6,6,$\infty$)&($\infty$,$\infty$,$\infty$)& & \\
\hline
(2,4,6)&(2,6,6)&(3,4,4)&(3,6,6)& & $\mbox{ l\hspace{-.47em}Q}$ \\ \hline
(2,3,8)&(2,4,8)&(2,6,8)&(2,8,8)&(3,3,4)& $\mbox{ l\hspace{-.47em}Q}(\sqrt{2})$ \\
(3,8,8)&(4,4,4)&(4,6,6)&(4,8,8)&  &    \\ \hline
(2,3,12)&(2,6,12)&(3,3,6)&(3,4,12)&(3,12,12)& $\mbox{ l\hspace{-.47em}Q}(\sqrt{3})$ \\
(6,6,6)& & & & & \\ \hline
(2,4,12)&(2,12,12)&(4,4,6)&(6,12,12)& & $\mbox{ l\hspace{-.47em}Q}(\sqrt{3})$ \\ \hline
(2,4,5)&(2,4,10)&(2,5,5)&(2,10,10)&(4,4,5)& $\mbox{ l\hspace{-.47em}Q}(\sqrt{5})$\\
(5,10,10)& & & & & \\ \hline
(2,5,6)&(3,5,5)& & & & $\mbox{ l\hspace{-.47em}Q}(\sqrt{5})$ \\ \hline
(2,3,10)&(2,5,10)&(3,3,5)&(5,5,5)& & $\mbox{ l\hspace{-.47em}Q}(\sqrt{5})$ \\ \hline
(3,4,6)& & & & & $\mbox{ l\hspace{-.47em}Q}(\sqrt{6})$ \\ \hline
(2,3,7)&(2,3,14)&(2,4,7)&(2,7,7)&(2,7,14)& $\mbox{ l\hspace{-.47em}Q}(\cos\pi/7)$ \\
(3,3,7)&(7,7,7)& & & & \\ \hline
(2,3,9)&(2,3,18)&(2,9,18)&(3,3,9)&(3,6,18)& $\mbox{ l\hspace{-.47em}Q}(\cos \pi/9)$ \\
(9,9,9)& & & & & \\ \hline
(2,4,18)&(2,18,18)&(4,4,9)&(9,18,18)& & $\mbox{ l\hspace{-.47em}Q}(\cos \pi/9)$ \\ \hline
(2,3,16)&(2,8,16)&(3,3,8)&(4,16,16)&(8,8,8)& $\mbox{ l\hspace{-.47em}Q}(\cos \pi/8)$ \\ \hline
(2,5,20)&(5,5,10)& & & & $\mbox{ l\hspace{-.47em}Q}(\cos \pi/10)$ \\ \hline
(2,3,24)&(2,12,24)&(3,3,12)&(3,8,24)&(6,24,24)& $\mbox{ l\hspace{-.47em}Q}(\cos \pi/12)$ \\
(12,12,12)& & & & & \\ \hline
(2,5,30)&(5,5,15)& & & & $\mbox{ l\hspace{-.47em}Q}(\cos \pi/15)$ \\ \hline
(2,3,30)&(2,15,30)&(3,3,15)&(3,10,30)&(15,15,15)&
$\mbox{ l\hspace{-.47em}Q}(\cos \pi/15)$ \\ \hline
(2,5,8)&(4,5,5)& & & & $\mbox{ l\hspace{-.47em}Q}(\sqrt{2},\sqrt{5})$ \\ \hline
(2,3,11)& & & & & $\mbox{ l\hspace{-.47em}Q}(\cos \pi/11)$ \\ \hline
\end{tabular}
\end{center}
\label{table1}
\end{table}

\subsection{Multiplicities of Periodic Orbits for General Arithmetic Groups}
\label{Multiplicities}

The geometrical length of the periodic orbit, $l$,  is connected with the
trace of class of conjugate matrices by (\ref{PeriodicLength}). When 
$l\to \infty$
$$ 
\exp \frac{l}{2} = |\mbox{Tr}(\gamma)|\;. 
$$
Let us prove that for an arithmetic group the number of possible
values of group matrix traces obeys  
$$ 
N(|\mbox{Tr}(\gamma)| \leq R ) \stackrel{R\to \infty}{\longrightarrow} 
C\cdot R 
$$
with a constant $C$ depending on the group. The traces of matrices of 
arithmetic groups are dispatched as usual integers among real numbers.

Let $\Gamma$ be an arithmetic group. 
The set of traces  $\{\mbox{Tr}(\gamma),\; \gamma \in \Gamma\}$ 
are  integers of an algebraic field $\mbox{I\hspace{-.15em}F}$
$$ 
t_0= \sum_{i=0}^{n-1} m_i \beta_i 
\label{betai}
$$
where $m_i$ are integers and $\beta_i$ are linearly independent elements of the
field. Consider the simplest case $\beta_i = \alpha^i$ then
$$ 
\mbox{Tr}(\gamma)\equiv t_0 = \sum_{i=0}^{n-1} m_i \alpha^{i} \;.
$$
For a field of degree $n$ there exist $n-1$ non-trivial
isomorphisms $\phi_k:\:\: \alpha \rightarrow \alpha_k$ where $\alpha_k$ 
is a root of the defining polynomial different from $\alpha$.

Suppose that all such transformations change $|\mbox{Tr}(\gamma)|$.
According to the criterion of Takeuchi all transformed traces satisfy 
$$ 
| t_k| \leq 2 
$$
where
$$
t_k \equiv \phi_k (\mbox{Tr}(\gamma)) = \sum_{i=0}^{n-1} m_i \alpha_k^i\;.
$$
Consider these equations as  transformations from variables $t_i$ to new  
variables $m_i$ \cite{Bolte}. The volume elements in these two representations are related as 
$$ 
\D t_0 \D t_1...\D t_{n-1} = |{\cal J}| \D m_0 \D m_1...\D  m_{n-1} 
$$
where
$$ 
{\cal J}= \mbox{det} \left( \frac{\partial t_j}{\partial m_k} \right ) 
$$
is called the discriminant of the field and in our case 
(when  $\beta_i =\alpha^i$)
$$ 
{\cal J}= \mbox{det}( \alpha^j_k)|_{k,j=0,...,n-1} = \prod_{i\neq
j}(\alpha_i-\alpha_j)\;.
$$
As $m_i$ are integers the volume of the smallest cell is one, and the 
total number of possible integer solutions is asymptotically 
$$ 
N(|\mbox{Tr}(\gamma)| \leq R )=N(|t_0|\leq R, |t_j|\leq 2) \simeq C \cdot R 
$$
where $C=2^n/\cal{J}$.

For any surface of finite area generated by  a discrete group
the total number of periodic orbits with length less that a given value is
asymptotically the following  
$$ 
N_{tot}(l<L)\stackrel{L\to \infty}{\longrightarrow} \frac{\E^{L}}{L}\;. 
$$
The number of periodic orbits with {\em different} lengths is the same as
the number of group matrix traces  
$$
 N_{\mbox{\scriptsize diff. lengths}}(l<L)\sim C \cdot \E^{L/2}
$$
Let $g(l)$ be the multiplicity of periodic orbits with length $l$. 
Then
$$
\sum_{l<L}g(l)=\frac{\E^{L}}{L} \mbox{ and } \sum_{l<L}1= C \E^{L/2} 
$$
where the summation is done over different lengths.   

Finally the {\em mean} multiplicity of arithmetic systems defined as in (\ref{MeanDefinition}) has the following asymptotics
\begin{equation}
\left\langle g\right \rangle =\frac{(\sum_{l<L}g(l))'}{(\sum_{l<L}1)'}\sim
\frac{2\E^{L/2}}{C L}\;.
\label{DifferentLengths}
\end{equation}
Thus we demonstrate that the arithmetic nature of arithmetic groups leads to 
exponential multiplicities of periodic orbit lengths.

For generic systems one usually does not expect any degeneracy of 
periodic orbit lengths  except the ones which follow from  exact symmetries of
the model. For example, systems with time-reversal invariance, in general,
should have the mean multiplicity equal to $2$, which corresponds to the same
geometrical periodic orbits spanned in two directions. Therefore, 
arithmetic systems are {\em very exceptional} in this respect as they
display exponentially large multiplicities of periodic orbit lengths.  
Notice, nevertheless,  that according to Horowitz-Randol theorem
\cite{Horowitz}, \cite{Randol} this degeneracy is unbounded for any 
surface generated by a discrete group.  However degeneracies of this theorem 
are much smaller than exponential.
                          
\section{Diagonal Approximation for Arithmetic Systems}\label{DiagonalArithmetic}

The large multiplicities of periodic orbit lengths  in arithmetical systems
seem to have no importance in classical mechanics.  
These systems are as chaotic  as any other models of free motion 
on  constant negative curvature surfaces with finite area. Nevertheless,  
the quantum spectra of these systems are anomalous: is it connected to these 
degeneracies? In this Section we estimate the quantum two-point correlation
form factor for arithmetic systems in the diagonal approximation as was done in 
Sect.~\ref{Diagonal} for generic chaotic systems.

Assume that there exist $g(l)$ periodic orbits with the same length $l$. 
Exactly as it was done in Sect.~\ref{Diagonal} one gets  the 
following expression for the diagonal approximation of the two-point 
correlation function 
\begin{equation}
R_2^{(diag)}(\epsilon)=\sum_{p,n}|A_{p,n}(l_p)|^{2}g(l_{p})
\E^{\imag nT_p\epsilon} +\mbox{c.c.}
\label{diagR}
\end{equation}
where the summation is done over all periodic orbits. The only difference 
with (\ref{R2diag}) is that in  Sect.~\ref{Diagonal} it was assumed 
that $g$ is a constant but here the multiplicity $g=g(l)$.

Define the two-point correlation form factor as the Fourier transform
of $R_2(\epsilon)$  
$$
K(t)=\int_{-\infty}^{+\infty}R_2(\epsilon)\E^{\imag t\epsilon}  \;.
$$
This definition differs from the previous one by the absence of the 
factor $2\pi$ in the exponent. For later purposes it is more convenient.

Equation (\ref{diagR}) leads to the following expression for the two-point correlation form factor in the diagonal approximation
\begin{equation}
K^{(diag)}(t)=2\pi \sum_{p,n}|A_{p,n}(l_p)|^{2}g(l_{p})
\delta \left (t-\frac{nl_p}{2k}\right )\;.
\label{diagK}
\end{equation}
From (\ref{DifferentLengths}) it follows that the mean multiplicity
of periodic orbit lengths for arithmetic systems is asymptotically
$$
\left \langle g(l_p)\right \rangle = \frac{2\E^{l_p/2}}{C l_p}
$$
with a model dependent constant $C$ (for the modular group $C=1$).

For any models generated by discrete groups the summation over all 
periodic orbits is asymptotically equals the integration with the following
measure
$$
\sum_{l_p}\to \int \frac{dl}{l}\E^{l}\;.
$$
Taking into account that when $l\to\infty$  the term with $n=1$
dominates and (see (\ref{Apn}))  
$$
A_{p,1}(l) \stackrel{l\to \infty}{\longrightarrow} \frac{l\E^{-l/2}}{4\pi k}
$$
one obtains that in the diagonal approximation
\begin{equation}
K^{(diag)}(t)\sim \frac{\E^{k t}}{ 2\pi k C}\;. 
\label{DiagArith}
\end{equation}
It means that the correlation form factor $K(t)$ for arithmetic systems 
grows much faster than was usually assumed and that for time of order of 
the Ehrenfest time it becomes of the order of 1.

The simplest approximation to the full form factor is the following 
$$  
K(t)=\left \{ \begin{array}{ll}K^{diag}(t)& \mbox{for}\:\: t<t^{*}\\
                               \bar{d}&\mbox{for}\:\: t>t^{*}
                               \end{array}\right . 
$$
where $t^{*}$ is defined by the requirement that $K^{diag}(t^{*})=\bar{d}$
$$ 
t^{*}\sim \frac{1}{k}\ln (2\pi k C\bar{d})\;. 
$$
For the true  Poisson statistics $K(t)$  always equals $\bar{d}$. For 
usual integrable systems $K(t)$ increases to this value during the 
time of the order of shortest periodic orbit periods, $t^{*}\sim 1/k$.
For arithmetic systems  $K(t)$ jumps to the universal saturation value  in a time of order of the Ehrenfest time which has an additional logarithm  of the momentum.  

Therefore, spectral statistics of arithmetic systems is much closer
to the Poisson prediction typical for integrable systems than to any of 
standard random matrix ensembles conjectured for generic ergodic systems.

\section{Exact Two-Point Correlation Function for the Modular Group}
\label{ExactModular}

The diagonal approximation gives quite  crude estimate of the form
factor. For the modular group it is possible to compute  explicitly 
the two-point correlation function \cite{BogomolnyLeyvraz}. 
The calculations are based on a generalization of the Hardy--Littlewood  method and depend strongly on the number-theoretical properties of the multiplicities of the periodic orbits of the modular group. In Sect.~\ref{Identities} using the Selberg trace formula the 
two-point correlation form factor is expressed through the two-point correlation function of multiplicities of periodic orbit lengths for the modular group. In Sect.~\ref{CorrelationMultiplicities} the latter is calculated by a certain generalization of the Hardy--Littlewood method. 
Quite tedious explicit formulas are given in Sect.~\ref{Explicit} and the final expression for the two-point correlation form factor is presented in Sect.~\ref{Formfactor}.    

\subsection{Basic Identities}\label{Identities}

The modular group has been considered in Sect.~\ref{Peculiarities}. It is the 
group of all $2\times2$ matrices with integer elements  and unit determinant. 
The periodic orbits of the modular group  correspond in a unique way to
the conjugacy classes of hyperbolic elements of the modular group. The length of 
periodic orbit $l_p$ is related with the trace of a representative matrix
 of the conjugacy class as follows
$$
|\mbox{Tr} M|=2\cosh l_p/2\;.
$$
As all  elements of modular group  matrices are integers the trace is also an 
integer
$$
|\mbox{Tr} M|=n\;.
$$
In Sect.~\ref{Peculiarities} it was demonstrated that the {\em mean} multiplicity 
of periodic orbit length for the modular group is  
$$
\left\langle{g(l)}\right\rangle=2{\E^{l/2}\over l}\;.
$$
Denote by $n$ the trace of a given conjugacy class and by $g(n)$
the number of distinct conjugacy classes corresponding to trace $n$.
As $n$ goes as $e^{L/2}$ when $n\to\infty$ one concludes that
\begin{equation}
\left\langle{g(n)}\right\rangle\stackrel{n\to \infty}{\longrightarrow}{n\over\ln n}\;.
\label{ModularMultiplicity}
\end{equation}
According to the Selberg trace formula  the density of eigenvalues for the modular surface $d(E)=\bar{d}(E)+d^{(osc)}(E)$ where the oscillating part of the density is represented by the following formal sum
$$
d^{(osc)}(E)={2\over\pi k}\sum_{n}g(n){\ln n\over n}\cos(2k\ln n)\;.
$$
From  (\ref{ModularMultiplicity}) it follows that mean value of $g(n)\ln n/n$
is one. Therefore we define
$$
\alpha(n)=g(n)\frac{\ln n}{n}\;,
$$
so
$$
d^{(osc)}(E)={1\over\pi k}\sum_{n}\alpha(n)\cos(2k\ln n)
$$
and $\left\langle{\alpha(n)}\right\rangle=1$.

As it was done in Sect.~\ref{Correlation} one gets
$$
R_2(\epsilon_1,\epsilon_2)=\bar{d}^2+R_2^{c}(\epsilon_1,\epsilon_2)
$$
where
\begin{eqnarray*}
R_2^{c}(\epsilon_1,\epsilon_2)&=
&{1\over (2\pi k)^2}\sum_{n_1,n_2}\alpha (n_1)\alpha (n_2)
\left \langle \E^{2\imag (k_1\ln n_1+k_2\ln n_2)}\right .+\nonumber \\
&+& \left .\E^{2\imag (k_1\ln n_1-k_2\ln n_2)}\right \rangle +\mbox{c.c.}
\end{eqnarray*}
and
$$
k_i\approx \sqrt{E+\epsilon_i}\stackrel{E\to\infty}{\longrightarrow} k+\epsilon_i/2k.
$$
As was discussed in Sect.~\ref{Correlation} due to the energy average 
the first term will be washed out and the second one gives contributions only when
$$
n_2=n_1+r\;\;\;\; \mbox{with }\; r\ll n_1\;.
$$
Finally $R_2^{c}(\epsilon_1,\epsilon_2)=\overline R_2(\epsilon)$  
where $\epsilon=\epsilon_1 - \epsilon_2$ and 
$$
\overline R_2(\epsilon)=
{1\over 4\pi^2k^2}\sum_{n}\sum_{r}
\alpha(n) \alpha(n+r)
\exp\left(-2\imag {kr\over n}+\imag \epsilon{\ln n\over k}\right)+\mbox{c.c.} \;.
$$
Let assume that the following mean value exists
$$
\gamma(r)=\lim_{N\to\infty}{1\over N}\sum_{n=1}^N \alpha(n)\alpha(n+r)\;.
$$
The dominant contribution to the two-point correlation function
corresponds to replace the product $\alpha(n)\alpha(n+r)$ by 
its mean value $\gamma (r)$
\begin{equation}
\overline R_2(\epsilon)\approx 
{1\over 4\pi^2k^2}\int_{n0}^{\infty}  dn \sum_{r=-\infty}^\infty
\gamma(r)e^{-2ikr/n} \exp\left(\imag \epsilon{\ln n\over k}\right)+\mbox{ c.c.}
\label{ModularR2}
\end{equation}
where we have used a continuum approximation for $n$ starting formally from a 
certain fixed $n_0\gg 1$, since only large values of $n$ make a significant contribution.

Define a (real) function $f(x)$ as follows
\begin{equation}
f(x)=\sum_{r=-\infty}^\infty \gamma(r)e^{-irx}\;.
\label{fx}
\end{equation}
This function has the meaning of the Fourier transform of the two-point 
correlation function for multiplicities of the modular group.

After the changing variable $n \to \E^{uk}$ in (\ref{ModularR2}) one gets
that the two-point correlation function for the modular group is expressed 
through $f(x)$ as follows
\begin{equation}
\overline R_2(\epsilon)\approx {1\over 2\pi^2k}\int_{0}^\infty 
\E^{ku }f(2ke^{-ku})\cos\epsilon u \ \D u
\label{Rmodular}
\end{equation}
and the two-point correlation form factor is
\begin{equation}
K(t)=\int_{-\infty}^\infty \overline R_2(\epsilon)
\E^{\imag \epsilon t}\D \epsilon
=\frac{1}{2\pi k}\E^{kt}f(2ke^{-kt})\;.
\label{Kmodular}
\end{equation}
Therefore  all non-trivial information is contained in functions $\gamma(r)$ or $f(x)$.

The simplest diagonal approximation 
is to assume that the $\alpha(n)$ are essentially uncorrelated, that is, 
$\gamma(r)$ is zero for $r\neq 0$. This gives for $f(x)$ a constant value 
which leads to an exponential growth of $K(t)$ as in (\ref{DiagArith}). But
from general considerations $K(t)$ obtained from a discrete spectrum
has to saturate to a constant value for $t\to\infty$, consequently,  the diagonal 
approximation cannot be correct for large $t$. 

\subsection{Two-Point Correlation Function of Multiplicities}
\label{CorrelationMultiplicities}
The purpose of this Section is to calculate the two-point correlation function
 of modular group multiplicities, $\gamma(r)$, whose Fourier harmonics
 according to (\ref{Kmodular}) determines the two-point correlation form factor.
 
The calculation will be done by a generalization of the Hardy-Littlewood method 
for prime pairs discussed in Sect.~\ref{HardyLittlewood}. As for primes one has 
to perform the three following steps.

\paragraph{The first step} 

Define the mean value of $\alpha(n)$ when $n$ runs over integers of the form $mq+r$ for fixed $q$ and $r<q$ in the following way
$$
\alpha(q;r)=\lim_{N\to\infty}{1\over
N}\sum_{m=0}^{N-1}\alpha(mq+r)\;.
$$
Since $\left\langle{\alpha(n)}\right\rangle=1$
$$
\sum_{r=0}^{q-1}\alpha(q;r)=q\;.
$$
Let  $M_q$ be the set of $2\times2$ matrices with entries being integers modulo $q$ and having determinant equals one modulo $q$.
These matrices form a group under multiplication modulo $q$ 
which is sometimes called the modulary group.

Define also $M_{q,r}$ to be the set of
elements of $M_q$ with trace equal to $r$ modulo $q$. One can prove 
\cite{BogomolnyLeyvraz} that
$$
\alpha(q;r)={q|M_{q,r}|\over|M_q|}
$$
where $|M|$ is the number of elements of a set $M$.

The intuitive meaning of this  result is the following: $g(n)$ is the
number of conjugacy classes of modular matrices of trace $n$. To each
modular matrix, one can associate an element of $M_q$ in a unique way
simply by taking the entries of the matrix modulo $q$. If $n$ is equal to
$r$ modulo $q$, then all these matrices will belong to $M_{q,r}$.
If we therefore assume that the matrices of the modular
group cover the set $M_q$ in some sense uniformly,
the result follows. More careful treatment has been performed in 
\cite{BogomolnyLeyvraz}.

\paragraph{Example}

Let us consider $q=2$. Integers modulo $2$ are $0$ and $1$. The group $M_2$ consists of  the following matrices
$$
\left (\begin{array}{cc}1&0\\0&1\end{array}\right ),
\left (\begin{array}{cc}1&1\\0&1\end{array}\right ),
\left (\begin{array}{cc}1&0\\1&1\end{array}\right ),
\left (\begin{array}{cc}0&1\\1&0\end{array}\right ),
\left (\begin{array}{cc}0&1\\1&1\end{array}\right ),
\left (\begin{array}{cc}1&1\\1&0\end{array}\right )\;.
$$
The dimension of the group $M_2$, i.e. the total number of matrices, $|M_2|=6$. 
Besides these matrices there are four matrices  with zero trace (mod 2), i.e. $|M_{2,0}|=4$, and two matrices with trace equals $1$ (mod 2), $|M_{2,1}|=2$. Therefore
$$
\alpha(2;0)=\frac{2\cdot 4}{6}=\frac{4}{3}\;,\;\;
\alpha(2;1)=\frac{2\cdot 2}{6}=\frac{2}{3}\;.
$$
\paragraph{The second step} 

Define as in the Hardy-Littlewood method the following function
$$
\Phi(z)=\sum_{n=0}^\infty \alpha(n)z^n.
$$
Since $\left\langle{\alpha(n)}\right\rangle=1$  the convergence radius of 
this series is equal to one.

The importance of this function follows from the integral
$$
J_r(e^{-u})=\E^{ru}\int_0^{2\pi}{\D \phi\over2\pi}\Phi^*\left(\E^{-u+\imag \phi}
\right)\Phi\left(\E^{-u-\imag \phi}\right)\E^{-\imag r\phi}=\sum_{n=1}^\infty
\alpha(n)\alpha(n+r)\E^{-2nu}
$$
whose right-hand side by a Tauberian theorem is connected  
with the two-point correlation function of multiplicities, $\gamma(r)$.

The essence of the Hardy--Littlewood approach  is the
investigation of the function $\Phi(z)$ when  
$z=\exp(-u+\imag \epsilon+2\pi \imag p/q)$
with $u\to 0$ and $\epsilon\to 0$, where $p$ and $q$ are co-prime integers.
The main step is then to write $n$ in the form $mq+r$ with $r$
lying between $0$ and $q-1$ and prove that in the expression for
$\Phi(z)$  the dominant contribution as $u$ and
$\epsilon$ go to zero will be given by the {\em mean} value of
$\alpha(mq+r)$, that is, one may substitute it by $\alpha(q;r)$.

Accepting this, one has that as $u\rightarrow 0$ and $\epsilon \rightarrow 0$
\begin{eqnarray*}
\Phi\left(\exp\left(-u+2\pi \imag  p/q+\imag \epsilon\right)\right)&=&\sum_{r=0}^{q-1}
\sum_{m=0}^\infty\alpha(mq+r)\E^{-(u-\imag \epsilon)(mq+r)}\E^{2\pi \imag rp/q}=
\nonumber\\
&=&\sum_{r=0}^{q-1}\alpha(q;r)\E^{2\pi \imag pr/q}{1\over q}
\int_0^\infty \D n\,\E^{-(u-\imag \epsilon)n}=
\nonumber\\
&=&{\beta(p,q)\over u-\imag \epsilon}
\end{eqnarray*}
where
$$
\beta(p,q)=q^{-1}\sum_{r=0}^{q-1}\alpha(q;r)\exp\left(2\pi \imag \frac{p}{q} r\right)\;.
$$
Hence  $\Phi(z)$ has a pole singularity at all rational points on the unit circle.

\paragraph{The third step} 

Divide the unit circle in intervals $I_{p,q}$ 
centered around $\exp(2\pi \imag p/q)$, where $p$ and $q$ are co-prime integers 
with $p<q$.  If one neglects all terms in each interval except the pole term
and extends the integration over $\epsilon$ to the whole line, one gets
\begin{eqnarray*}
J_r(\E^{-u})&=&\E^{ru}\sum_{(p,q)=1}\int_{-\infty}^\infty{\D\epsilon\over2\pi}
{|\beta(p,q)|^2\over u^2+\epsilon^2}\E^{\imag r(2\pi p/q+\epsilon)}=
\nonumber\\
&=&{1\over2u}\sum_{(p,q)=1}|\beta(p,q)|^2\exp\left(2\pi \imag  \frac{p}{q} r\right)\;.
\end{eqnarray*}
Finally one obtains that
$$
\gamma(r)=\sum_{(p,q)=1}|\beta(p,q)|^2\exp\left(2\pi \imag \frac{p}{q} r\right)\;.
$$
The sum is performed over all $q$, and $p$  co-prime to $q$ with $0<p<q$.

This is the two-point correlation function of {\em multiplicities}
of the periodic orbits for the modular group . All other quantities
of interest can be obtained from it. In particular, the function
$f(x)$ (\ref{fx}) is given by
$$
f(x)=2\pi\sum_{(p,q)=1}|\beta(p,q)|^2\delta \left (x-2\pi \frac{p}{q}\right )
$$
where the summation is done over all $p$ and $q$ co-prime, without
the restriction $p<q$.   

According to (\ref{Rmodular}) and (\ref{Kmodular}) the knowledge of $f(x)$ determines immediately the two-point correction function  the form factor of modular domain eigenvalues.

\subsection{Explicit Formulas}\label{Explicit}

Let us define the so-called Kloosterman sums 
$$
S(n,m;c)=\sum_{(d,c)=1}\exp\left({2\pi \imag \over c}(nd+md^{-1})\right)
$$
where the summation is taken over all $d<c$ co-prime with $c$ and $d^{-1}$ is an integer modulo $c$ which obeys $d^{-1}d=1$ (mod $c$).

One can show (see \cite{BogomolnyLeyvraz}) that $\beta(p,q)$ can be expressed 
through these sums in the following way
$$
\beta(p,q)=\frac{1}{q^2\prod_{\omega |q} (1-\omega^{-2})}S(p,p;q)
$$
where $\omega$ are the prime divisors of $q$.

The function $\gamma(r)$ can be written as 
$$
\gamma(r)=\sum_{n=1}^{\infty}A_r(n)
$$
where $A_r(q)$ is given by
$$
A_r(q)=\sum_{p:(p,q)=1}\left|\beta(p,q)\right|^2
\exp\left(2\pi \imag r \frac{p}{q} \right)\;.
$$
One can prove that $A_r(q)$ is  multiplicative function of $q$, i.e. 
$A_r(n_1n_2)=A_r(n_1)A_r(n_2)$ provided $(n_1n_2)=1$, therefore one needs to know only its values on powers of primes and $\gamma(r)$ can be rewritten as the infinite product over all prime numbers
$$
\gamma(r)=\prod_p (1+\sum_{k=1}^{\infty} A_r(p^k))
$$
To present a closed expression for $A_r(q)$ let us introduce  the standard
definition of the Legendre symbol
$$ 
\left(a\over q\right )=\left \{\begin{array}{rl} 1\;, & \mbox{ if } a
\equiv x^2\; (\mbox{mod }q) 
\mbox{ has a solution} a \not\equiv 0\; (\mbox{mod }q)\\  
0\;, & \mbox{ if } a\equiv 0 \;(\mbox{mod }q)\\
-1\;,& \mbox{ otherwise}\end{array}\right . \;.
$$
The meaning of this symbol is perhaps best understood by saying that the 
number of {\em distinct} solutions of the equation 
$x^2\equiv a \;(\mbox{mod }q)$ is $1+(a/q)$.

A fairly tedious evaluation of $A_r(q)$ (see \cite{BogomolnyLeyvraz} for details)  
gives the following formulas.
 
Let $q=p^n$ where $p$ is an odd prime. Then for $n=1$ one has
$$
A_r(p)={1\over(p^2-1)^2}\left[p\sum_{x=0}^{p-1}
\left((x^2-4)((x+r)^2-4 \over p \right )-1\right]\;.
$$
For $n\geq2$ we have, letting $t$ be an arbitrary non-zero
number modulo $p$,
$$
A_r(p^n)={1\over p^{2n}(1-p^{-2})}\left\{
\begin{array}{rlr}
2(1-1/p)&,\;r\equiv0&\pmod{p^n}\\
-2/p&,\;r\equiv tp^{n-1}&\pmod{p^n}\\
\epsilon(n,p)(1-1/p)&,\;r\equiv\pm4&\pmod{p^n}\\
-\epsilon(n,p)/p&,\;r\equiv\pm4+tp^{n-1}&\pmod{p^n}
\end{array}
\right.
$$
where $\epsilon(n,p)$ takes the value $-1$ if $n$ is odd and $p$ is
of the form $4k+3$ and is equal to $1$ in all other cases.
For $p=2$, we list down individual cases for low powers
and eventually state a general rule
\begin{eqnarray*}
A_r(2)&=&\frac{1}{9}\left\{
\begin{array}{rl}
1&,\;r\equiv0\pmod2\\
-1&,\;r\equiv1\pmod2
\end{array}\right.\;,\\
A_r(4)&=&\frac{1}{18}\left\{
\begin{array}{rl}
1&,\;r\equiv0\pmod4\\
-1&,\;r\equiv2\pmod4
\end{array}\right.\;,\\
A_r(8)&=&0\;,\\
A_r(16)&=&\frac{1}{9\cdot16}\left\{
\begin{array}{rl}
1&,\;r\equiv0\pmod{16}\\
-1&,\;r\equiv8\pmod{16}
\end{array}\right.\;,\\
A_r(32)&=&0\;,
\end{eqnarray*}
and finally, for the general case $n\geq 6$
$$
A_r(2^n)=\frac{1}{9\cdot2^{2n-4}}\left\{
\begin{array}{rlr}
2&,\;r\equiv0&\pmod{2^n}\\
-2&,\;r\equiv2^{n-1}&\pmod{2^n}\\
1&,\;r\equiv\pm(4+2^{n-2})&\pmod{2^n}\\
-1&,\;r\equiv\pm(4+2^{n-2}+2^{n-1})&\pmod{2^n}\\
\end{array}\right. \;.
$$
All terms not explicitly shown equal zero. 
In \cite{Peter} these formulas were proved by a different method.

\subsection{Two-Point Form Factor}\label{Formfactor}

These formulas give the explicit expression for the two--point
correlation form factor
$$
K(t)={1\over 2\pi^2k}\sum_{(p,q)=1}\left|{q\over p}\beta(p,q)\right|^2
\delta(t-t_{p,q})\;.
$$
where
$$t_{p,q}={1\over k}\ln{kq\over\pi p}\;.$$
In the limit $k\rightarrow \infty$ and $t$ fixed, the dominant contribution
comes from terms with $p/q\ll 1$. Smoothing over such values one can show (see 
\cite{BogomolnyLeyvraz}) that in  this limit $K(t)$ tends
to the constant Poisson value
$$
K(t)=\frac{A}{4\pi}
$$
where $A=\pi/3$ is the area of the fundamental region of the modular group.
For small $t$ (of the order of the Ehrenfest time $\ln k/k$) $K(t)$ has number--theoretical oscillations due to cumulative contributions of degenerate periodic
orbits. For very small values of $t$ (of the order of $1/k$) the two-point
form factor has $\delta$ peaks  connected with short periodic orbits.

Though the modular group is by no means a generic system, it is the first
ergodic dynamical system for which it was possible to compute explicitly the
distribution of the energy levels.

\section{Hecke operators}\label{Hecke}

Arithmetic groups have many interesting properties. 
In particular, for all arithmetic groups it is possible to construct an 
infinite number of mutually commuting operators which commute also with the 
Laplace--Beltrami operator. These operators are of pure arithmetic origin 
and are called the {\em Hecke operators} \cite{Hecke}, \cite{Terras}.

In a certain sense these operators permit to 'understand' why arithmetic 
systems have the Poisson statistics typical only for integrable systems. 
The point is that integrable systems are systems with sufficiently large 
number of independent commuting operators and Hecke operators may be viewed 
as a manifestation of a kind of arithmetic integrability of arithmetic systems 
which does the Poisson statistics for these models natural \cite{Bolte}.
Unfortunately, precise relations along this line seem to be impossible. 

Let us consider informally the construction of Hecke operators for the modular 
group.   
Choose  two matrices $A$ and $B$ from the modular group with the same trace. 
As they have the same trace and determinant, they have the same eigenvalues 
and there exists a matrix $\gamma$ such that
\begin{equation}
\gamma A \gamma^{-1}=B\;\mbox{ or }\;\;\gamma A=B\gamma \;
\mbox{ and }\;\det(\gamma)\neq 0\;.
\label{gam}
\end{equation}
If $A$ and $B$ are not congugated in the modular group, $\gamma\not\in$
PSL(2, $\mbox{Z\hspace{-.3em}Z}$). But matrix $\gamma$ can be chosen as 
a matrix with integer elements but with the determinant $\neq 1$.

\vspace{1ex}

{\em Example.}\\
Consider the following simple matrices
$$A=\left (\begin{array}{cc}3&1\\2&1\end{array}\right )\;,
\;\;\;B=\left (\begin{array}{cc}3&2\\1&1\end{array}\right )\;.
$$
General form of matrices $\gamma$ which obey (\ref{gam}) is 
$$
\gamma=\left (\begin{array}{cc}2\alpha+2\beta&\alpha \\ 
\alpha&\beta \end{array}\right )
$$
with arbitrary $\alpha$ and $\beta$.

It is clear that there exists no
$\gamma \in$ PSL(2, $\mbox{Z\hspace{-.3em}Z})$ but choosing different
integer values of $\alpha$ and $\beta$ one can construct an
infinite number of integer matrices with determinant $\neq 1$ which obeys 
(\ref{gam}). For example,
$$
\gamma=\left (\begin{array}{cc}0&1\\1&1\end{array}\right )\;\;(\det=-1)\;,\;
\gamma=\left (\begin{array}{cc}4&1\\1&1\end{array}\right )\;\;(\det=3)\ldots
\;.
$$

\vspace{1ex}

\noindent
These considerations demonstrate that when dealing with the modular group it 
is quite natural to consider matrices with integer elements but with the 
determinant different from one
$$
M_p= \left \{ \left (\begin{array}{cc}a&b\\c&d\end{array}\right )\;
a,b,c,d \mbox{ integers}, ad-bc=p\right \}\;.
$$
Matrices $M_p$ with $p\neq 1$ do not form a group because their product has 
not the same form.

A matrix $m_p\in M_p$ can uniquely be represented in the form
\begin{equation}
m_p=\mu \alpha_p
\label{mp}
\end{equation}
where $\mu\in$ PSL(2, $\mbox{Z\hspace{-.3em}Z})$ and $\alpha_p$ is one of
matrices from the following finite set
\begin{equation}
\alpha_p=\left \{ \left (\begin{array}{cc}a&b\\0&d\end{array}\right )\;
a,b,d \mbox{ integers }\;,\; ad=p\;,\; d>0\;,\; 0\leq b\leq d-1\right \} \;.
\label{lemma}
\end{equation}
Instead of proving this fact  let us  transform a simple
matrix    
$$
m_3=\left (\begin{array}{cc}4&1\\1&1\end{array}\right )
$$
to the form (\ref{mp}). General proof (see e.g. \cite{Terras}) follows the
same steps.\\
First, it is necessary to find a matrix 
$$
\mu'=\left (\begin{array}{cc}\mu_1&\mu_2\\ \mu_3&\mu_4\end{array}\right )
$$
such that (i) $\det \mu'=1$ and (ii)
$$
\left (\begin{array}{cc}\mu_1&\mu_2\\ \mu_3&\mu_4\end{array}\right )
\left (\begin{array}{cc}4&1\\1&1\end{array}\right )=
\left (\begin{array}{cc}a&b\\0&d\end{array}\right )\;.
$$
The condition of the zero of low-left element  gives the equation
$4\mu_3+\mu_4=0$ and because  $\mu_4$ and $\mu_3$ are coprime they can
be chosen as follows: $\mu_4=4$ and $\mu_3=-1$. Unit
determinant condition gives $\mu_1=k$ and $\mu_2=1-4k$ with an arbitrary
integer $k$. Finally,  $b=1-3k$ and  the smallest positive $b$ modulo $3$ 
corresponds to $k=0$. Hence, our matrix $m_3$ has the following representation  
$$
\left (\begin{array}{cc}4&1\\1&1\end{array}\right )=
\left (\begin{array}{cr}4&-1\\1&0\end{array}\right )  
  \left (\begin{array}{cc}1&1\\0&3\end{array}\right )\;.
$$
An important property of the set $M_p$ is that when one multiplies a matrix
from this set by a matrix from the modular group the resulting matrix also
belongs to $M_p$
$$
M_pg=M_p\; \mbox{ for all } g\in \mbox{PSL(2, Z\hspace{-.3em}Z})\;.
$$
Let $\Psi(z)$ be  an {\em automorphic function} of the modular group, i.e.
$$
\Psi(gz)=\Psi(z)\;,\;\;\mbox{ for all }g\in \mbox{ PSL}(2,
\mbox{Z\hspace{-.3em}Z})\;.
$$
Then it is easy to see that the function
$$
\Psi'(z)\equiv (T_p\Psi)(z)=\frac{1}{\sqrt{p}}\sum_{a,b,d}
\Psi(\frac{az+g}{d})
$$
where the summation is performed over all $ad=p, d>0, 0\leq b\leq d-1$ 
will also be an
automorphic function for the modular group. This is a consequence of the
fact that in the right-hand side of this expression there is effectively the
summation over all matrices from $M_p$. As $M_p$ goes not change after
multiplication by a modular group matrix $\Psi'(z)$ is an automorphic
function for the modular group. 
$(T_p\Psi)(z)$ is a kind of symmetrization of $\Psi(z)$ over images of $z$ by
all elements of $M_p$ and  the operators $T_p$ are called Hecke operators.

These operators form a {\em commutative algebra} with the following product
(see e.g. \cite{Terras})
\begin{equation}
T_nT_m=\sum_{d|(n,m)}T_{nm/d^2}
\label{tnm}
\end{equation}
where the summation is done over all divisors of the greatest common divisor
of $m$ and $n$.
The most important case corresponds to Hecke operators with prime indices
because all the others can be obtained from (\ref{tnm}).

When $p$ is a prime number 
$$
(T_p\Psi)(z)=\frac{1}{\sqrt{p}}\left [ \Psi(pz)+\sum_{0\leq
  j<p}\Psi\left (\frac{z+j}{p}\right )\right ]\;.
$$
Since Hecke operators involve only fractional transformations all them 
commute with the Laplace--Beltrami operator.
Consequently, if $\Psi(x,y)$ is an eigenfunction of the Laplace--Beltrami
operator, then $(T_p\Psi)(x,y)$ will also be an eigenfunction with the same 
eigenvalue. If there is no spectral
degeneracy (which strongly suggested by numerics) every eigenfunction of the
Laplace--Beltrami operator is in the same time an eigenfunction of all
Hecke operators
\begin{equation}
(T_p\Psi_n)(x,y)=\lambda_p(n)\Psi_n(x,y)\;.
\label{lambdaHecke}
\end{equation}
It is known (see e.g. \cite{Terras}) that eigenfunctions of the
Laplace--Beltrami operator for the modular group have the following  
Fourier expansion 
$$
\Psi_n(x,y)=y^{1/2}\sum_{p=-\infty}^{\infty} c_p(n) K_{s_n-1/2}(2\pi p y)e^{2\pi i p x}
$$
where the eigenvalue of the Laplace--Beltrami operator $E_n=s_n(s_n-1)$ and 
$K_{\nu}(x)$ is the Hankel function.

One has $z=x+iy$ and $(az+b)/d=(ax+b)/d+iay/d$, therefore
$$
(T_m\Psi_n)(x,y)=\frac{1}{\sqrt{m}}\sum_{a,b,d}
\left (\frac{ay}{d}\right )^{1/2}
\sum_{p}c_p(n)
  K_{s_n-1/2}\left (2\pi p\frac{ay}{d}\right )e^{2\pi i p(ax+b)/d}
$$
where the first summation is performed over all $a,b,d$ as in (\ref{lemma}).

The summation over $b$ gives zero if $d$ is not divide $p$. Otherwise
$$
(T_m\Psi_n)(x,y)=y^{1/2}\sum_{d|p,d|m}c_p(n)K_{s_n-1/2}(2\pi y pm/d^2)
e^{2\pi ipmx/d^2}\;.
$$
Let $k=m/d$ and $u=pm/d^2$. Then $p=mu/k^2$ and
$$
(T_m\Psi_n)(x,y)=y^{1/2}\sum_{u}\sum_{k|(m,u)} c_{mu/k^2}(n)K_{s_n-1/2}(2uy)
e^{2\pi iux}\;.
$$
If $T_m\Psi_n=\lambda_m(n)\Psi_n$ then by comparing the first Fourier
coefficient one gets
$$
c_m(n)=\lambda_m(n)c_1\;.
$$
Assuming $c_1\neq 0$ and using a convenient normalization $c_1=1$ one
concludes that {\em eigenvalues of the Hecke operators coincide with the Fourier 
coefficients}. 

We note also that similarly to the construction of the Selberg trace formula
one can build the trace formulas for Hecke operators (see e.g. \cite{BGGS}
and references therein). Such trace formula schematically has the form (cf.
(\ref{SelbergTrace})) 
\begin{eqnarray*}
\sum_{n}\lambda_p(n)h(k_n)&=&
\frac{1}{\sqrt{p}}\sum_{\mbox{\scriptsize hyperbolic}}
\frac{l_p}{2\sinh(L_p/2)} g(L_p)\\
&+&\mbox{smooth, parabolic, and elliptic terms}\;.
\end{eqnarray*}
Here $h(k)$ is a test function like in Sect.~\ref{Selberg} and $g(l)$ is its
Fourier transform.  
In the left-hand side the summation is performed over all
eigenvalues $E_n=k_n^2+1/4$ of the
Laplace--Beltrami operator and $\lambda_p(n)$ is the eigenvalue of the
Hecke operator $T_p$ (\ref{lambdaHecke}) applied to the eigenfunction of the 
Laplace--Beltrami
operator with eigenvalue $E_n$. In the right-hand side the summation is done
over all `hyperbolic' matrices from $M_p$ with $\mbox{ Tr }m_p\neq p+1$.
$L_p$ is the `length' associated with matrix $m_p$
$$
2\cosh (L_p/2)=|\mbox{ Tr }m_p|/\sqrt{p}
$$
and $l_p$ is the minimal length of  modular group matrices commuting with
$m_p$.

\section{The Jacquet--Langlands Correspondence}\label{JL}

Another curious fact about arithmetic groups is the 
Jacquet--Langlands correspondence (see \cite{Hejhal}) which claims
that for a arithmetic group derived from a quaternion group over 
$\mbox{l\hspace{-.47em}Q}$  (with a finite fundamental domain) one can find 
a subgroup of the modular group (with infinite fundamental domain) in
such a way that amongst all automorphic eigenvalues of the Laplace--Beltrami 
operator for this modular group subgroup one can find all eigenvalues of a 
compact arithmetic group.

The simplest arithmetic group $\Gamma$ derived from quaternion algebra over
$\mbox{l\hspace{-.47em}Q}$ with division is 
(see Sect.~\ref{QuaternionAlgebras})
$$
\Gamma= \left( \begin{array}{cc} k_1+k_2\sqrt{a}&k_3 +k_4\sqrt{a} \\
                   b (k_3 -k_4\sqrt{a}) & k_1-k_2\sqrt{a}  \end{array}
		 \right )
$$
where $b$ is a prime number, $a$ is an integer such that the equation
$x^2\equiv a\;(\mbox{ mod }b)$ has no integer solution (e.g. $a=3$, $b=5$), 
and   integers $k_i$ are such that
$$
\det (\gamma)=k_1^2-a k_2^2-bk_3^2+ab k_4^2=1\;.
$$
Denote $z=x+iy$, $\tau=u+iv$ $(y,v>0)$
and define for {\em all} $n_j$
$$
\alpha = n_1+n_2\sqrt{a}\;,\;\beta=n_3 +n_4\sqrt{a}\;,\;
$$
$$
\gamma= b (n_3 -n_4\sqrt{a})\;,\;\delta= n_1-n_2\sqrt{a}\;.
$$
Fix an arbitrary $z_0$ and compute the following kernel
$$
\Phi(\tau,z)=\sum_{n_j=-\infty}^{+\infty} 
\exp K(\tau,z)
$$
where
$$
K(\tau,z)= -\pi {\mbox Im }\tau 
\frac{|\alpha z_0+\beta-z(\gamma  \bar{z_0}+\delta)|^2}{\mbox{ Im }z
  {\mbox Im }z_0 }+
 2\pi i\bar{\tau}(\alpha\delta-\beta\gamma)\;.
$$
Here $\bar{z}$ is the complex conjugate of $z$.

Let $\psi_n(z)$ be an eigenfunction of the Laplace--Beltrami operator
automorphic with respect to the quaternion group $\Gamma$. It means that
\begin{itemize}
\item $\left (\Delta_{L-B}+E_n\right )\psi_n(z)=0\;,$
\item $\psi_n(Mz)=\psi_n(z)$ for all $M\in \Gamma$.
\end{itemize}
Then the function
$$
\Psi(\tau)={\mbox Im }\tau \int_{{\cal D}}
\Phi(\tau, z)\psi_n(z)\frac{dxdy}{y^2}
$$
where the integral is taken over the fundamental domain of the group $\Gamma$
is an eigenfunction the Laplace--Beltrami operator with the same eigenvalue
$E_n$ but 
automorphic with respect to the {\em congruence} subgroup of the modular
group $\Gamma_0(4ab)$ where
$$
\Gamma_0(N)=\left (\begin{array}{cc}m&n\\k&l\end{array} \right )
  \in \mbox{SL}(2, \mbox{Z\hspace{-.3em}Z})
$$
with an additional condition that
$$
 k\equiv 0 \;(\mbox{ mod } N)\;.
$$
Direct (but tedious) proof of this statement can be found in \cite{Hejhal}. 

\section{Non-arithmetic Triangles}\label{Nonarithmetic}

In the  precedent Section we have seen that arithmetic systems have the
Poisson spectral statistics. But what is about non-arithmetic models?

Let us consider, as example, the so-called {\em Hecke} triangles which are
hyperbolic triangles with angles  $(0,\pi/2,\pi/n)$.  All of them
tessellate the upper half-plane and  are fundamental domains of the
discrete groups generated by reflections across their sides.
The modular billiard is a part of them corresponding to $n=3$. Similar to it
they all have an infinite cusp.

According to Table~\ref{table1} the Hecke triangles are arithmetic only 
for $n=3, 4, 6,\infty$. All these arithmetic  triangles
have  an exponential degeneracies of  periodic orbit lengths  which leads to 
the Poisson-like statistics of energy levels. 

\begin{figure}
\center
\includegraphics[height=8cm, angle=-90]{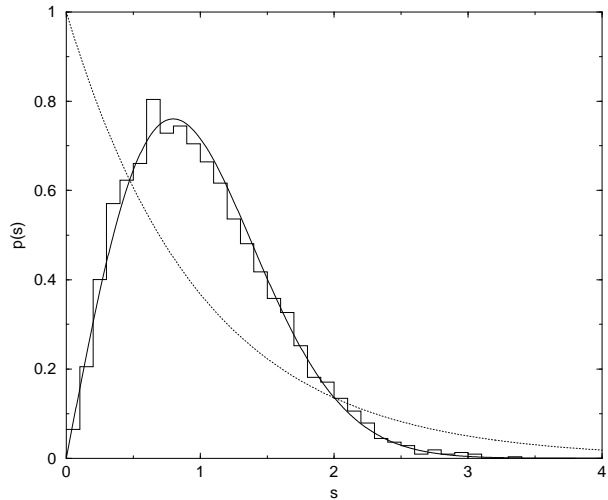}
\caption{
The nearest-neighbor distribution of 6000 energy levels for the non-arithmetic
 Hecke triangular billiard with $n=5$. The solid line -- the GOE prediction.
 Dotted line -- the Poisson result.}
\label{Hecke5}
\end{figure}

The simplest non-arithmetic Hecke triangle is the one with $n=5$. At
Fig.~\ref{Hecke5} we present the results of numerical calculations of the
nearest-neighbor distribution for 6000 first energy levels for this triangle
with the Dirichlet boundary conditions. For others Hecke triangles one gets
similar pictures.  It is clearly seen that numerics
agrees very well with the predictions of the Gaussian Orthogonal ensembles
of random matrices as it should be for generic chaotic models. 

But what are  the multiplicities of periodic orbit lengths for
non-arithmetic Hecke triangles? 
As these model are not-arithmetic,  one would expect that their length 
multiplicities should be equal to two as for generic time-reversal invariant
systems. Nevertheless numerical calculations (see \cite{BGGS} for details) 
demonstrated that this is not always the case. 
\begin{figure}
\center
\includegraphics[height=8cm, angle=-90]{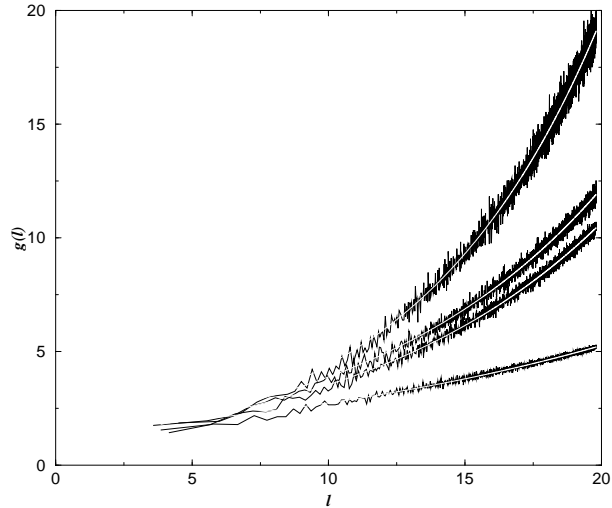}
\caption{Mean length multiplicities of periodic orbits for
the Hecke triangles with  (from top to bottom) $n=12$, $n=5$, $n=8$, and 
$n=10$. White lines are numerical fits (\ref{fits}).}
\label{pl}
\end{figure}
At Fig.~\ref{pl} we present the numerically computed mean length
multiplicities for the Hecke triangles  with $n=5,\;8,\;10,\;12$ for lengths
$l<20$. White lines indicate a two-parameter fit to these data in the form
$\bar{g}(l)=a_ne^{b_nl}$
\begin{eqnarray}
n&=&5\; \;:\; \bar{g}(l)\approx 1.235e^{.114 l}\;,\nonumber\\
n&=&8\; \;:\; \bar{g}(l)\approx 1.095e^{.114 l}\;,\label{fits}\\
n&=&10\;:\; \bar{g}(l)\approx 1.143e^{.065 l}\;,\nonumber\\
n&=&12\;:\; \bar{g}(l)\approx .986e^{.150 l}\;.\nonumber
\end{eqnarray}
These expressions fit numerical data in the given interval of lengths quite
well and indicate that for, at least, certain Hecke triangles mean length
multiplicity increases exponentially. We stress that (\ref{fits})  are only 
the best least-square numerical fits and no attempts were made to determine 
the accuracy of coefficients. 

The discussion of the origin of such unexpected multiplicities for 
non-arithmetic triangles is beyond the scope of these lectures 
(on this subject see \cite{unpublished}). However it is of interest to 
understand why exponentially large multiplicities of periodic orbit 
lengths do not contradict the observed GOE behaviour of spectral 
statistics (cf. Fig.~\ref{Hecke5}).

Assume that a system has an exponentially large number of periodic
orbits with the same length $l$ increasing as 
$$
g(l)\sim \frac{\E^{\lambda l}}{l}
$$
with $\lambda \leq 1/2$. 
 
Let us repeat the arguments of Sect.~\ref{Criterion} for this case with 
exact degeneracicies. In Sect.~\ref{Criterion} it was demonstrated that 
periodic orbits with different lengths can be treated in the diagonal approximation
if 
\begin{equation}
l_{p_1}-l_{p_2}\gg \frac{k}{\Delta E}
\label{lp1lp2}
\end{equation}
where $k$ is the momentum and $\Delta E$ is the width of the energy average
inherent in the definition of correlation functions of dynamical systems.

As the density of orbits with {\em different} lengths is
$$
\rho_{\mbox{\scriptsize diff. lengths}}\approx \frac{\E^{l}}{g(l)l}
\sim \E^{(1-\lambda)l}
$$
it follows that the inequality (\ref{lp1lp2}) is valid till maximal length
\begin{equation}
l_{m}\sim \frac{1}{1-\lambda}\ln \frac{\Delta E}{k}
\sim \frac{1}{1-\lambda}\ln k\;.
\label{maxl}
\end{equation}
Notice that due to assumed large multiplicity $l_{m}$ is different 
from (\ref{LMax}). 

From (\ref{diagK}) it follows that the two-point correlation form factor
in the diagonal approximation up to numerical factor is
$$
K(t) \sim \frac{k}{l}|A(l)|^2g(l)\E^{l}
$$
where $t=l/2k$ and $A(l)\sim l\E^{l/2}/k$. Combining all terms together 
one obtains that during the maximal time of applicability of the 
diagonal approximation $t_m=l_m /2k$ with $l_m$ from (\ref{maxl})
the form factor increases till
$$
K(t_m)\sim \frac{\E^{\lambda l_m}}{k}\sim k^{(2\lambda -1)/(1-\lambda)}\;.
$$
Hence, if $\lambda=1/2$ as for arithmetic systems the two-point correlation
form factor during the time of validity of the diagonal approximation 
increases till a constant value of the order of 1. But if $\lambda<1/2$
the form factor for the time of validity can reach only a value
of the order of $k^{-\nu}$ with $\nu= (1-2\lambda)/(1-\lambda)>0$. As 
$k\to\infty$ this value tends to zero and no apparent contradiction 
with standard random matrix ensembles can be derived within the diagonal
approximation.   

\section{Summary}

Arithmetic groups are a special sub-class of discrete groups characterized
by the existence of a representation by matrices with integer elements. A
readable mathematical review of such groups is given in \cite{Katok}. There
are two types of arithmetic groups. The first includes groups commensurable
with the modular group and having non-compact fundamental domains with infinite
cusps. The second type of compact arithmetic groups combines groups
commensurable with groups derived from  quaternion algebras with division.
These groups have finite fundamental domains.

From classical viewpoint the free motion on surfaces generated by arithmetic
groupes is as chaotic as for any hyperbolic surfaces. But quantum mechanics 
on these arithmetic surfaces is very special. In particular, spectral
statistics of the Laplace--Beltrami operator automorphic with respect to
arithmetic group is described by the Poisson statistics typical for integrable
systems and not by the random matrix statistics typical for chaotic models.

The origin of this peculiarity can be traced to the existence in arithmetic
systems of a very large number of periodic orbits with exactly the same
length. For all arithmetic groups the mean multiplicity of periodic
orbits with length $l$ behaves like $\E^{l/2}/l$. This has to be compared
with the total density of periodic orbits which for all discrete groups is
$\E^l/l$. It is the cumulative effect of the interference of many periodic
orbits with the same length which changes drastically the spectral statistics.

In the diagonal approximation the two-point correlation form factor $K(t)$
for arithmetic systems at small $t$ increases exponentially like $\E^{kt}/k$ and
during the Ehrenfest time (which is the limit of applicability of the diagonal
approximation) reaches a constant value.

More detailed information can be obtained for the modular group where it is
possible to compute the two-point correlation form factor  analytically. The
final answer is 
$$
K(t)={1\over 2\pi^2k}\sum_{(p,q)=1}\left|{q\over p}\beta(p,q)\right|^2
\delta(t-t_{p,q})
$$
where 
$$
t_{p,q}={1\over k}\ln{kq\over\pi p}
$$
and $\beta(p,q)$ is a number-theoretical function given in Sect.~\ref{Explicit}.

This formula means that the two-point form factor for the modular group is a
sum over $\delta$-functions at  special points $t_{p,q}$ situated in a
vicinity of the Ehrenfest time. The set of $\delta$-functions is dense but the
largest peaks correspond to the smallest ratios $p/q$. Nevertheless, small peaks
with $p/q\ll 1$ are much more numerous and integrally they dominate. In
the limit $t$ fixed and $k\to\infty$ $K(t)\to \bar{d}$ thus confirming the Poisson
nature of the spectral statistics of the modular group. 

Arithmetic groups have many interesting properties. Hecke operators and  
the Jacquet--Langlands correspondence are the most remarkable. 
\pagebreak

\vspace{1ex}

{\bf Acknowledgement} 

\vspace{1ex}

It is a pleasure to thank Charles Schmit for his aid in the preparation of
these lectures and A.M. Odlyzko for presenting numerical data of the two-point
correlation function of Riemann zeros.

\end{document}